\documentclass[twocolumn]{aastex7}

\begin{document}

\title{An All-sky Survey of White Dwarf Merger Remnants: Far-UV is the Key}

\author[orcid=0000-0001-6098-2235,sname='Kilic']{Mukremin Kilic} 
\affiliation{Homer L. Dodge Department of Physics and Astronomy, University of Oklahoma, 440 W. Brooks St., Norman, OK, 73019 USA}
\email[show]{kilic@ou.edu}

\author[orcid=0000-0003-2368-345X,sname='Bergeron']{Pierre Bergeron}
\affiliation{D\'epartement de Physique, Universit\'e de Montr\'eal, C.P. 6128, Succ. Centre-Ville, Montr\'eal, QC H3C 3J7, Canada}
\email[show]{bergeron@astro.umontreal.ca}

\author[orcid=0000-0002-4462-2341,sname='Brown']{Warren R.\ Brown}
\affiliation{Center for Astrophysics, Harvard \& Smithsonian, 60 Garden Street, Cambridge, MA 02138 USA}
\email{wbrown@cfa.harvard.edu}

\author[orcid=0000-0002-9632-1436,sname='Blouin']{Simon Blouin}
\affiliation{Department of Physics and Astronomy, University of Victoria, Victoria BC V8W 2Y2, Canada}
\email{sblouin@uvic.ca}

\author[orcid=0009-0009-9105-7865,sname='Jewett']{Gracyn Jewett}
\affiliation{Homer L. Dodge Department of Physics and Astronomy, University of Oklahoma, 440 W. Brooks St., Norman, OK, 73019 USA}
\email{gjewett@ou.edu}

\author[orcid=0000-0001-7143-0890,sname='Moss']{Adam Moss}
\affiliation{Homer L. Dodge Department of Physics and Astronomy, University of Oklahoma, 440 W. Brooks St., Norman, OK, 73019 USA}
\email{Adam.G.Moss-1@ou.edu}

\author[orcid=0000-0003-4609-4500,sname='Dufour']{Patrick Dufour}
\affiliation{D\'epartement de Physique, Universit\'e de Montr\'eal, C.P. 6128, Succ. Centre-Ville, Montr\'eal, QC H3C 3J7, Canada}
\email{patrick.dufour@umontreal.ca}

\author[orcid=0000-0002-7729-484X,sname='Vincent']{Olivier Vincent}
\affiliation{D\'epartement de Physique, Universit\'e de Montr\'eal, C.P. 6128, Succ. Centre-Ville, Montr\'eal, QC H3C 3J7, Canada}
\email{o.vincent@umontreal.ca}

\begin{abstract}

The majority of merging white dwarfs leave behind a white dwarf remnant.
Hot/warm DQ white dwarfs with carbon-rich atmospheres have high masses and unusual kinematics. All evidence
points to a merger origin. Here, we demonstrate that far-UV + optical photometry provides an efficient way to identify these merger remnants.
We take advantage of this photometric selection to identify 167 candidates in the GALEX All-Sky Imaging Survey footprint, and provide
follow-up spectroscopy.  Out of the 140 with spectral classifications, we identify 75 warm DQ white dwarfs with $T_{\rm eff}>10,000$ K, nearly
tripling the number of such objects known. Our sample includes 13 DAQ white dwarfs with spectra dominated by hydrogen and (weaker) carbon lines.
Ten of these are new discoveries, including the hottest DAQ known to date with
$T_{\rm eff}\approx23,000$ K and $M=1.31~M_{\odot}$. We provide a model atmosphere analysis of all warm DQ white dwarfs found, and present their
temperature and mass distributions. The sample mean and standard deviation are $T_{\rm eff} = 14,560 \pm 1970$ K and $M=1.11 \pm 0.09~M_{\odot}$. Warm DQs
are roughly twice as massive as the classical DQs found at cooler temperatures. All warm DQs are found on or near the crystallization sequence.
Even though their estimated cooling ages are of order 1 Gyr, their kinematics indicate an origin in the thick disk or halo. Hence, they are likely
stuck on the crystallization sequence for $\sim$10 Gyr due to significant cooling delays from distillation of neutron-rich impurities. Future all-sky
far-UV surveys like UVEX have the potential to significantly expand this sample.

\end{abstract}

% https://astrothesaurus.org

\section{Introduction}

Stellar mergers are relatively common; binary population synthesis calculations predict that about one in four white dwarfs
in the solar neighborhood come from mergers \citep{temmink20}. More importantly, the predicted merger fraction increases significantly
for higher mass white dwarfs, and it may be as high as $\sim$40\% for $M_{\rm WD}\geq0.9~M_{\odot}$, depending on the efficiency
of the common envelope ejection. The majority of these mergers are non-explosive, hence they leave behind a remnant that eventually turns into a single white dwarf.

Merger remnants may display rapid rotation \citep{schwab21,caiazzo21}, magnetism \citep{garciaberro12,bagnulo21},
or unusual composition \citep[e.g.,][]{kilic23a,caiazzo23,jewett24}. Hot and warm DQ white dwarfs with carbon-rich atmospheres top the list of potential
merger remnants due to their unusual atmospheric composition, kinematics, and high masses
\citep{dunlap15,coutu19,kawka23}. 

G35-26 and G227-5 were the first discoveries of hot white dwarfs that primarily show atomic carbon lines in their spectra \citep{liebert83,wegner85}.
With a significantly larger sample of spectroscopically confirmed white dwarfs from the Sloan Digital Sky Survey Data Release 4 \citep{eisenstein06}, 
\citet{dufour07,dufour08} identified 9 hot DQs with $T_{\rm eff}\approx18,000-24,000$ K and  atmospheres primarily composed of carbon. 
Carbon-atmosphere white dwarfs are rare; \citet{koester19} found only 26 warm DQs in a sample of 20,088 spectroscopically confirmed white dwarfs
in the SDSS Data Release 14 \citep{kepler19}. All 26 warm DQs in that study are relatively massive with $M= 0.86-1.19~M_{\odot}$.

Gaia Data Release 2 \citep{gaia18} made it possible to create volume-limited samples of massive white dwarfs. \citet{jewett24} obtained 
spectroscopy of all candidates with $M>0.9~M_{\odot}$ and $T_{\rm eff}\geq11,000$ K in the Montreal White Dwarf Database\footnote{https://www.montrealwhitedwarfdatabase.org/} \citep[MWDD,][]{dufour07}
100 pc sample and the Pan-STARRS footprint. Among the 204 objects in their sample, 20 stars (or nearly 10\%) turned out to be warm DQs,
including several DAQ white dwarfs that show hydrogen and carbon absorption lines \citep{liebert83,hollands20,kilic24}. The discovery of the DAQ
J0655+2939 in their paper is especially interesting; this object was identified as an outlier in GALEX far-UV (FUV) data by \citet{wall23}. J0655+2939
is about two times fainter than expected in the FUV compared to the predictions for a pure H atmosphere model. Through a high quality spectrum
obtained at the 6.5m MMT, \citet{jewett24} detected weak atomic carbon lines and demonstrated that this is indeed a DAQ white dwarf with a hydrogen and
carbon atmosphere. 

\begin{figure}
\includegraphics[width=3.4in]{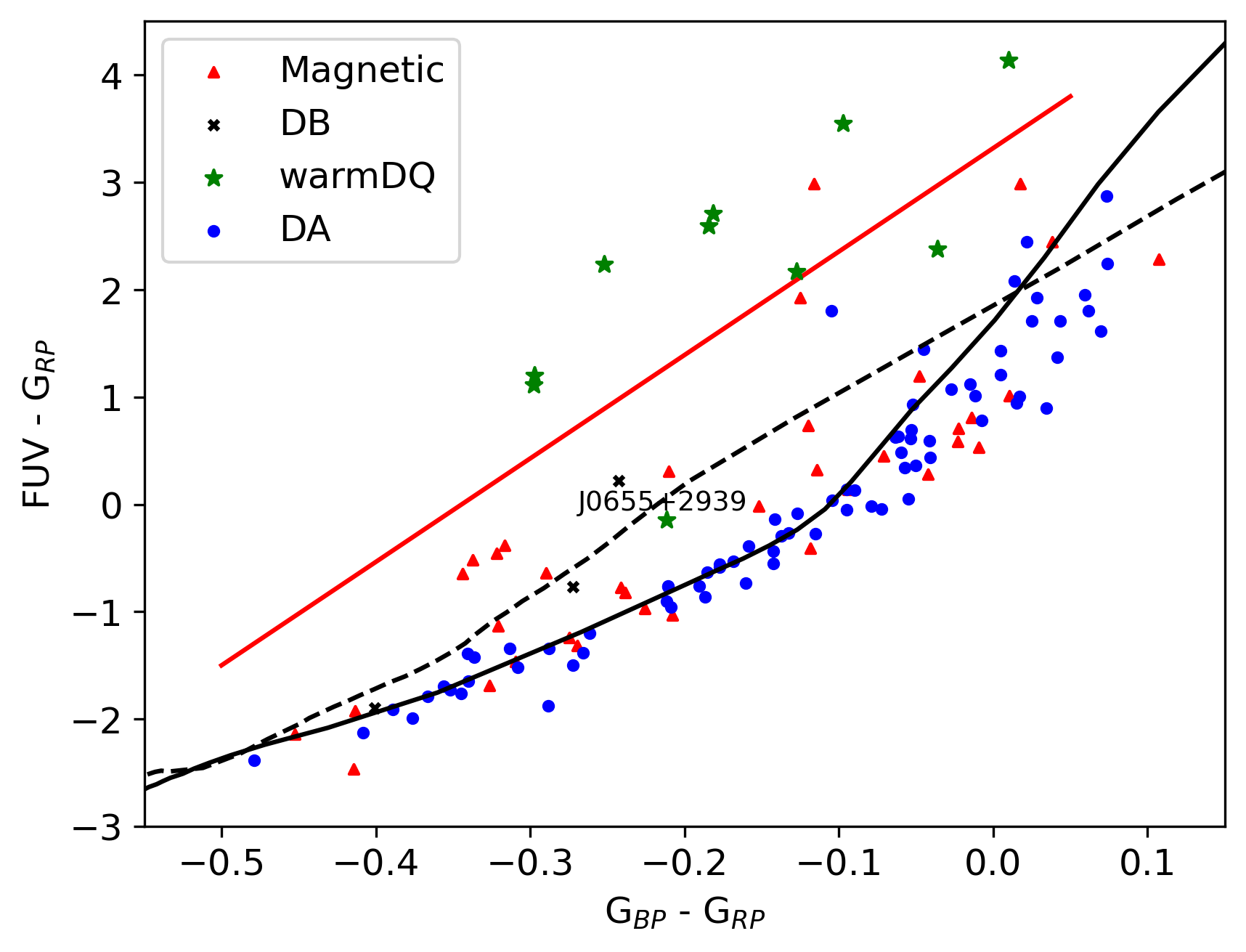}
\caption{UV-optical color-color diagram of the massive white dwarf sample from \citet{jewett24}. DA, DB, magnetic white dwarfs, and warm DQs
are represented by different symbols. Solid black and dashed lines show the evolutionary sequences for $M=1.1~M_{\odot}$ pure H and pure He atmosphere white dwarfs, respectively. Warm DQs stand out in their FUV colors. The red line shows the color-selection used in our study, which
is discussed in Figure \ref{figsample}. }
\label{figgra} 
\end{figure}

Figure \ref{figgra} shows an ultraviolet-optical color-color diagram of the massive white dwarf sample from \citet{jewett24} based on Gaia and GALEX
photometry. Because this study was restricted to $T_{\rm eff}\geq11,000$ K, the objects are limited to $G_{\rm BP}-G_{\rm RP}\leq0.1$. Solid black and dashed
lines show the cooling sequences for $M=1.1~M_{\odot}$ pure H and pure He atmosphere white dwarfs, respectively.
In addition to DA white dwarfs, many of the magnetic white dwarfs also land on the pure H atmosphere sequence. However, some of the magnetic white dwarfs
land on or close to the pure He atmosphere evolutionary sequence, likely because strong fields shift and distort their absorption lines such that their broad-band colors are impacted. More interestingly, warm DQs (green stars) stand out in their colors. For example, J0655+2939 (labeled in the Figure) is offset by about a magnitude from typical
DA white dwarfs at the same temperature (or $G_{\rm BP} - G_{\rm RP}$ color), but the other warm DQs in this figure are even more strongly offset in their FUV
colors due to the additional opacity from carbon. The red line shows the target selection used in this study, which is discussed below. Eight out of
nine stars above the red line are indeed warm DQs.  

\begin{figure}
\includegraphics[width=3.4in]{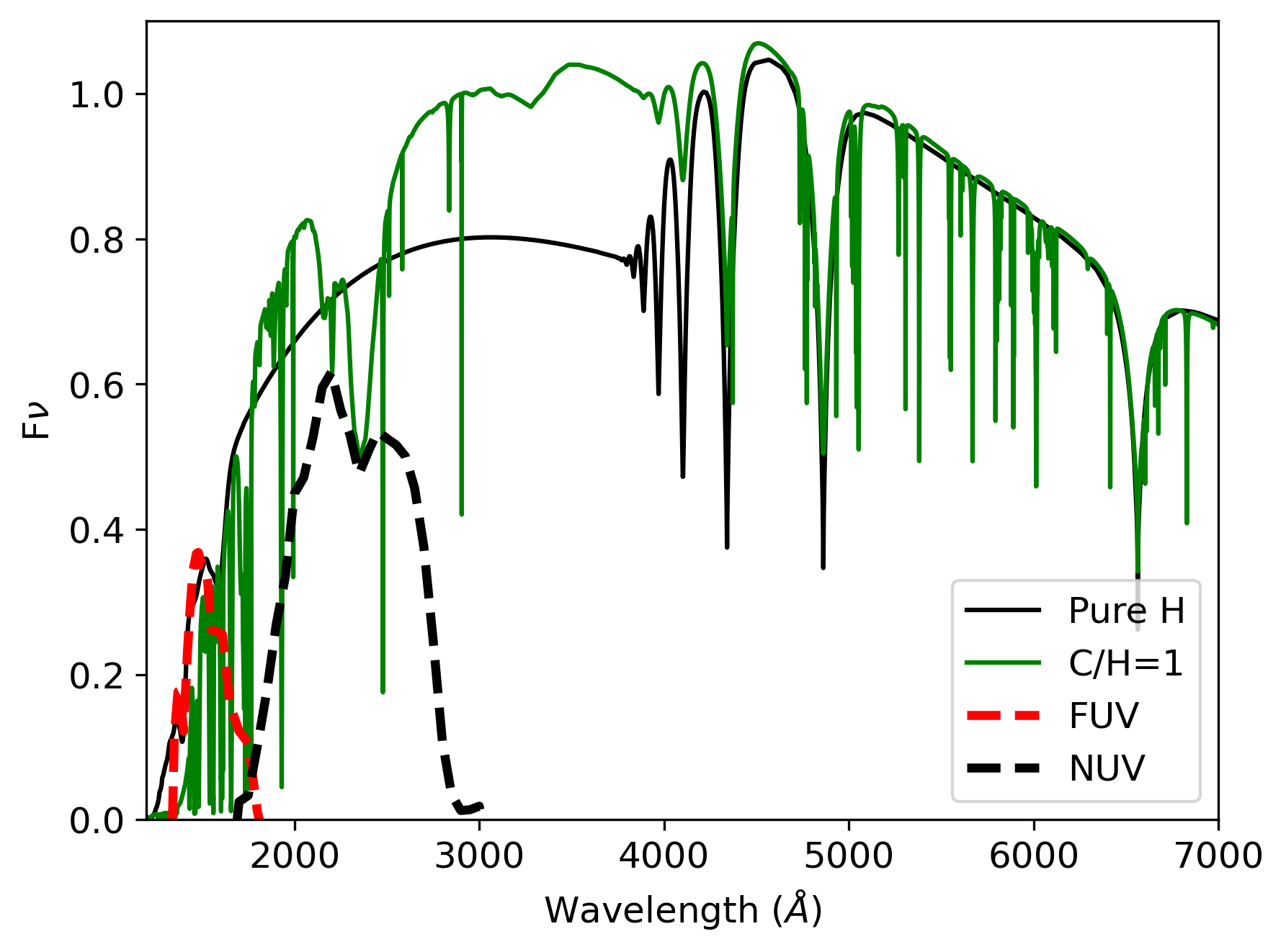}
\caption{Model spectra for $T_{\rm eff}=13,000$ K and $\log{g}=9.0$ white dwarfs with a pure H (solid black)
and mixed C/H composition (C/H = 1, green). The dotted lines show the transmission profiles of the GALEX
FUV and NUV bands.}
\label{figmodel} 
\end{figure}

Figure \ref{figmodel} shows the model spectra for $T_{\rm eff}=13,000$ K and $\log{g}=9.0$ white dwarfs with a pure H (solid black line)
and a mixed C/H atmosphere with equal amounts of carbon and hydrogen (green line). Besides the narrow atomic carbon features, the spectra have remarkably similar continua
in the optical. However, carbon has numerous absorption lines in the UV, especially in the FUV. The dashed lines show the GALEX FUV and NUV
filter transmission profiles. Clearly, the additional carbon opacity in the FUV makes warm DQ white dwarfs significantly fainter than expected compared
to carbon-free models. This is the reason why FUV photometry is key for identifying hot/warm DQs through photometry \citep[also see Figure 4 in][]{dufour08}. 

Here we take advantage of this
photometric selection to perform an all-sky survey of warm DQ white dwarfs in the GALEX All-Sky Imaging Survey footprint. Section \ref{sample}
presents our target selection and observations, whereas Section \ref{model} provides the details of our model atmosphere analysis. We discuss
the entire sample of warm DQs found in our survey in Section \ref{res}, and present their temperature and mass distributions, and our conclusions in
Section \ref{con}.

\section{All-Sky Survey Sample Selection and Observations}
\label{sample}

To create a clean white dwarf sample where we can constrain the physical parameters of each object reliably, we start with all objects with
probability of being a white dwarf $P_{\rm WD}>0.9$ and a distance $\leq300$ pc from the Gaia EDR3 white dwarf catalog of \citet{gentile21}.
GALEX obtained FUV and NUV observations of 24,790 square degrees as part of its All-Sky Imaging survey between 2003 and 2009. We
propagate the Gaia DR3 positions back to the GALEX epoch based on Gaia astrometry. We then search the Revised Catalog of GALEX
Ultraviolet Sources \citep[GUVcat,][]{bianchi17} for matching sources within 3 arcsec of the predicted positions. We also cross-match
this sample with the SDSS, Pan-STARRS, and SkyMapper surveys to collect additional photometry for our model
atmosphere analysis. 

\begin{figure}
\includegraphics[width=3.4in]{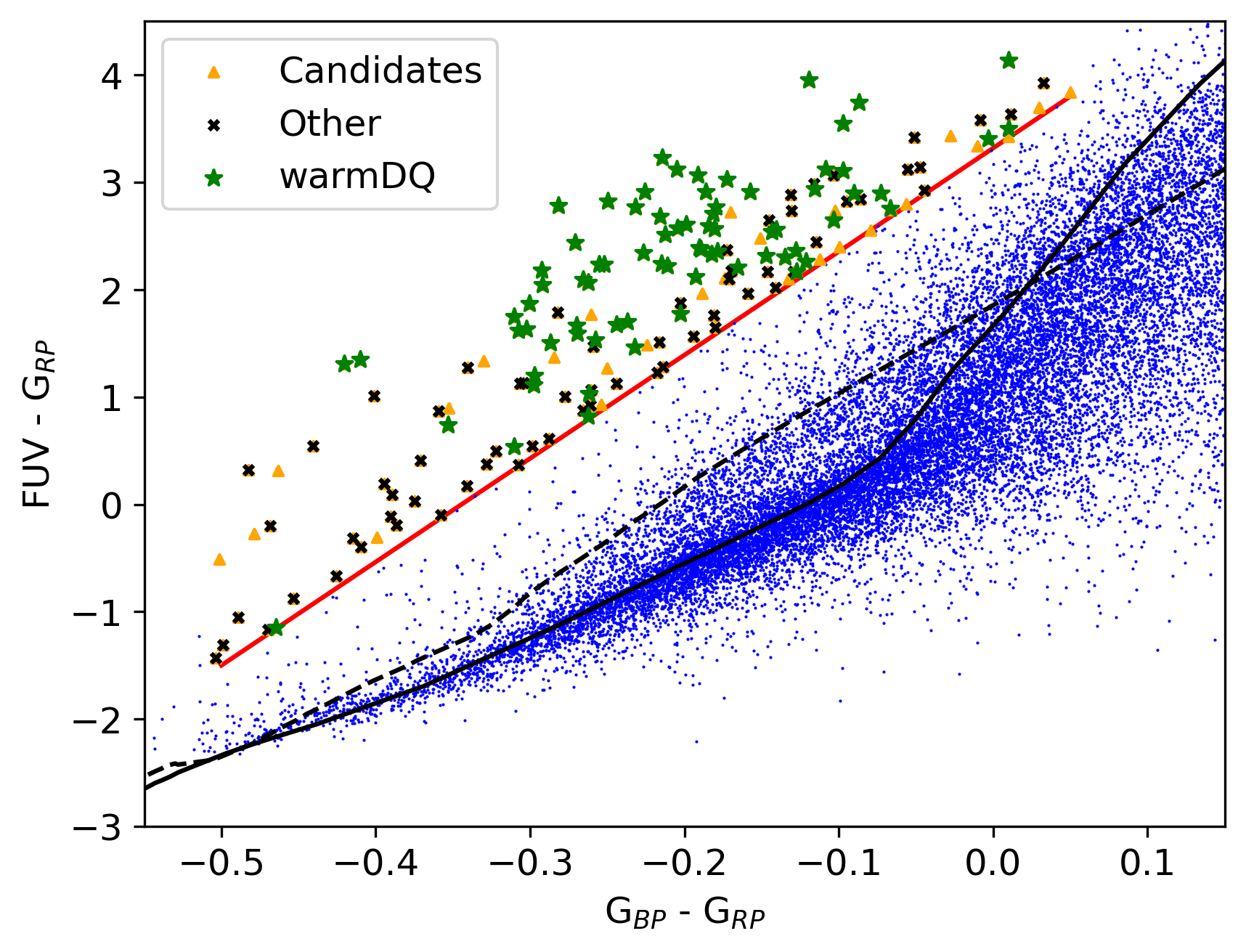}
\caption{Gaia-GALEX color-color diagram of all white dwarfs with $P_{\rm WD}>0.9$ from the \citet{gentile21} catalog with distance $\leq300$ pc
and GALEX FUV photometry. Solid black and dashed lines show the evolutionary sequences for $M=0.6~M_{\odot}$ pure H and pure He atmosphere white dwarfs, respectively. 
Sources above the red line are selected for spectroscopic follow-up. Spectroscopically confirmed warm DQs are
marked by green stars. Crosses mark the remaining targets, non-DQs, including magnetic white dwarfs.}
\label{figsample} 
\end{figure}

Figure \ref{figsample} shows a Gaia-GALEX color-color diagram of this sample (blue points). Given the location of the warm DQs in the 100
pc sample shown in Figure \ref{figgra}, and in order to limit the sample size to less than 200 targets for follow-up observations, we chose targets
with ${\rm FUV}-G_{\rm RP} > (9.63636\ (G_{\rm BP} - G_{\rm RP}) + 3.318182)$  and $G_{\rm BP} - G_{\rm RP}=-0.5$ to +0.05. This selection is not meant to find all warm
DQs in the GALEX All-Sky Imaging Footprint, as it is a formidable task. Instead, it is designed to explore the use of GALEX + Gaia photometry
to identify the most significant outliers in the FUV, and significantly increase the current sample of warm DQs known.

Our selection results in a sample of 167 candidates, including 46 with spectral classifications
available in the literature: 41 of these have their spectra available in the MWDD. These 46 previously
known white dwarfs include 25 warm DQs. The remaining 21 targets are a mix of normal or magnetic DA, DB, DZ, and He-DA white dwarfs. This
leaves us with 121 targets for follow-up spectroscopy. 

We obtained follow-up optical spectroscopy of 40 candidates using the 8m Gemini North and South telescopes equipped
with the Gemini Multi-Object Spectrograph (GMOS) as part of the queue programs GS-2023B-Q-329 (2 targets), GS-2024A-Q-330 (2 targets),
GS-2024A-Q-331 (1 target), GS-2024B-Q-305 (13 targets), GS-2024B-Q-405 (1 target), GS-2025A-Q-205 (2 targets), and GN-2025A-Q-205 (19
targets). We used the B600 grating for GS-2023B-Q-329, but switched to the B480 grating for the other programs.
With a 1$\arcsec$ slit, the B600 and B480 gratings provide spectra with resolving powers of $R=844$ and 761, respectively.

We obtained follow-up spectroscopy of 22 candidates using the 6.5-m MMT equipped with the Blue Channel Spectrograph \citep{schmidt89},
500 l mm$^{-1}$ grating, and a $1.25\arcsec$ slit. This setup provides spectra over the wavelength range 3700 - 6850 \AA\ with a spectral
resolution of 4.8 \AA.  
We obtained spectroscopy of 21 additional targets at the 6.5m Magellan telescope with the MagE spectrograph. We used the $0.85\arcsec$ slit, providing
wavelength coverage from about 3400 \AA\ to 9400 \AA\ with a resolving power of $R = 4800$.
The reduced spectra from our Gemini, MMT, and Magellan programs are publicly available on the MWDD.
For 11 targets, we make use of optical spectra provided by Data Release 1 of the Dark Energy Spectroscopic Instrument \citep[DESI,][]{desi25}. DESI  provides spectra over the wavelength range 3600-9800 \AA\ with a dispersion of 0.8 \AA\ per pixel. 

In total, we were able to obtain spectral classification for 140 of the 167 stars in our sample, or 84\%. 
We find 75 warm DQ white dwarfs, including 13 DAQ white dwarfs where the Balmer
lines are stronger than the carbon lines in the optical spectra. Note that 10 of these DAQs are new discoveries. Table \ref{tabdq} provides observational
parameters and spectral types for our warm DQ sample. These objects are also marked by green stars in
Figure \ref{figsample}.  Our survey has a 54\% success rate in identifying warm DQs, which demonstrates the power of FUV photometry. Through our survey, we nearly tripled the number of warm DQs known with $T_{\rm eff}>10,000$ K.

\begin{deluxetable*}{lrrccclc}
\tabletypesize{\tiny}
\tablecolumns{8} \tablewidth{0pt}
\tablecaption{Warm DQ white dwarfs in our sample.\label{tabdq}} 
\tablehead{\colhead{Name} & \colhead{Gaia DR3 SourceID} & \colhead{$d$ (pc)} & \colhead{$G_{\rm BP} - G_{\rm RP}$} & \colhead{FUV$-$NUV} & \colhead{$\zeta^a$} & \colhead{Type} & \colhead{Source}}
% &  & (pc) & (mag) & (mag) &  & & }
\startdata
WDJ000638.90+081247.04 & 2752056099423455744 & 78.1 & +0.01 & 2.93 & 12.96 & DQA & MMT \\
WDJ001423.65+302214.88 & 2860750657849167232 & 229.3 & $-$0.21 & 2.26 & 13.39 & DAQ & MMT \\
WDJ003658.72$-$232257.31 & 2348262202173337216 & 183.7 & $-$0.18 & 2.44 & 13.28 & DQ & \nodata \\
WDJ004430.27$-$451348.26 & 4979462766611010304 & 160.4 & $-$0.20 & 2.10 & 13.00 & DAQ & Gemini \\
WDJ004527.50$-$233629.29 & 2348747743931814656 & 47.2 & +0.01 & 3.32 & 13.26 & DQ & \nodata \\
WDJ011550.57+502545.98 & 403724902897548160 & 140.9 & $-$0.27 & 1.93 & 13.16 & DQA & MMT  \\
WDJ011629.20$-$150147.25 & 2454979769973042176 & 204.9 & $-$0.20 & 2.39 & 13.21 & DQA & DESI \\
WDJ014656.90$-$042613.50 & 2480243386083359360 & 108.6 & $-$0.00 & 2.82 & 12.97 & DQA & DESI \\
WDJ015756.65$-$051541.73 & 2491726994761425792 & 242.9 & $-$0.23 & 2.85 & 13.30 & DQ: & DESI \\
WDJ020549.70+205707.96 & 94276941624384000 & 85.5 & $-$0.30 & 1.81 & 13.08 & DAQ & MWDD \\
WDJ023633.74+250348.86 & 126173494772886144 & 176.2 & $-$0.21 & 3.27 & 12.91 & DQA & SDSS \\
WDJ024459.25+423011.43 & 337111540683814528 & 108.0 & $-$0.19 & 2.95 & 13.12 & DQA & MMT \\
WDJ024524.49$-$251209.70 & 5076733910323983872 & 191.8 & $-$0.20 & 2.97 & 13.09 & DQ & Magellan \\
WDJ025719.30$-$614033.08 & 4721812110929376768 & 103.5 & $-$0.16 & 2.64 & 13.13 & DQA & Gemini \\
WDJ030446.58$-$642628.33 & 4720049658213446400 & 129.7 & $-$0.14 & 2.46 & 12.56 & DAQ & Magellan \\
WDJ031054.74+220137.77 & 108837907254684672 & 107.2 & $-$0.19 & 2.42 & 13.08 & DQA & MMT \\
WDJ032521.11+254038.62 & 117261167051122688 & 128.9 & $-$0.29 & 2.35 & 13.21 & DAQ & MMT \\
WDJ033845.40$-$534428.60 & 4732621341021719552 & 220.4 & $-$0.27 & 2.50 & 12.82 & DQA & Gemini \\
WDJ034140.15$-$252505.02 & 5082168658860941824 & 297.4 & $-$0.26 & 2.10 & 13.20 & DQA & Magellan \\
WDJ042728.55$-$320603.04 & 4872425442789515776 & 169.7 & $-$0.18 & 2.19 & 13.08 & DQA & Gemini \\
WDJ044413.03$-$431509.09 & 4814772069550031104 & 154.7 & $-$0.23 & 1.86 & 12.97 & DAQ & Gemini \\
WDJ053407.26$-$412843.94 & 4806168665940960128 & 224.5 & $-$0.26 & 1.98 & 13.54 & DQA & Gemini \\
WDJ082329.59$-$085629.47 & 5753067293803129728 & 138.8 & $-$0.24 & 2.15 & 13.11 & DAQ & Gemini \\
WDJ083135.57$-$223133.63 & 5702793425999272576 & 81.9 & $-$0.13 & 2.42 & 13.02 & DAQ & \citet{kilic24} \\
WDJ090157.91+575135.91 & 1037553185478639360 & 154.5 & $-$0.21 & 2.65 & 12.98 & DQZA & SDSS \\
WDJ090229.67+303958.30 & 711615367289960320 & 191.0 & $-$0.18 & 2.54 & 12.88 & DQA & SDSS \\
WDJ091723.70+654107.17 & 1067868335963881216 & 212.3 & $-$0.21 & 2.57 & 13.08 & DAQ & MMT \\
WDJ091851.74$-$052536.93 & 5758460978157729408 & 182.6 & $-$0.12 & 2.64 & 13.00 & DQA & Gemini \\
WDJ093026.42+641436.95 & 1064480774998488960 & 254.0 & $-$0.26 & 1.23 & 13.35 & DQ & MMT \\
WDJ095837.16+585303.16 & 1049528378933767424 & 177.7 & $-$0.26 & 1.62 & 13.02 & DAQ & SDSS \\
WDJ100911.54$-$215821.87 & 5666458346271348992 & 205.6 & $-$0.30 & 2.31 & 12.97 & DAQ & Magellan \\
WDJ103655.39+652252.01 & 1059498612934930176 & 173.0 & $-$0.27 & 2.76 & 12.83 & DQH & SDSS \\
WDJ104906.61+165923.72 & 3982534641696914688 & 195.5 & $-$0.15 & 2.10 & 13.12 & DQA & SDSS \\
WDJ110058.04+175807.15 & 3983742627018262656 & 150.6 & $-$0.11 & 2.83 & 12.94 & DQA & SDSS \\
WDJ111811.89$-$392705.93 & 5395968560168183040 & 83.1 & $-$0.09 & 2.90 & 12.81 & DQA & Gemini \\
WDJ112538.72$-$311247.73 & 3481710531219239296 & 137.4 & $-$0.14 & 2.40 & 12.89 & DQA & Magellan \\
WDJ113918.84$-$391857.31 & 5384539476100968320 & 149.9 & $-$0.35 & 1.47 & 13.14 & DAQ & Gemini \\
WDJ120331.90+645101.41 & 1585063422960992256 & 87.1 & $-$0.10 & 3.26 & 13.04 & DQA & SDSS \\
WDJ123549.68$-$380203.98 & 6156465156413726464 & 140.4 & $-$0.23 & 2.44 & 13.12 & DQZ & Magellan \\
WDJ125938.32$-$260158.23 & 3497239930370525056 & 154.1 & $-$0.19 & 2.35 & 13.11 & DQA & Magellan \\
WDJ130717.12$-$375839.08 & 6142428481734335360 & 193.6 & $-$0.19 & 2.19 & 13.07 & DQA & Gemini \\
WDJ133151.40+372755.15 & 1475194238223608064 & 135.2 & $-$0.27 & 2.04 & 13.09 & DQA & SDSS \\
WDJ133941.33+015749.19 & 3663639779599974784 & 254.1 & $-$0.18 & 2.65 & 12.62 & DQAH & Gemini \\
WDJ143437.84+225859.68 & 1242732500581829632 & 195.6 & $-$0.29 & 2.43 & 13.07 & DQA & SDSS \\
WDJ143534.18+043425.70 & 3668901977825959040 & 264.7 & $-$0.24 & 2.10 & 12.64 & DQA & SDSS \\
WDJ144854.90+051903.80 & 1159147901516125440 & 120.3 & $-$0.31 & 2.07 & 13.12 & DQA & SDSS \\
WDJ145524.89+420910.81 & 1489435937460963712 & 269.1 & $-$0.22 & 2.76 & 12.89 & DQA & SDSS \\
WDJ153450.85+354034.09 & 1374533368881401728 & 258.8 & $-$0.29 & 1.89 & 12.47 & DQA & SDSS \\
WDJ160514.47$-$075528.74 & 4348642017695591552 & 146.6 & $-$0.13 & 1.97 & 13.06 & DQA & Gemini \\
WDJ162236.25+300455.29 & 1318204460477280512 & 74.8 & $-$0.25 & 2.58 & 13.00 & DQA & SDSS \\
WDJ171034.72$-$200541.95 & 4128167950420485632 & 86.5 & $-$0.18 & 2.61 & 13.02 & DQA & \citet{jewett24} \\
WDJ175253.90$-$663452.57 & 5812676526434412928 & 53.3 & $-$0.12 & 3.55 & 13.10 & DQ & Gemini \\
WDJ175631.55+372827.41 & 4609840522521845504 & 206.1 & $-$0.12 & 2.07 & 12.95 &  DQ & Gemini \\
WDJ175821.12+590644.92 & 1422782516088307840 & 95.8 & $-$0.30 & 1.62 & 12.82 & DQH & \citet{jewett24} \\
WDJ182531.11$-$514834.11 & 6654039930861530368 & 174.6 & $-$0.31 & 1.85 & 13.39 & DQA & Magellan \\
WDJ190619.92+202137.67 & 4519466580903738880 & 113.3 & $-$0.20 & 2.51 & 13.16 & DQA & Gemini \\
WDJ190956.00$-$533734.45 & 6644415080946230656 & 266.0 & $-$0.41 & 1.02 & 13.62 & DQA & Magellan \\
WDJ192555.20$-$034626.55 & 4213471120498390784 & 56.6 & $-$0.18 & 2.70 & 13.02 & DQA & \citet{jewett24} \\
WDJ193431.84$-$653025.80 & 6440255975194724736 & 97.1 & $-$0.19 & 2.96 & 13.12 & DQA & Magellan \\
WDJ194919.09$-$420025.11 & 6688580710687236352 & 172.3 & $-$0.26 & 2.48 & 12.90 & DQA & Magellan \\
WDJ203508.73$-$574026.84 & 6467865296283423488 & 150.9 & $-$0.28 & 3.01 & 13.02 & DQA & Gemini \\
WDJ205700.44$-$342556.37 & 6780482118784364032 & 128.8 & $-$0.46 & 0.16 & 13.36 & DAQ & Gemini \\
WDJ210901.84+255821.12 & 1841750421125652864 & 128.8 & $-$0.17 & 2.27 & 13.14 & DQA & MMT \\
WDJ214023.96$-$363757.44 & 6589369272547881856 & 39.8 & $-$0.09 & 3.24 & 13.07 & DQ & \citet{bergeron21} \\
WDJ214027.28$-$085115.90 & 6893810809483853824 & 247.0 & $-$0.30 & 1.69 & 13.07 & DQ & \nodata \\
WDJ214713.66+111005.68 & 1765682152266062976 & 218.6 & $-$0.14 & 2.68 & 13.05 & DQA & SDSS \\
WDJ215755.84$-$114844.06 & 2613599200047182848 & 256.0 & $-$0.31 & 1.34 & 12.34 & DQ & Gemini \\
WDJ222658.76$-$375629.39 & 6594864764806659712 & 140.5 & $-$0.10 & 2.72 & 13.21 & DQA & Gemini \\
WDJ223421.25+144521.41 & 2732910539671785600 & 259.6 & $-$0.07 & 2.51 & 12.92 & DQA & MMT \\
WDJ224234.11$-$480404.19 & 6518961598785128320 & 194.2 & $-$0.17 & 3.12 & 13.12 & DQA & Gemini \\
WDJ225635.07$-$213015.24 & 2396977477886031616 & 213.1 & $-$0.42 & 2.00 & 13.23 & DQA & Gemini \\
WDJ230930.47+282540.41 & 2845705039520675712 & 219.1 & $-$0.23 & 2.61 & 13.19 & DQA & MMT \\
WDJ233410.23$-$262257.02 & 2332601308302888064 & 146.0 & $-$0.07 & 2.51 & 12.33 & DQH & Magellan \\
WDJ234501.17$-$062543.22 & 2439748991307934464 & 111.7 & $-$0.25 & 2.89 & 13.13 & DQA & DESI \\
WDJ234933.31$-$232036.52 & 2339287644669893760 & 121.4 & $-$0.10 & 2.55 & 13.01 & DQA & Magellan
\enddata
\tablecomments{$\ ^a\ \zeta=M_G - 1.2\ (G_{\rm BP} - G_{\rm RP})$ \citep{camisassa21}}
\end{deluxetable*}

\section{Model Atmosphere Analysis}
\label{model}

\begin{figure*}
\centering
\includegraphics[width=3.0in]{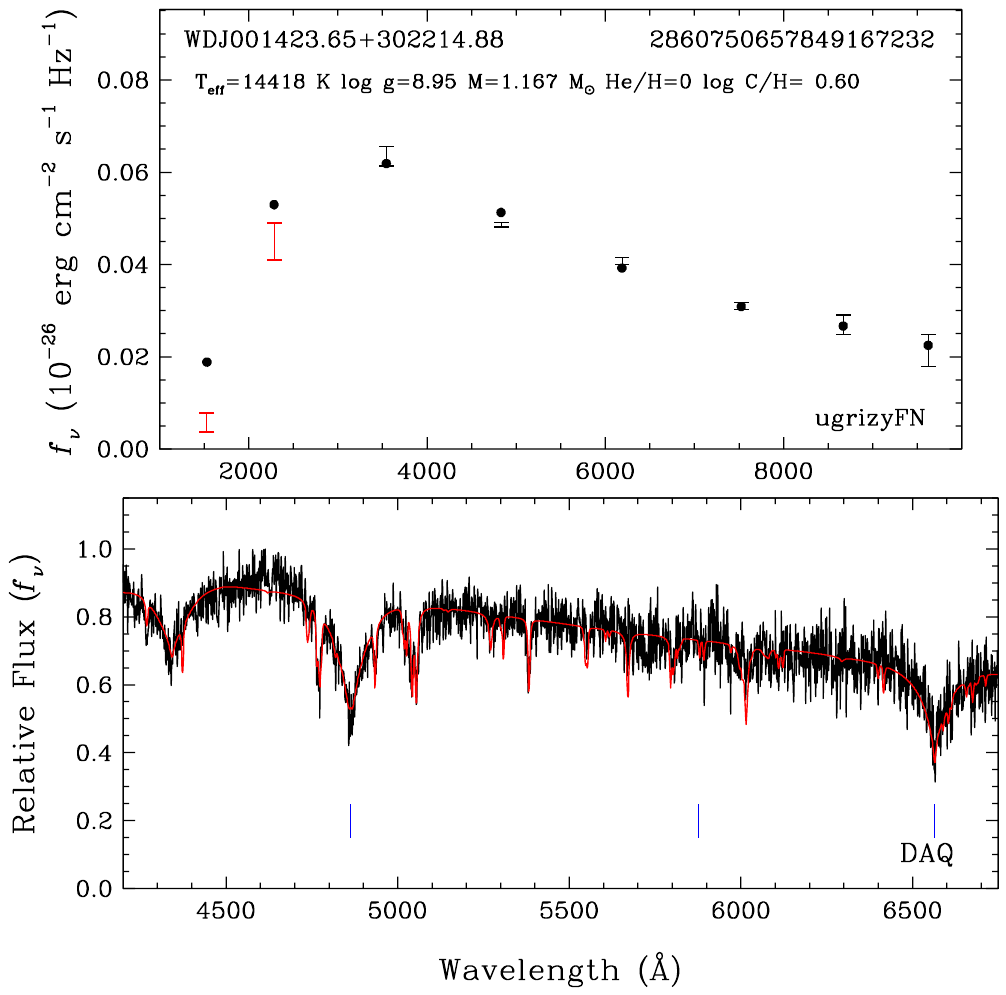}
\includegraphics[width=3.0in]{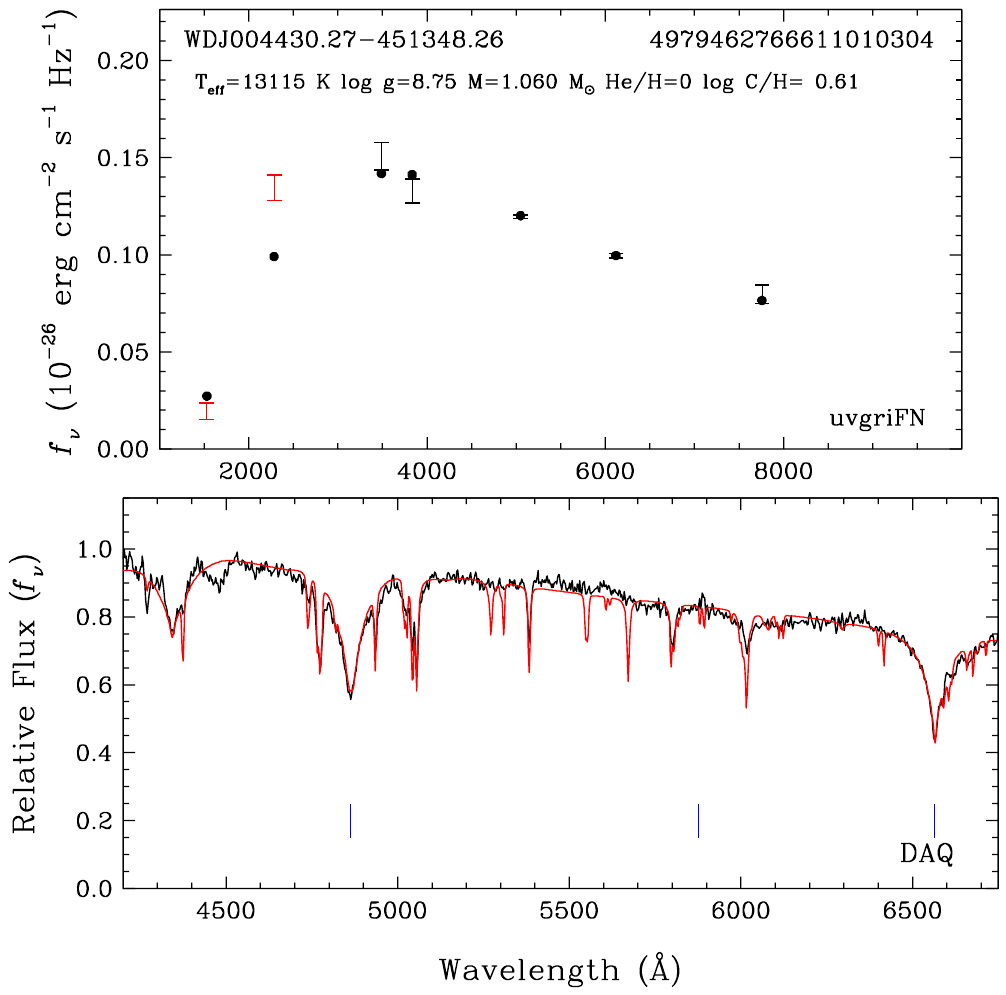}
\includegraphics[width=3.0in]{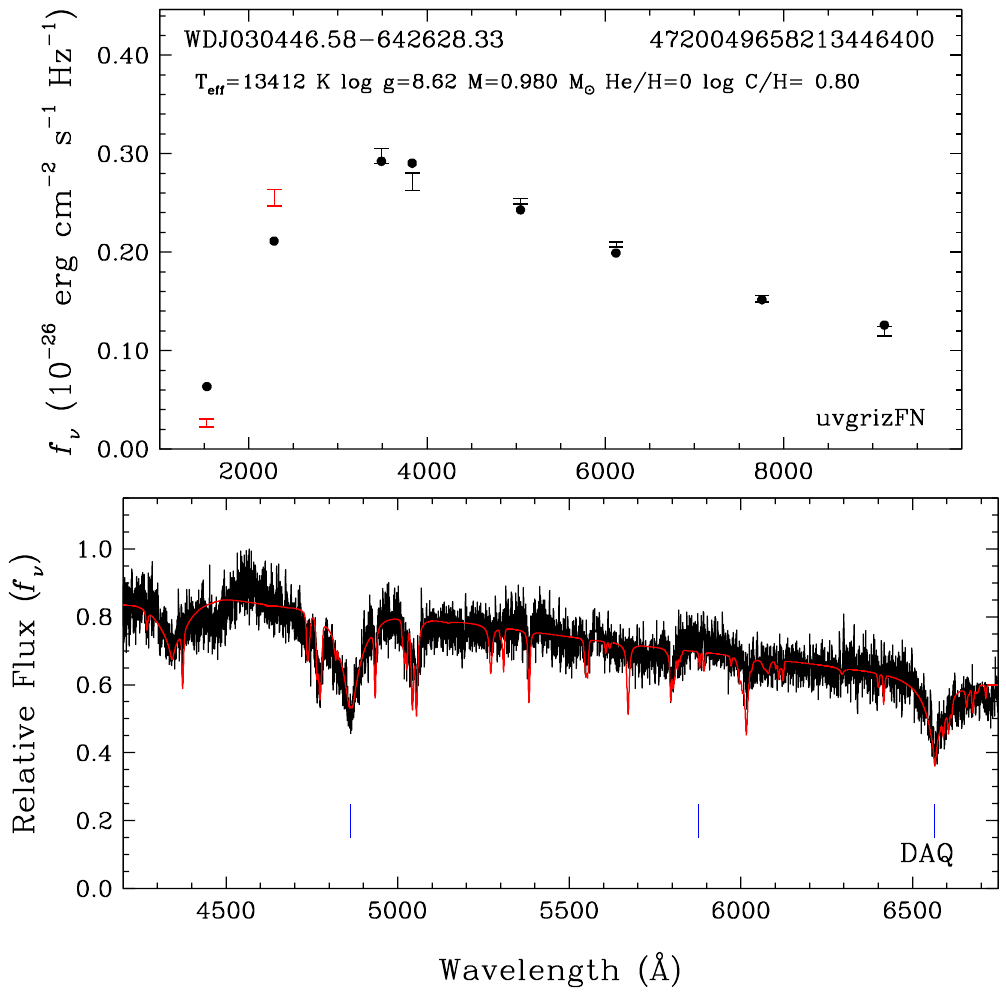}
\includegraphics[width=3.0in]{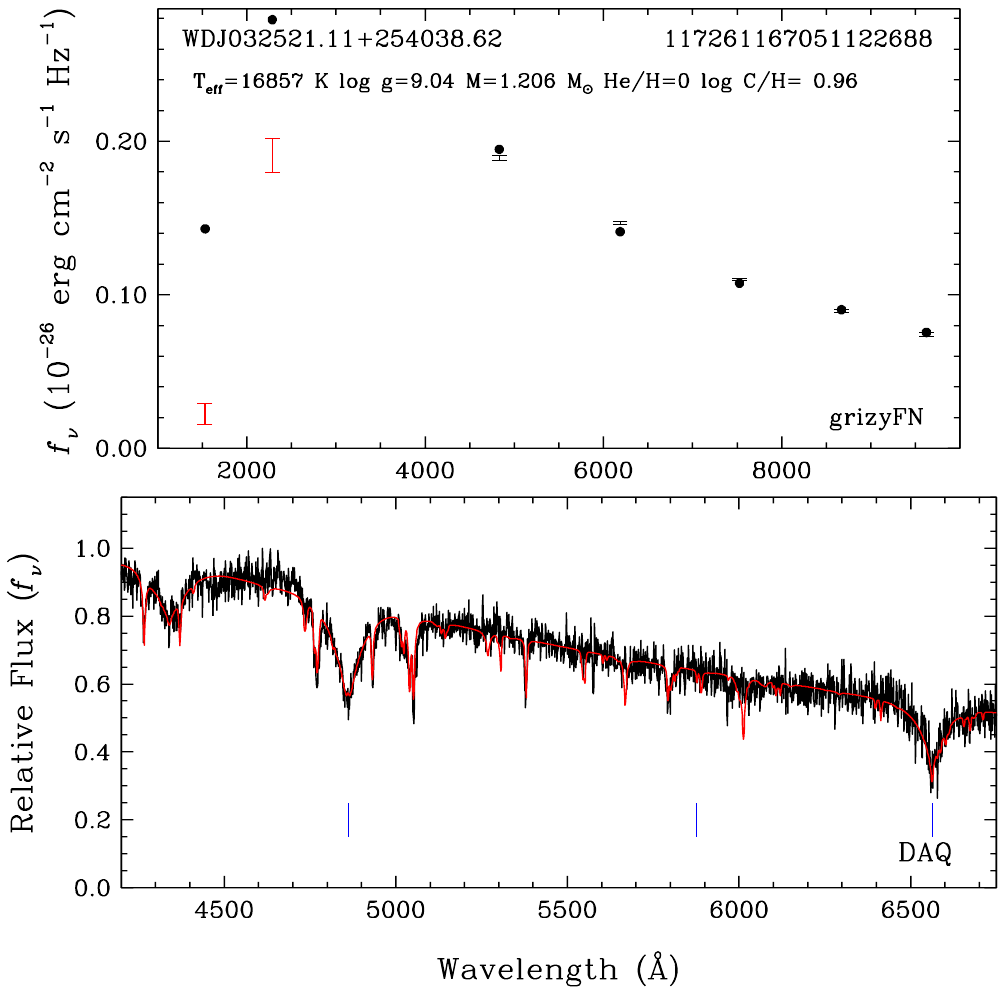}
\includegraphics[width=3.0in]{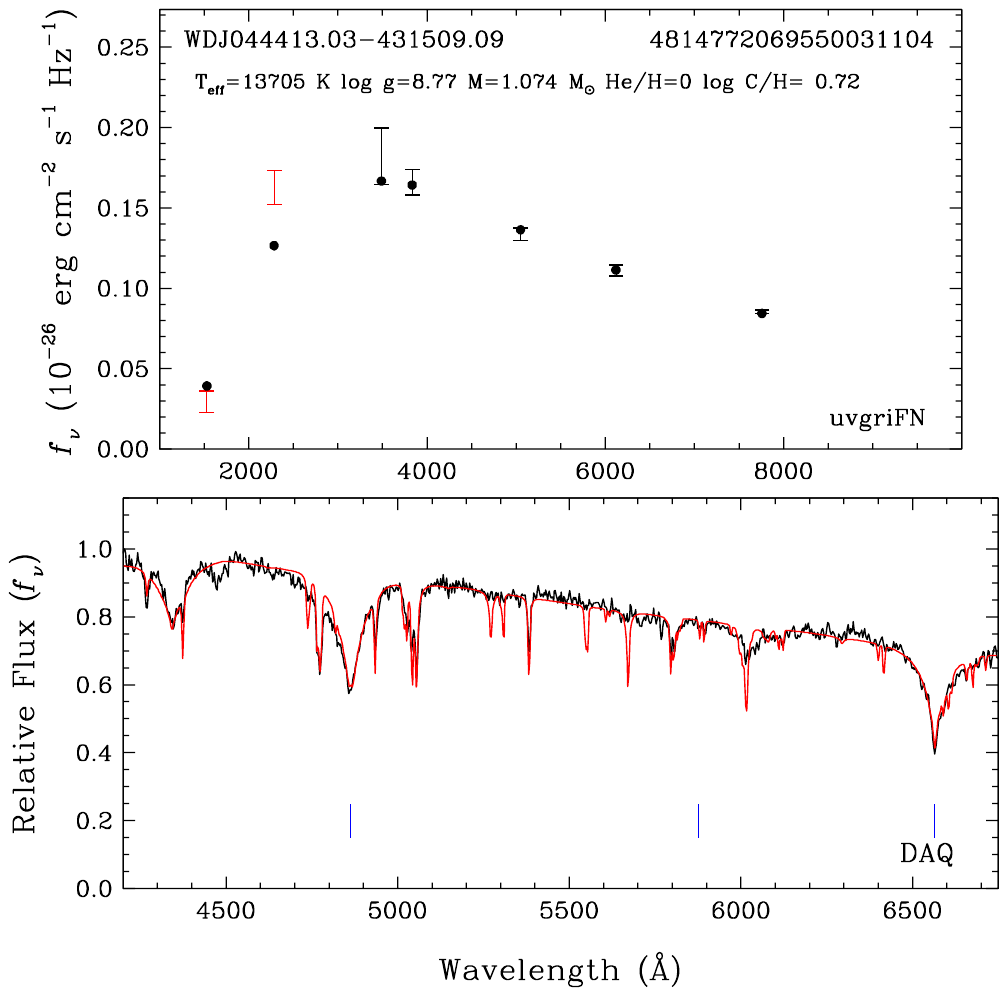}
\includegraphics[width=3.0in]{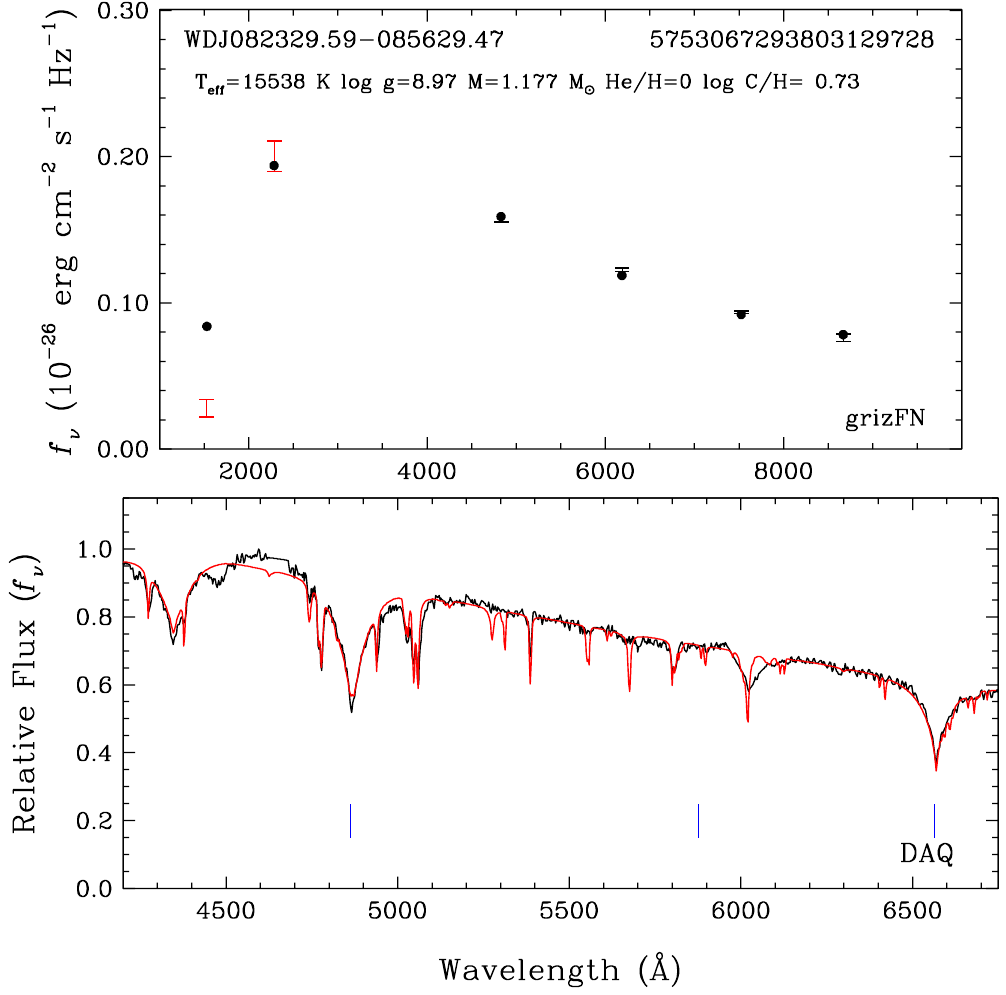}
\caption{Model atmosphere fits to ten of the newly discovered DAQ white dwarfs. The top and bottom panels show the photometric and the spectroscopic fits, respectively. The best-fitting model parameters are presented in the top panel, which also includes the Gaia DR3 Source ID, object name, and the photometry
displayed in the panel. The bandpasses displayed in red are not included in the fits. Blue tick marks here and in the following figures mark the locations of H$\beta$, \ion{He}{1} $\lambda$5876, and H$\alpha$ from
left to right. Note that none of the warm DQs in our sample show any evidence of \ion{He}{1} $\lambda$5876.}
\label{figdaq} 
\end{figure*}

\addtocounter{figure}{-1}
\begin{figure*}
\centering
\includegraphics[width=3in]{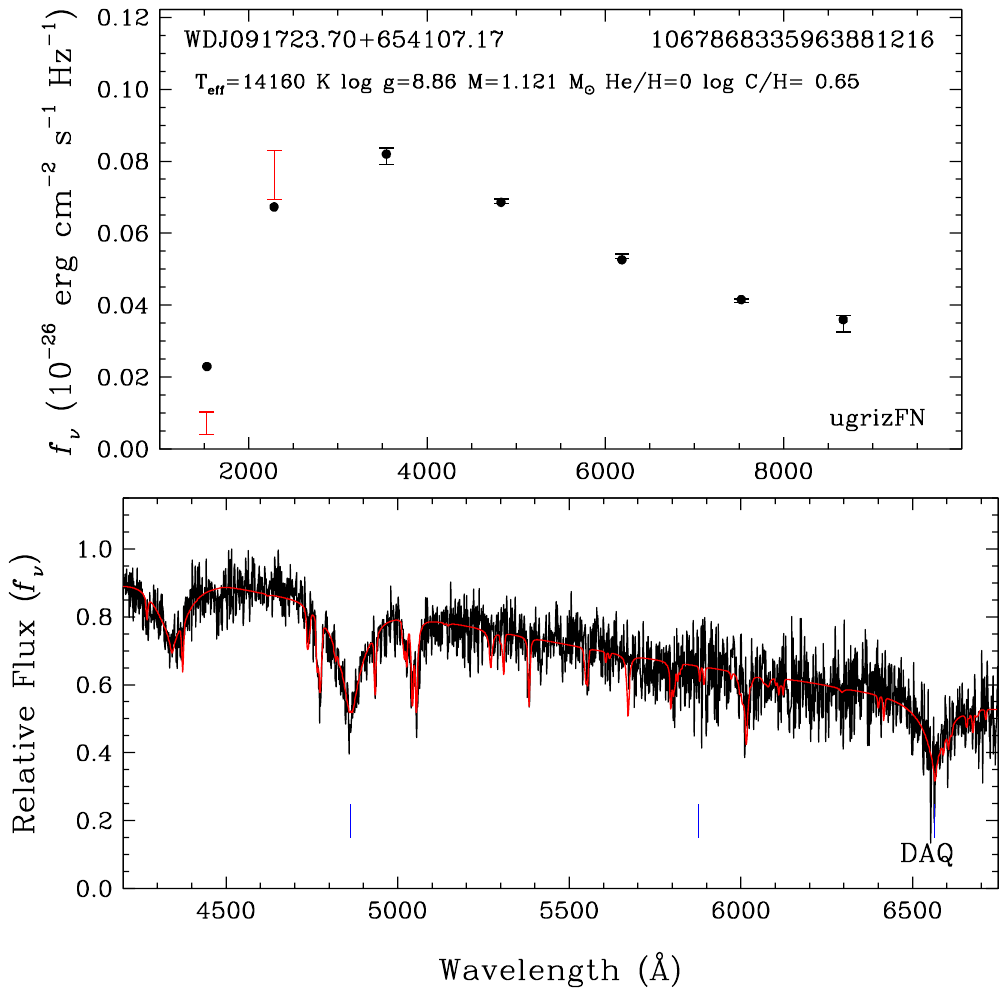}
\includegraphics[width=3in]{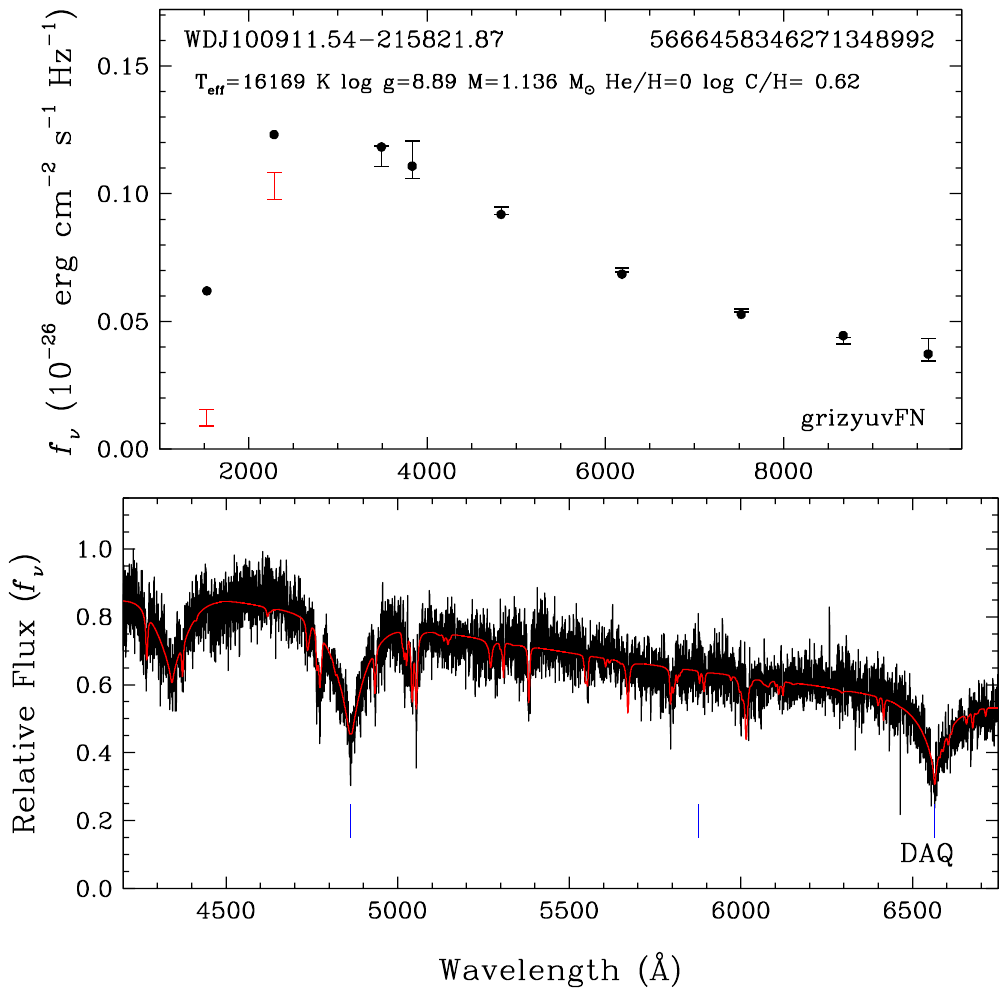}
\includegraphics[width=3in]{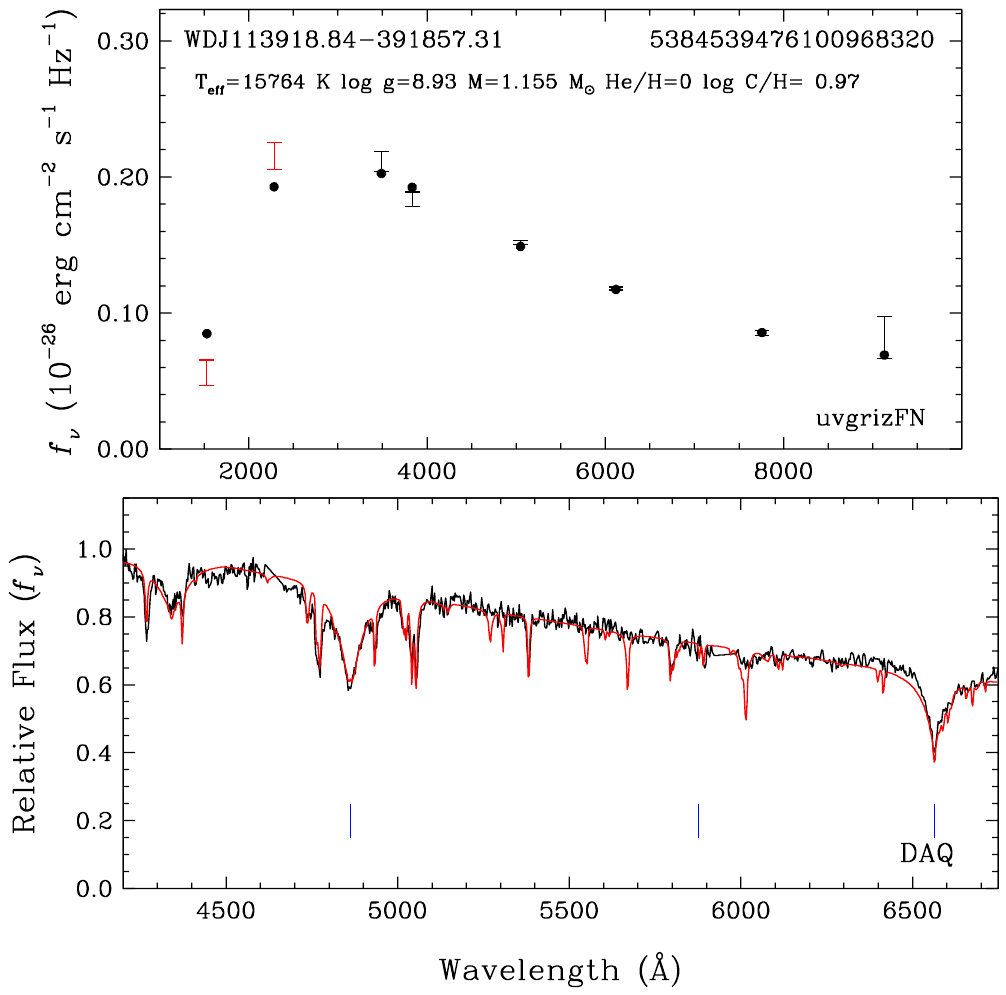}
\includegraphics[width=3in]{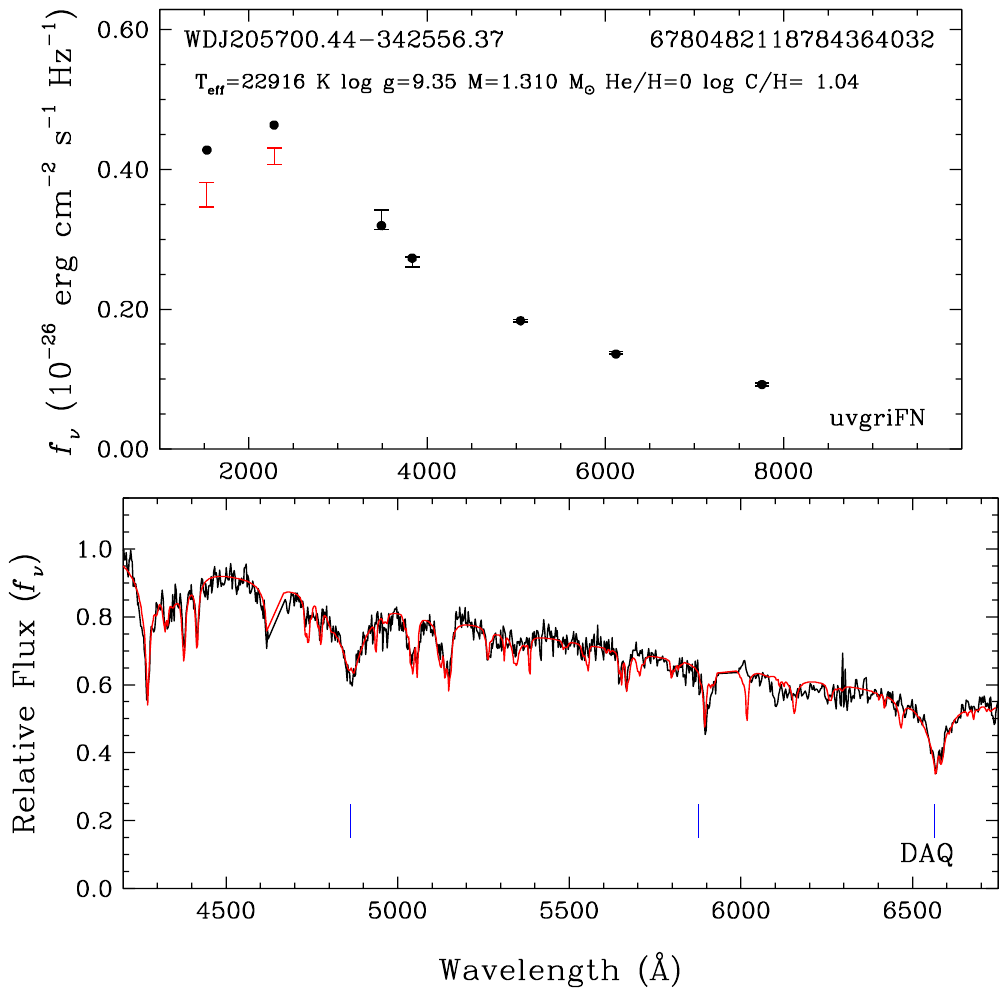}
\caption{continued.}
\end{figure*}

We use the photometric and spectroscopic methods to constrain the physical parameters of each system \citep{bergeron19}. We concentrate on the model fits to the warm DQ white dwarfs here, but we also discuss the fits to the rest of the white dwarfs in our sample in the appendix, see section \ref{app}.

We use the SDSS $u$ and Pan-STARRS $grizy$ photometry (if available) along with the Gaia DR3 parallaxes to constrain
the effective temperature and the solid angle. For southern targets, we instead rely on SkyMapper $uvgriz$ data. In a few cases where
photometry is available only in a few bands, we also use the Gaia photometry to help constrain the parameters. 
We use the geometric distances from \citet{bailer21} and the reddening values from \citet{gentile21}. Given the distance measurements,
we constrain the radius of each star directly, and use the evolutionary models for white dwarfs to constrain the mass. 

We convert the observed magnitudes into average fluxes, and compare with the synthetic fluxes calculated from model atmospheres with the
appropriate chemical composition. A $\chi^2$ value is defined in terms of the difference between the observed and model fluxes over all
bandpasses, properly weighted by the photometric uncertainties, which is then minimized using the nonlinear least-squares method of
Levenberg-Marquardt \citep{press86} to obtain the best fitting parameters and uncertainties. Based on the abundances derived from the spectroscopic
fit, we repeat the photometric and spectroscopic fits until a consistent solution is found.

We rely on new model atmosphere grids with improved calculations for the Stark broadening of the neutral carbon lines. These models
supersede the ones used in \citet{kilic24}. We rely on four sets of models. The first set is for C+H atmospheres (no helium), and it covers
the range $T_{\rm eff} = 11,000\ (500)\ 23,000$ K, $\log{g} = 8.0\ (0.5)\ 9.5$, and $\log {\rm C/H} = -1.5\ (0.5)\ 3.0$, where the numbers
in parentheses indicate the step size. The second set is for C+He atmospheres (no hydrogen), and it covers the range
$T_{\rm eff} = 9000\ (500)\ 16,000$ K, $\log{g} = 7.5\ (0.5)\ 9.5$, and $\log {\rm C/He} = -5.0\ (0.5)\ -1.0$.
The third set is for C+He+H atmospheres, and it covers the range $T_{\rm eff} = 11,000\ (500)\ 19,000$ K, $\log{g} = 8.0\ (0.5)\ 9.5$,
$\log {\rm H/He} = -4.0\ (1.0)\ 0.0$, and $\log {\rm C/He} = -2.0\ (0.5)\ 1.0$. Finally, the last set is for pure C atmospheres, and it covers
the range $T_{\rm eff} = 11,000\ (500)\ 19,000$ K and $\log{g} = 7.5\ (0.5)\ 9.5$. Note that the pure C atmosphere models are used for comparison
only, and they rely on our older calculations of Stark broadening of the neutral carbon lines. 
Some of the warm DQs in our sample are magnetic, but we simply fit them with non-magnetic models.

We rely on the CO core evolutionary models from \citet{bedard20} with $q({\rm He})\equiv \log M_{\rm He}/M_{\star}=10^{-2}$ and $q({\rm H})=10^{-10}$, which are 
appropriate for He- and thin H-atmosphere white dwarfs. Admittedly, the He mass fraction is relatively large in these evolutionary sequences, but that has
minimal impact on the radius and mass measurements for our targets.
\citet{koester19}, \citet{hollands20}, and \citet{kilic24} discuss the problems with the
accuracy of the oscillator strengths for the carbon lines, especially in the blue. The same problems persist for the current analysis, where
the depths of some of the carbon lines cannot be reproduced by the models. Similar to the analysis presented in \citet{kilic24}, here we exclude the carbon lines with
quality flags D and E in the NIST database, and two additional absorption features in the models near 5268 and 5668 \AA\ that are not observed.

\section{Results: 71 warm DQs}
\label{res}

\subsection{DAQ}

Prior to our targeted search for warm DQs, there were only 6 DAQ white dwarfs known \citep{liebert83,hollands20,kilic24,jewett24}. 
These stars display hydrogen lines that are stronger than the carbon features in their spectra, hence the DAQ classification. However, \citet{kilic24}
demonstrated that the distinction between DAQ and warm DQ/DQA white dwarfs is superficial, as there is a range of hydrogen abundances among
the warm DQ population. Nevertheless, given the relatively small number of DAQ white dwarfs known, we start with
the discussion of the newly identified DAQ white dwarfs.

\begin{figure}
\includegraphics[width=3in]{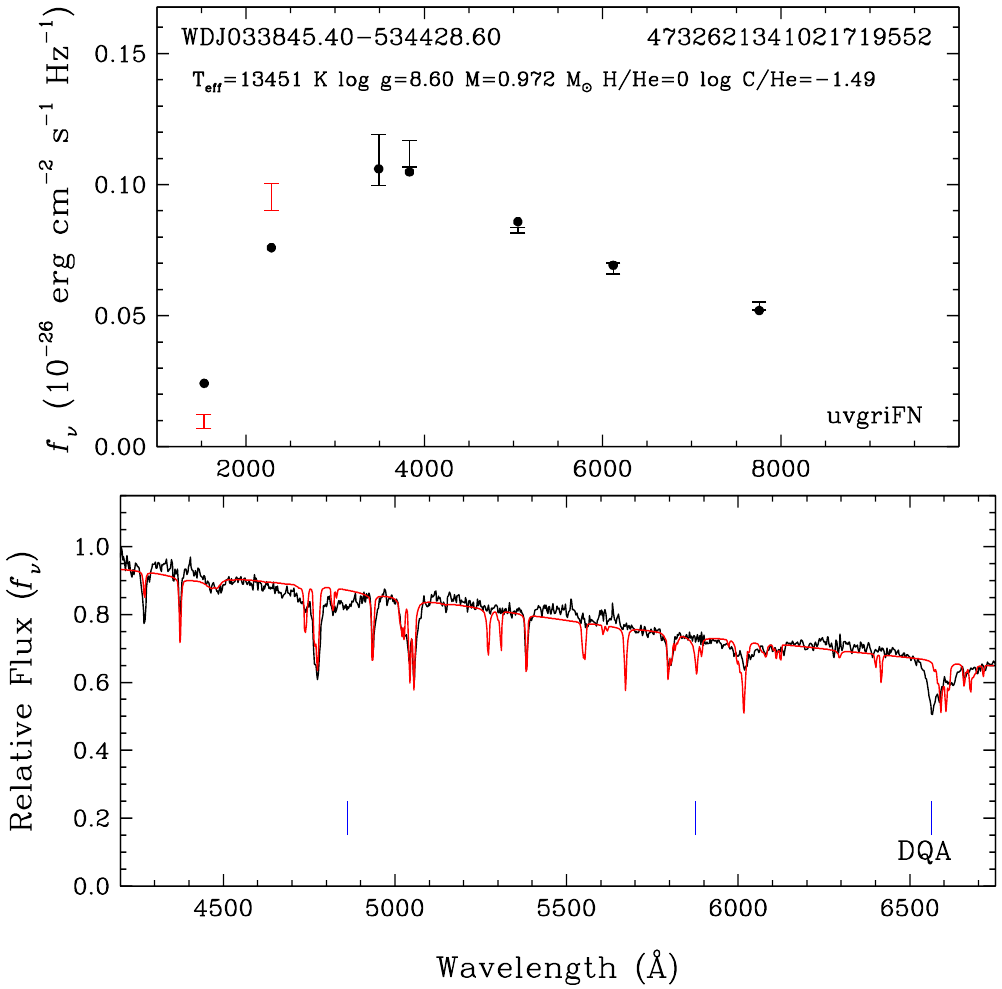}
\includegraphics[width=3in]{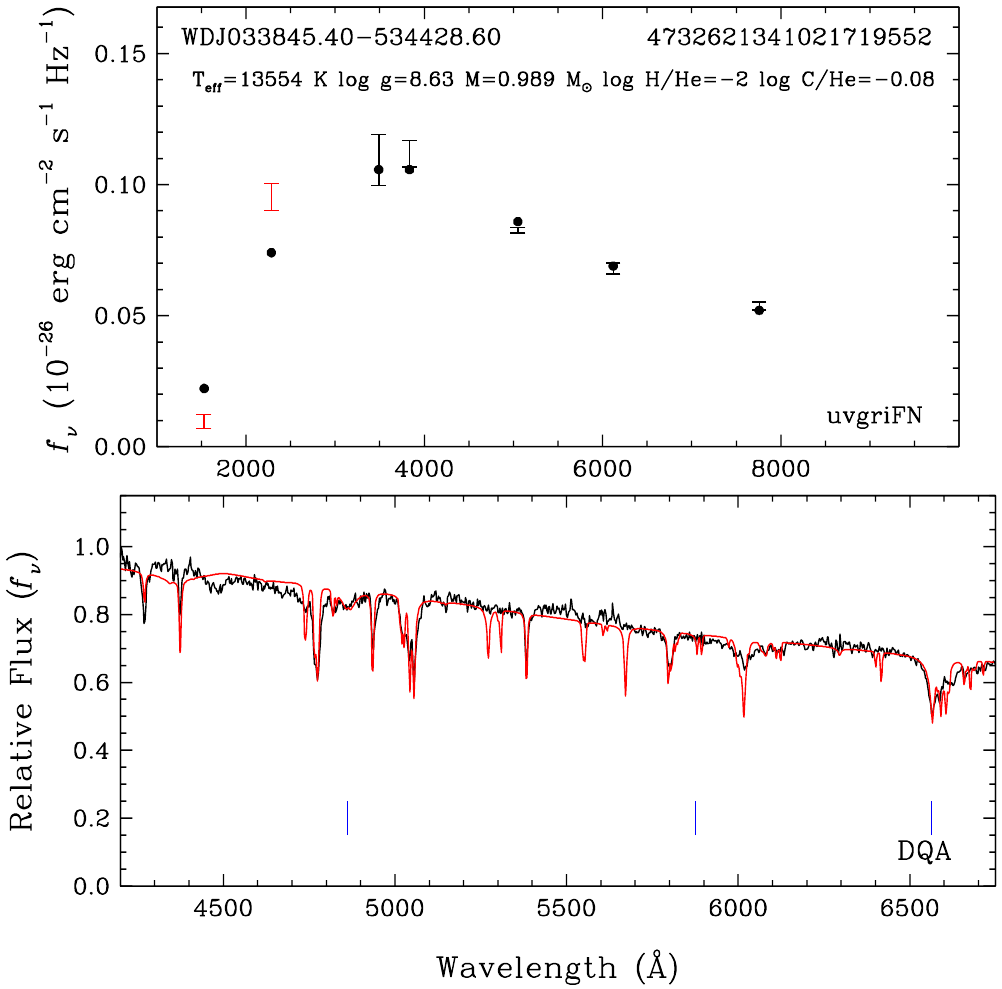}
\includegraphics[width=3in]{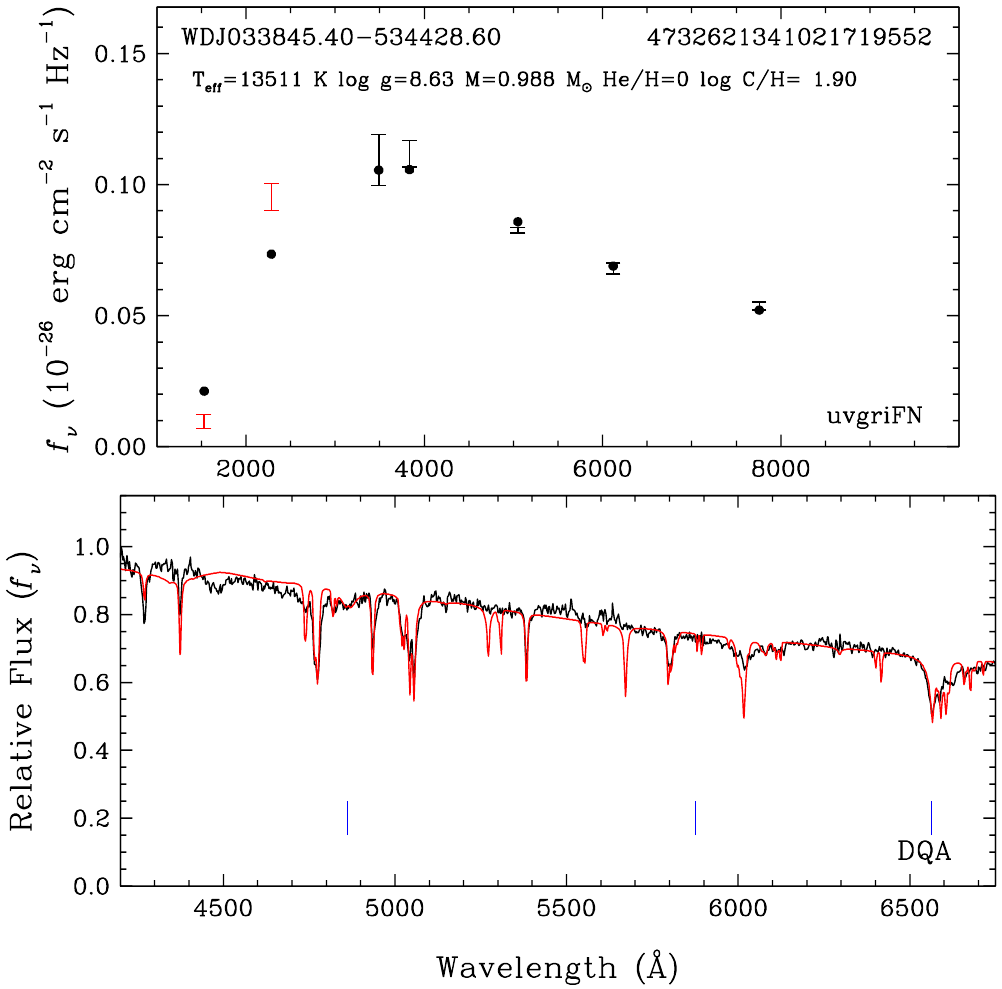}
\caption{Model fits to the DQA white dwarf J0338$-$5344 under the assumption of a C+He (top, note the relatively strong \ion{He}{1} $\lambda$5876 predicted
here), C+He+H (middle), and a C+H atmosphere (bottom panels). The symbols are the same as in Figure \ref{figdaq}.}
\label{figexplore} 
\end{figure}

There are 13 DAQ white dwarfs in our sample, including ten newly discovered systems.
Figure \ref{figdaq} shows the model fits to these newly discovered DAQs. The top and bottom panels show our photometric and
spectroscopic fits, respectively. The top panel includes GALEX FUV and NUV photometry for comparison. For nine of these targets, the best-fitting
models have $T_{\rm eff}=13,115-16,857$ K, $M=0.98-1.21~M_{\odot}$, and $\log {\rm C/H}$ = +0.60 to +0.97. Hence, they are all massive and
they have atmospheres dominated by carbon. 

The one outlier in this sample, J2057$-$3425, is significantly hotter. It is the bluest confirmed DQ in Figure \ref{figsample}. We initially classified this object
as a hot DQ, but then realized that it is essentially a hotter version of the other DAQs found in our sample. J2057$-$3425 displays H$\alpha$ and
H$\beta$ lines that are stronger than the nearby \ion{C}{1} lines, though it also displays several relatively strong \ion{C}{2} lines, e.g. $\lambda4267$,
$\lambda5145$, and $\lambda5890$. J2057$-$3425 becomes the hottest and most massive DAQ known, with $T_{\rm eff}=22,916 \pm 1117$ K and $M=1.310_{-0.015}^{+0.017}~M_{\odot}$. 
\citet{dufour08} found evidence of hydrogen in two of the nine hot DQs in their sample.
%with $\log$ C/H $\approx1.7$. 
J2057$-$3425 has about five times more hydrogen than those two hot DQs, making it an outlier among both the DAQ sample (due to its temperature) and hot DQs (due to its  hydrogen abundance).

\subsection{DQA}

There are 61 warm DQ/DQA white dwarfs in our sample that display atomic carbon lines, and one DQ that shows molecular carbon features. The majority of warm DQs display evidence of hydrogen in their spectra, but never helium. Traditionally, the small flux depression near 4470 \AA\ is used as an
indication of He-dominated atmospheres in these stars \citep{koester19}. However,  it is impossible to hide the stronger \ion{He}{1} absorption feature at
5876 \AA\ and still see the 4471 \AA\ line. None of the warm DQs in our sample show any evidence of a 5876 \AA\ line due to \ion{He}{1}. 
\citet{hollands20} demonstrated that the 4470 \AA\ feature in these stars is most likely due to a carbon triplet \citep[see the discussion in Section 3 of][]{kilic24}. 

We inspect each model fit with and without hydrogen (using the sets of model grids described in Section \ref{model}) to decide if an object shows clear evidence of H$\alpha$ and/or H$\beta$, and decide on
the DQ or DQA spectral type based on visual inspection. Figure \ref{figexplore} shows the model fits to one of the DQA in our sample, J0338$-$5344, as
an example. The top panels show the fits assuming C+He atmospheres, and no hydrogen. The best-fitting model predicts a relatively strong \ion{He}{1} 
$\lambda$5876 line that is not observed, and it also does not predict an H$\alpha$ feature, which is clearly present in the observed spectrum. Hence,
we can rule out H-free models. 

We thoroughly investigated all the fits with varying H/He abundances. The middle and bottom panels in Figure \ref{figexplore} provide an example of these results. The \ion{He}{1}
$\lambda5876$ line is visible for $\log$ H/He $\leq -3$, but not at $\log$ H/He $=-2$ or higher. This sets a lower limit on the hydrogen abundance of $\log$ H/He $\geq-2$, which is a relatively weak limit. For instance, there are no (or very few) DBs with such large H/He ratios
\citep[see Figure 8 in][and references therein]{bedard23}. So, there could be a relatively large amount of helium present in this star, but it remains invisible for $\log$ H/He $\geq-2$.

The middle panels in Figure \ref{figexplore} show that a model with $\log$ H/He = $-2$ and $\log$ C/He = $-0.08$ provides an excellent fit to the
observed spectrum. The bottom panels show the fits for a helium-free atmosphere, where a model with $\log$ C/H = 1.9 also provides an
excellent fit. Regardless of the helium abundance,
we can safely constrain the C/H ratio to $\log$ C/H = 1.9, as the fits in both middle and bottom panels have the same ratio. 
In addition, the amount of helium present in these atmospheres have a minimal impact on the physical parameters of warm DQs, including
their effective temperatures and masses.

Modeling the interior structure of the DAQ white dwarf J0551+4135 with thin He and H layers, \citet{hollands20} found that chemical
diffusion can erase the helium buffer and create a hydrogen-carbon interface. This allows convection to dredge-up carbon from the interior, without any
signatures of helium in the spectrum. Similarly, \citet{kilic24} proposed a scenario where a massive DA white dwarf with a thin hydrogen atmosphere goes
through convective dilution by the deeper C/He-rich envelope as it cools off, gradually turning it into a DAQ and eventually a warm DQA as hydrogen gets diluted
further by the underlying C/He convective envelope.
It is possible to fit all warm DQs with carbon and hydrogen atmospheres. We just cannot tell if there is any helium, due to the lack of spectral features from helium. Therefore, we use the same model grid used for the DAQ white dwarfs, with zero helium, to fit the spectra of the warm DQ and DQA white dwarfs in our sample. 

\begin{figure}
\includegraphics[width=3.0in]{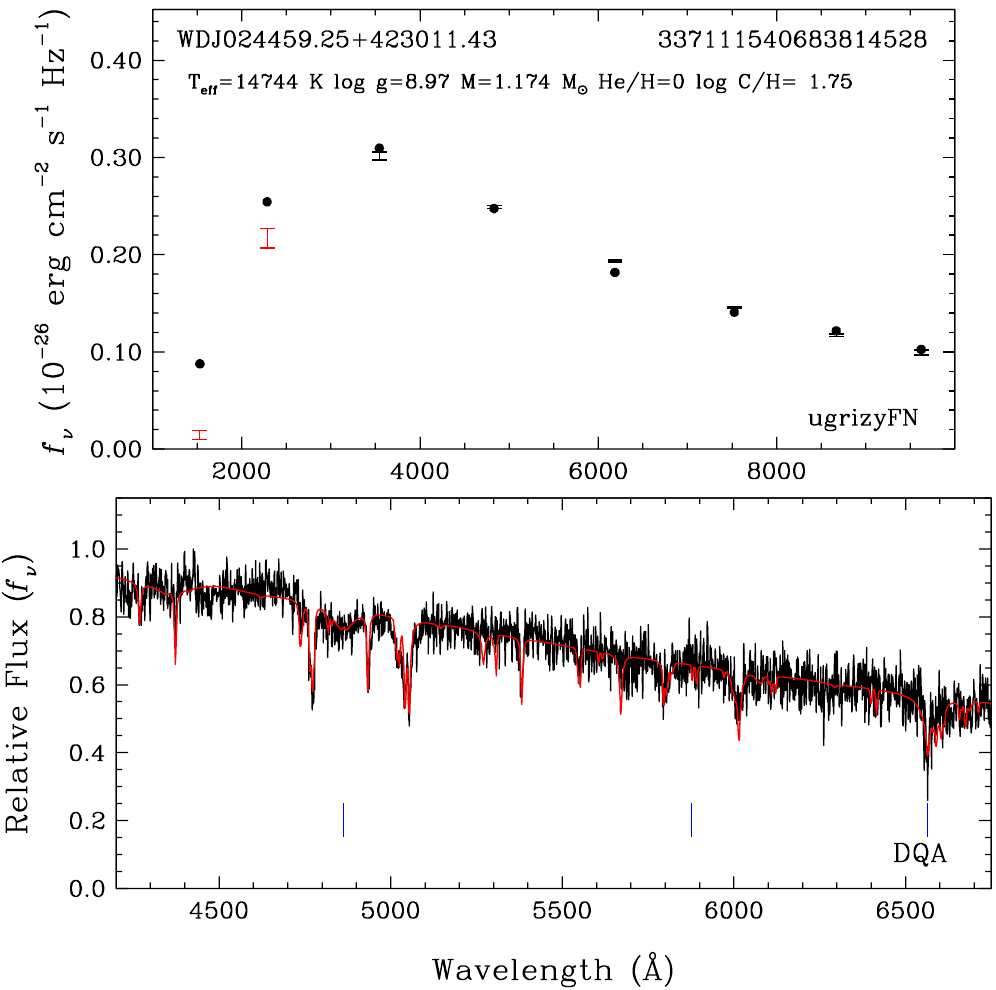}
\includegraphics[width=3.0in]{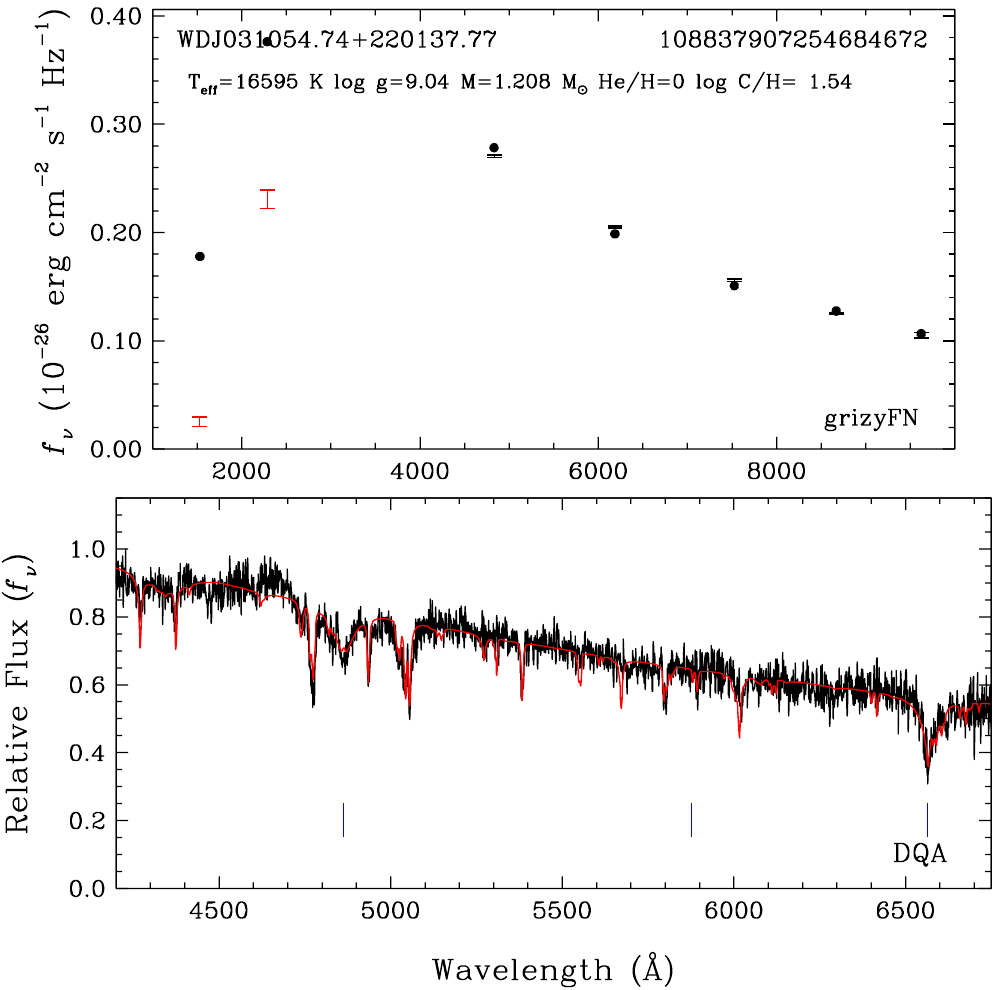}
\includegraphics[width=3.0in]{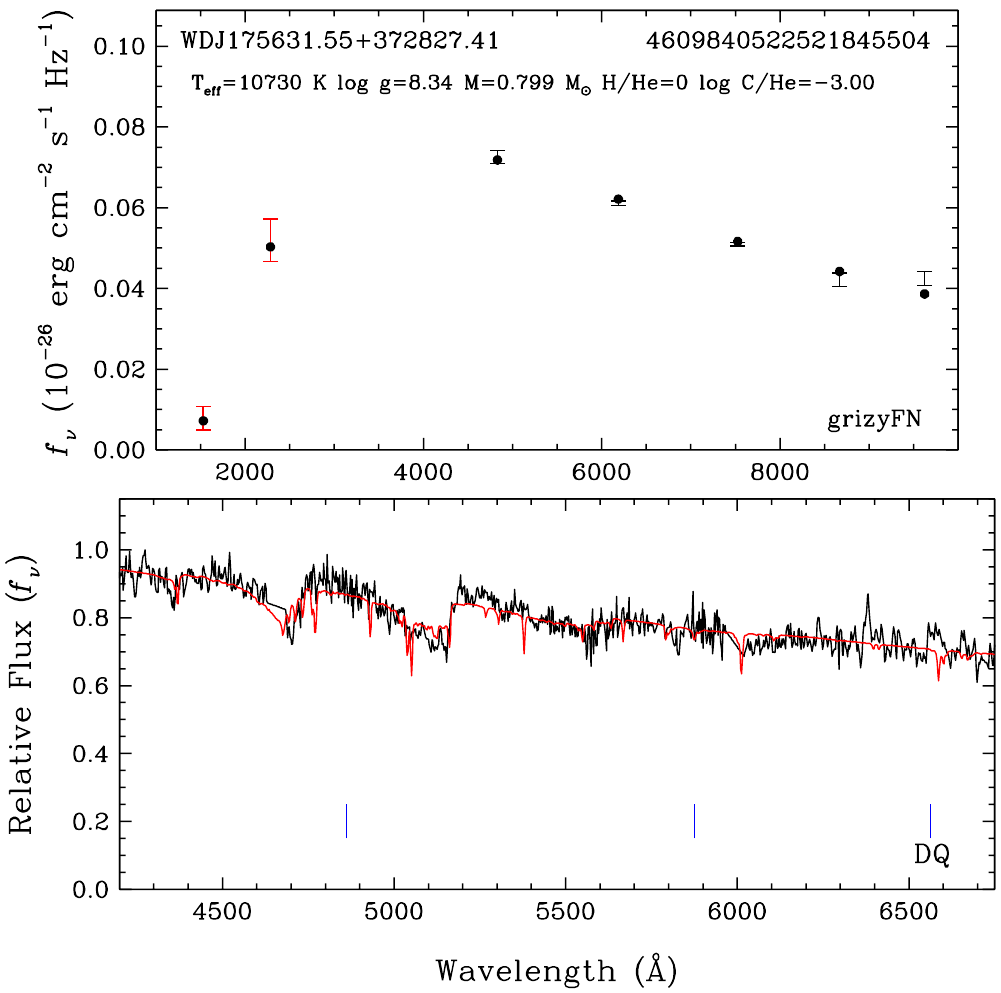}
\caption{Model fits to three DQ white dwarfs. Symbols are the same as in Figure \ref{figdaq}. J1756+3728 (bottom) is the coolest
DQ found in our survey, and it shows C$_2$ Swan bands.}
\label{figdq} 
\end{figure}

Figure \ref{figdq} shows additional fits to three other warm DQs. J0244+4230 (top) is similar to J0338$-$5344 (Figure \ref{figexplore}), where the presence of hydrogen can be inferred from the narrow H$\alpha$ absorption feature seen, as models
without hydrogen do not predict a feature at 6563 \AA. Models with $T_{\rm eff} = 14,744 \pm 233$ K, $M=1.174_{-0.011}^{+0.010}~M_{\odot}$, and
$\log$ C/H = 1.75 provide an excellent match to the spectral energy distribution of this white dwarf.

The middle panels in Figure \ref{figdq} display the model fits to a slightly hotter DQA, J0310+2201. This star shows clear signatures of H$\beta$ and
H$\alpha$ absorption. A model with $T_{\rm eff} = 16,595 \pm 246$ K, $M=1.208_{-0.009}^{+0.010}~M_{\odot}$, and $\log$ C/H = +1.54 provides an excellent
match to the optical photometry and spectrum of this object. However, the best-fitting model appears to have issues matching the UV photometry. This
problem is not unique to J0310+2201. Indeed, the majority of the hottest DQ/DQA white dwarfs in our sample show a similar problem in the UV.
We suspect that inaccurate carbon opacities cause this discrepancy between the observed and predicted UV photometry.

The bottom panels in Figure \ref{figdq} display the model fits to the coolest DQ found in our survey. Unlike the other 70 warm DQs in our sample that show
atomic carbon, J1756+3728 is cool enough to show molecular carbon bands. He-free atmosphere models cannot match the observed spectrum
in this case, and instead we find a best-fitting atmosphere model with  $T_{\rm eff} = 10,730 \pm 260$ K, $M=0.799_{-0.047}^{+0.041}~M_{\odot}$, and
$\log$ C/He = $-3.0$. J1756+3728 is significantly cooler and less massive than the rest of our warm DQ sample. 

\begin{figure}
\center
\includegraphics[width=3.4in]{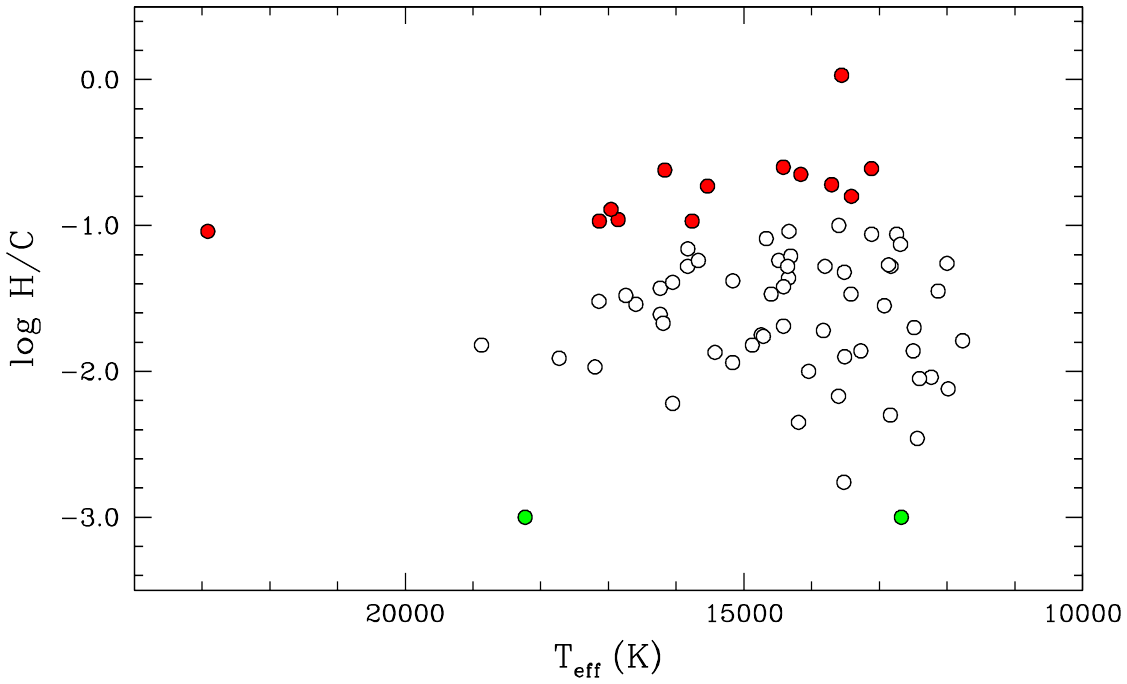}
\caption{H/C ratio versus effective temperature for DAQ (red dots) and DQ/DQA (white dots) in our sample.
Green dots mark the two sources where $\log$ H/C was fixed at $-3$, which is the lower limit of our C+H atmosphere model
grid. Note that the plot presents $\log$ H/C, instead of $\log$ C/H as shown
in the model fits, since hydrogen is always a trace element in our sample.}
\label{fighc}
\end{figure}

\begin{figure*}
\includegraphics[width=3.4in]{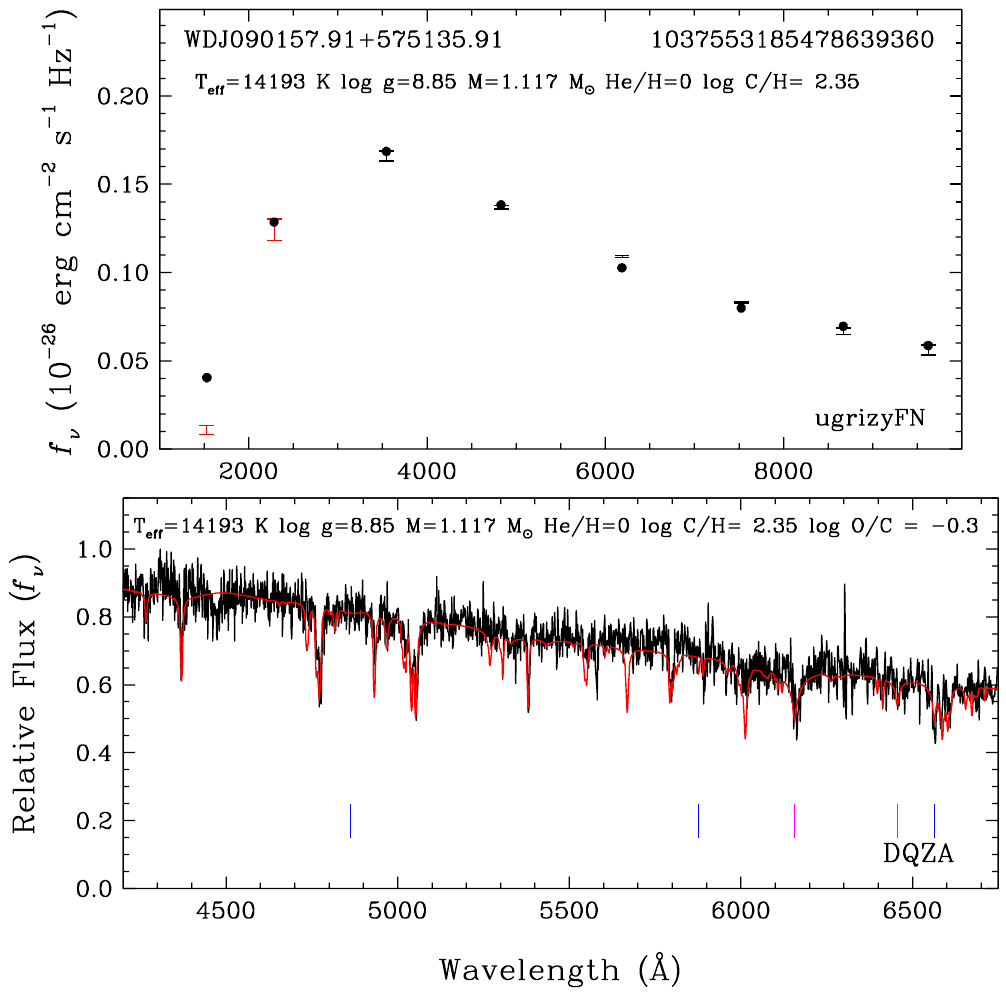}
\includegraphics[width=3.4in]{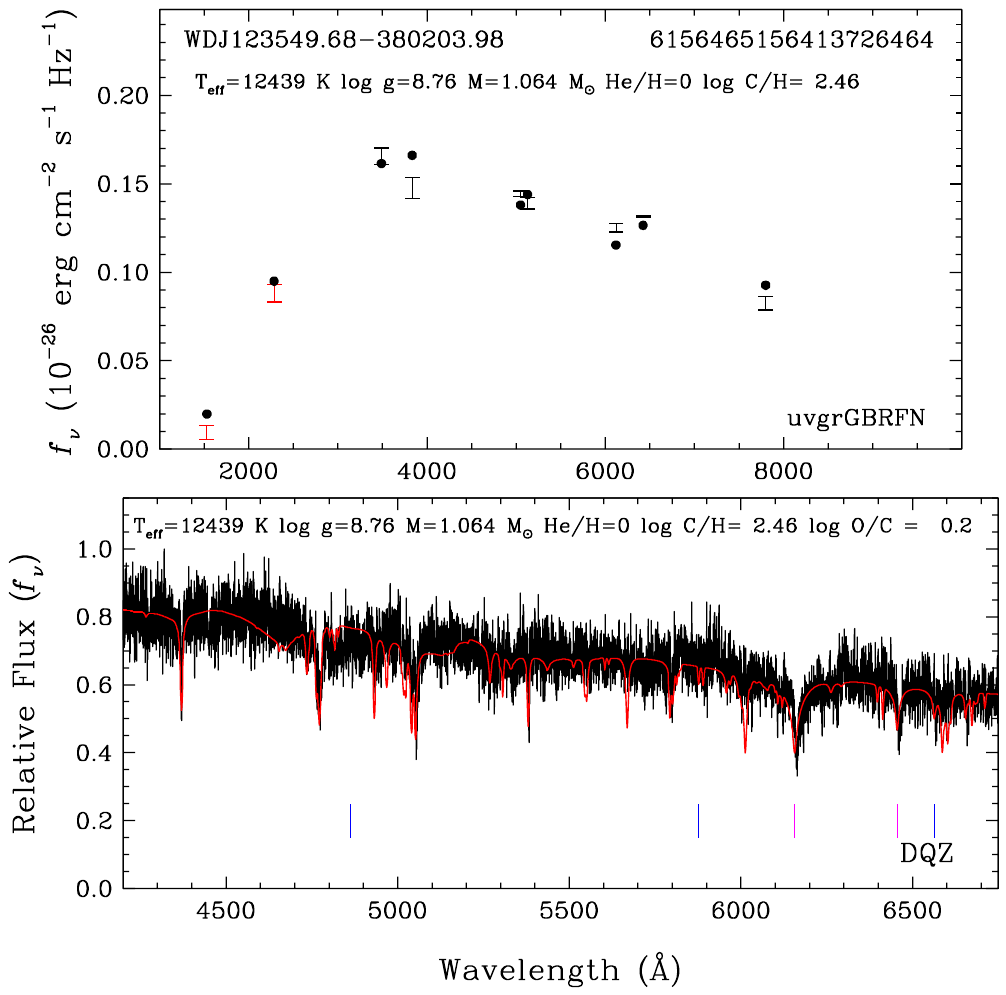}
\caption{Model atmosphere fits to the carbon and oxygen atmosphere white dwarfs J0901+5751 and J1235$-$3802. Note the relatively strong \ion{O}{1} $\lambda$6156 line in both spectra. J1235$-$3802 also shows \ion{O}{1} $\lambda$6456. Oxygen lines are marked by the magenta tick marks.}
\label{figox} 
\end{figure*}

Figure \ref{fighc} shows the H/C ratio versus effective temperature for our sample. Note that the plot presents $\log$ H/C, instead of $\log$ C/H as shown
in the model fits, since hydrogen is always a trace element in our sample.
Warm DQ white dwarfs have a broad range of H/C ratios, and
DAQ white dwarfs represent the most hydrogen-rich stars in this population with $\log$ H/C $\gtrsim-1.0$. However, the difference between
DAQ and DQA white dwarfs is clearly superficial, as they are the same population of stars with slightly different H/C ratios. 
 
We also highlight two DQ stars (J1752$-$6634 and J2157$-$1148, green points) where we had to force a solution at $\log$ H/C $=-3.0$, which is the
boundary of our C+H atmosphere model grid. The H/C ratios in these stars are probably lower because their model fits are identical to spectroscopic fits
using pure carbon models (not shown here). At such low H/C ratios, hydrogen plays no role in the spectrum. Hence, we cannot tell how much hydrogen there is in the atmosphere. Alternatively,
instead of hydrogen, there could be helium in the atmosphere. We could also achieve good fits using trace amounts of helium. However, helium also plays no role
in the spectrum (there are no helium lines observed). Hence, the most C-rich stars in our sample may or may not have trace amounts of hydrogen and
helium. These stars could be related to hot DQs \citep{dufour13}, most of which have pure C atmospheres \citep{dufour08}. With $T_{\rm eff} = 18,230 \pm 547$ K and $M=1.036_{-0.049}^{+0.044}~M_{\odot}$, J2157$-$1148 especially has very similar parameters to those of hot DQs. 

\subsection{Oxygen-rich DQs}

\begin{figure*}
\includegraphics[width=3.4in]{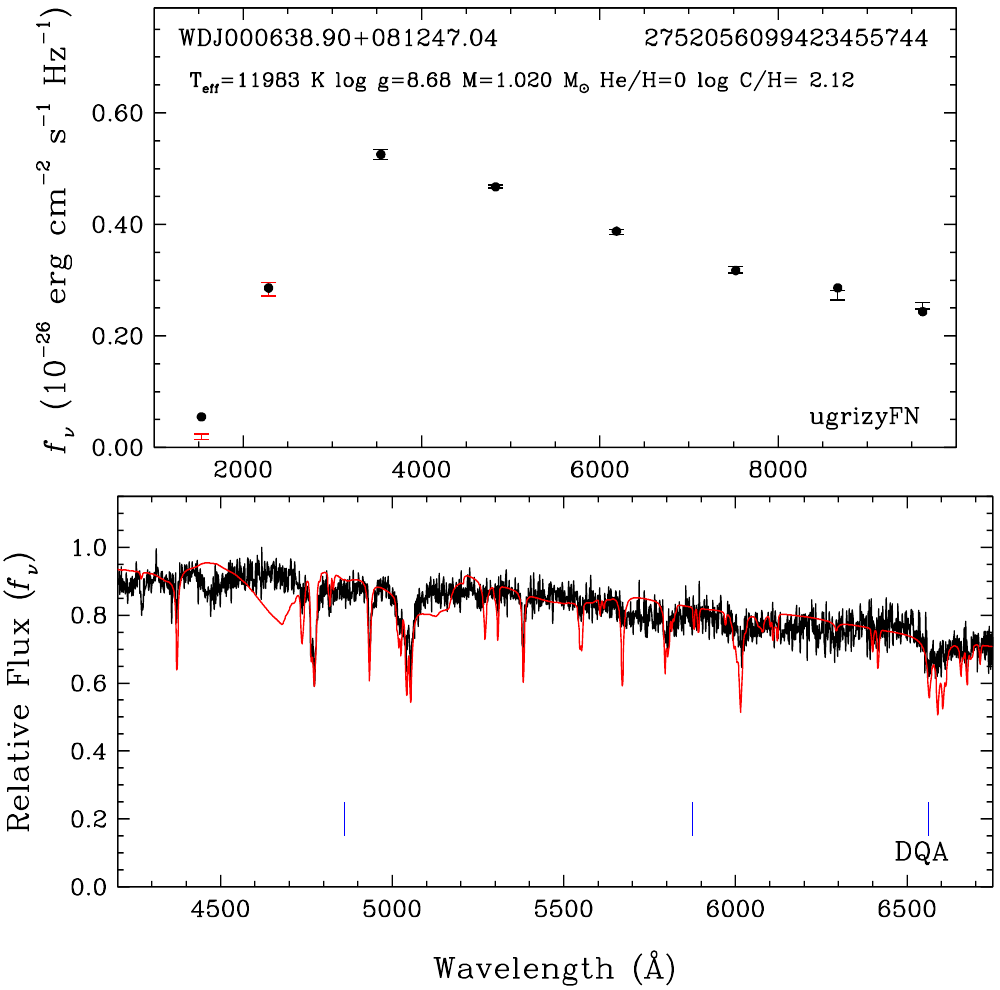}
\includegraphics[width=3.4in]{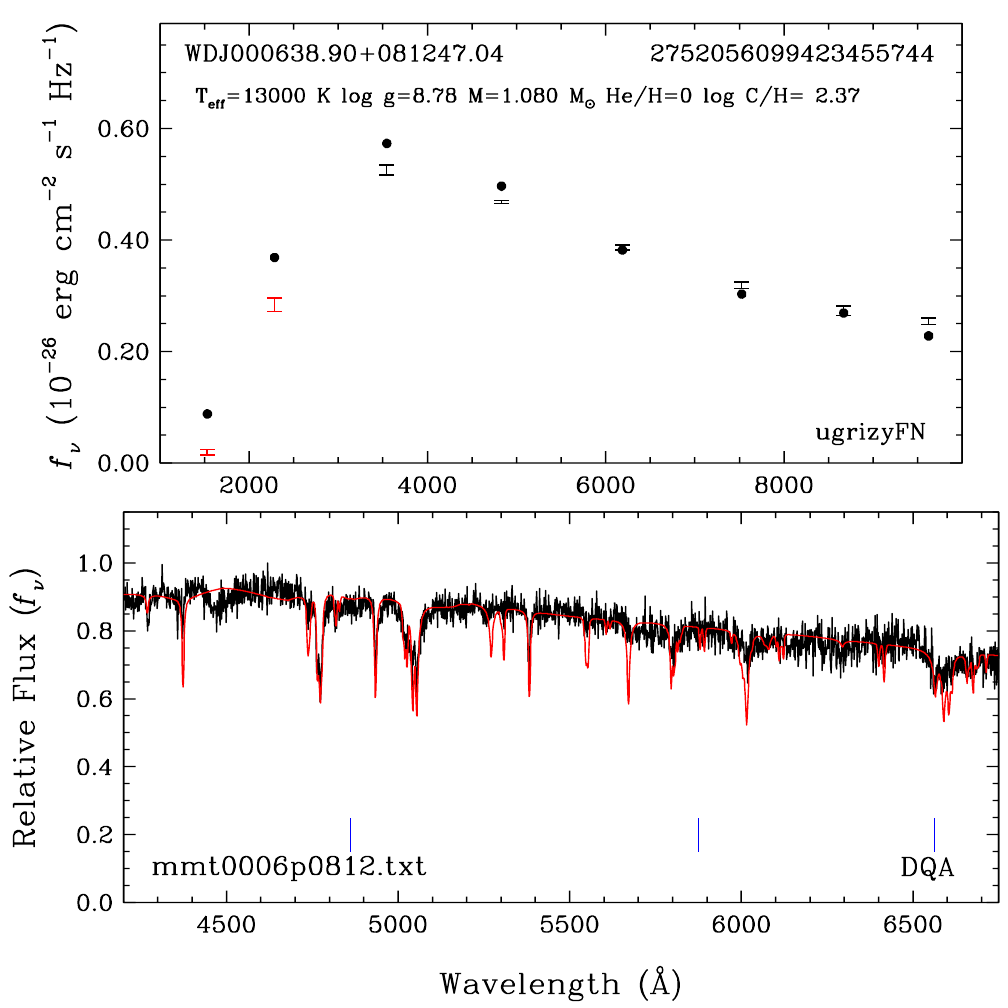}
\caption{Model fits to J0006+0812. The left panels show the results from the photometric fits, where the best-fitting model (red line) also predicts
C$_2$ Swan bands near 4737 and 5165 \AA. The right panels show a model fit where we arbitrarily increased the effective temperature to 13,000 K.}
\label{fig0006} 
\end{figure*}

Analyzing the DQ white dwarfs in the MWDD with Gaia DR2 parallaxes, \citet{coutu19} found
three oxygen-rich objects: J0901+5751, J0922+2928, and J1423+5729. The first object, J0901+5751, is warm and massive, and has GALEX
photometry available, and it is included in our sample. We identify an additional oxygen-rich warm DQ, J1235$-3802$ with even stronger oxygen lines
than J0901+5751. J2256$-$2130 may also show an oxygen line at 6156 \AA, but its spectrum is too noisy for confirmation. Given the presence
of oxygen in the spectrum, we classify these targets as DQZ, but note that there are no other metal lines detected in their optical spectra.

\citet{coutu19} provided a model atmosphere analysis of J0901+5751, but did not fit for the oxygen abundance in this star. 
Figure \ref{figox} shows our model fits. The top left panel shows the photometric fit for J0901+5751, which is based on models with no oxygen. The
bottom left panel shows the spectroscopic fit with oxygen added in, but with all the other parameters (temperature, mass, and C/H ratio) remaining the
same. We explored model fits with and without oxygen, and found that the continuum in the optical and the depth of the carbon lines are not affected
by the addition of oxygen. Hence, the temperature, mass, and C/H ratio derived from O-free models should be reliable. For J0901+5751, the best-fitting
model has $T_{\rm eff} = 14,193 \pm 271$ K, $M=1.117_{-0.017}^{+0.018}~M_{\odot}$, $\log$ C/H = 2.35, and $\log$ O/C = $-0.3$.

J1235$-$3802 is cooler with even stronger oxygen lines in its spectrum. The best-fitting model in this case has $T_{\rm eff} = 12,439 \pm 305$ K, $M=1.064_{-0.023}^{+0.025}~M_{\odot}$, $\log$ C/H = 2.46, and $\log$ O/C = +0.2. Hence, both stars have atmospheres dominated by
carbon and oxygen with trace amounts of hydrogen.
\citet{hollands20} constrained the oxygen abundance in the DAQ white dwarf J0551+4135 to  $\log$ O/H $<-4.5$ at 99 \% confidence level, for which
they find $\log$ C/H $=-0.83$. Hence, oxygen-rich DQs like J0901+5751 and J1235$-$3802 are unusual among the warm DQ population. 

Table \ref{tabpardq} presents the physical parameters for 72 warm DQs in our sample. Three previously known warm DQs  do not
have spectra available in the MWDD, and therefore they are missing from this table. In addition, one of the targets, J0146$-$0426 has an unrelated
nearby Gaia source (source ID 2480243390378263168) that contaminates its photometry in the $izy$ filters. We could not obtain
a good fit to its spectrum with parameters obtained from photometry for this target. We could only achieve a good fit by forcing the
effective temperature to 12,000 K. This fit is included in Table \ref{tabpardq}, but with an uncertain error in temperature.

\subsection{Caveats}

For warm DQs with effective temperatures around 12,000 K, our best-fitting models also predict molecular features (from C$_2$) that are not observed.
Figure \ref{fig0006} left panel shows our model fits to J0006+0812, which has $T_{\rm eff} = 11,983 \pm 100$ K and $M=1.020_{-0.016}^{+0.015}~M_{\odot}$. This model matches
the atomic C features reasonably well, but it also predicts C$_2$ absorption features at 4737 and 5165 \AA, which are not present in the observed
spectrum. There is a similar problem observed in classical DQs; \citet{dufour05} and \citet{coutu19}
showed that molecular and atomic carbon absorption features generally cannot be reproduced simultaneously when both are visible. 
\citet[][see their Figure 10]{kilic25} demonstrate that the best-fitting temperature is too low in those DQs, and that the spectral features
are reproduced better by increasing the effective temperature by about 1000 K. 

The right panels in Figure \ref{fig0006} show the results from a similar experiment, where we arbitrarily increased the effective temperature by about
1000 K for J0006+0812. Even though this leads to a worse photometric fit, this model provides a much better spectroscopic fit, and it does not
predict molecular features. We thus observe a similar problem here between C and C$_2$ dissociation balance. This is likely due to problems with
our poor understanding of carbon opacity in the UV. Carbon has so many lines in the UV that likely affect the entire spectrum in a way that we obtain
the wrong photometric temperature and C/C$_2$ balance. Hence, a caveat in our analysis is that the temperature scale of these cooler (warm) DQs 
remains uncertain.

\section{Discussion and Future Prospects}
\label{con}

\subsection{FUV is the Key}

Our survey is the first exploratory survey of its kind in taking advantage of FUV photometry to identify warm DQ white dwarfs.
Out of the 140 candidates in our sample with spectral classifications, 75 are warm DQ white dwarfs. This efficiency, 54\%, compares with the
0.1\% efficiency of the un-targeted SDSS DR14 spectroscopy for finding warm DQs \citep{koester19}. Based on the lessons learned from
our spectroscopic survey, it may be possible to design an even more efficient selection by targeting candidates in specific color
regions. 

\begin{figure}
\includegraphics[width=3.4in]{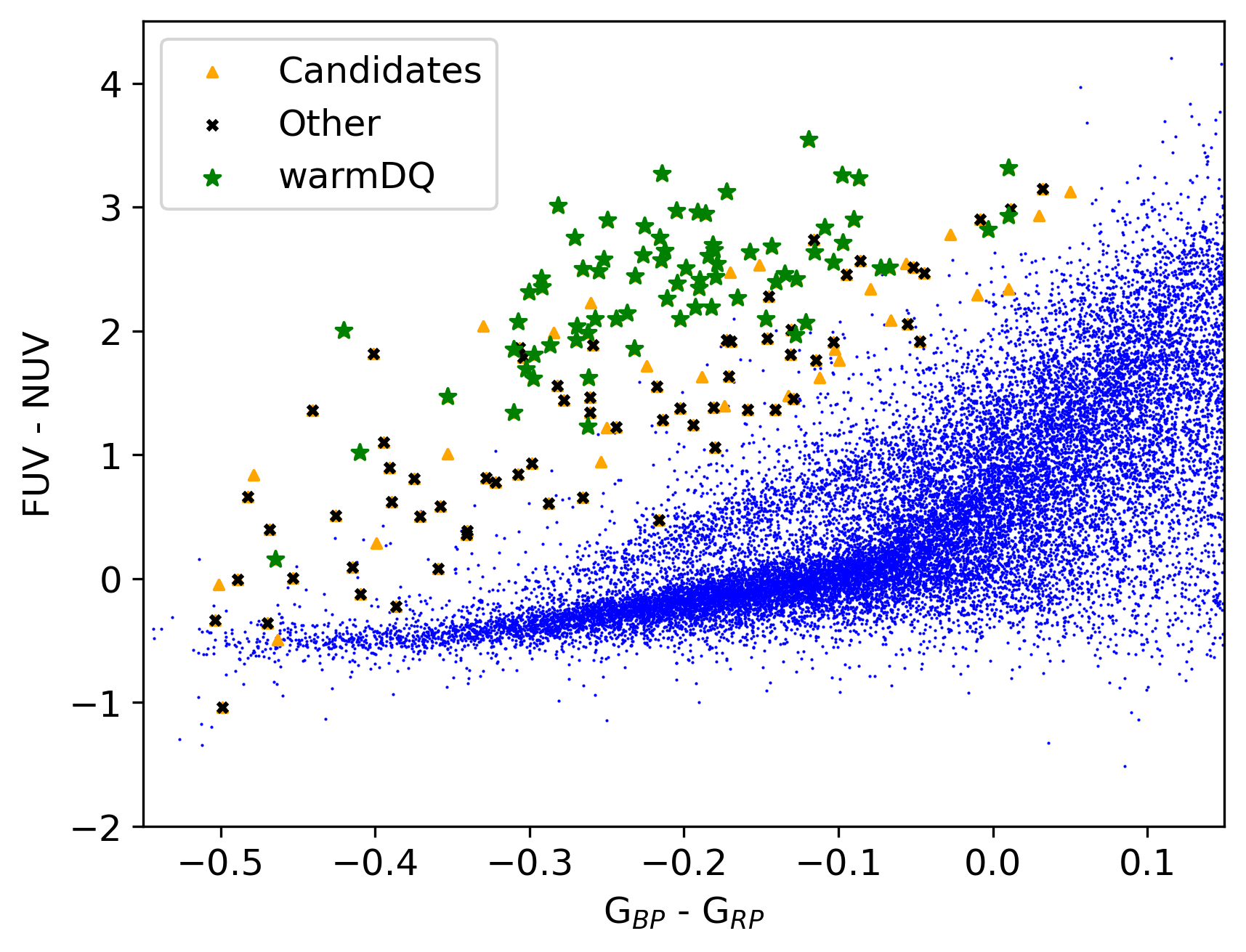}
\caption{${\rm FUV-NUV}$ versus $G_{\rm BP} - G_{\rm RP}$ color-color diagram of our sample. Symbols are the same as in Figure \ref{figsample}.}
\label{figfuvnuv} 
\end{figure}

Figure \ref{figfuvnuv} shows an ${\rm FUV-NUV}$ versus $G_{\rm BP} - G_{\rm RP}$ color-color diagram of our sample. Warm DQs are found between
$G_{\rm BP} - G_{\rm RP}=-0.46$ to +0.01 and ${\rm FUV-NUV}=0.16$ to 3.55. However, all but one of them have ${\rm FUV-NUV}>1.0$. The sample
mean and standard deviation are  $G_{\rm BP} - G_{\rm RP}=-0.21 \pm 0.09$ and ${\rm FUV-NUV}=2.38\pm0.56$. Hence, a future survey targeting
a larger sample of white dwarfs with these colors may be even more successful in finding larger samples of warm DQ white dwarfs.

\begin{figure}
\includegraphics[width=3.4in]{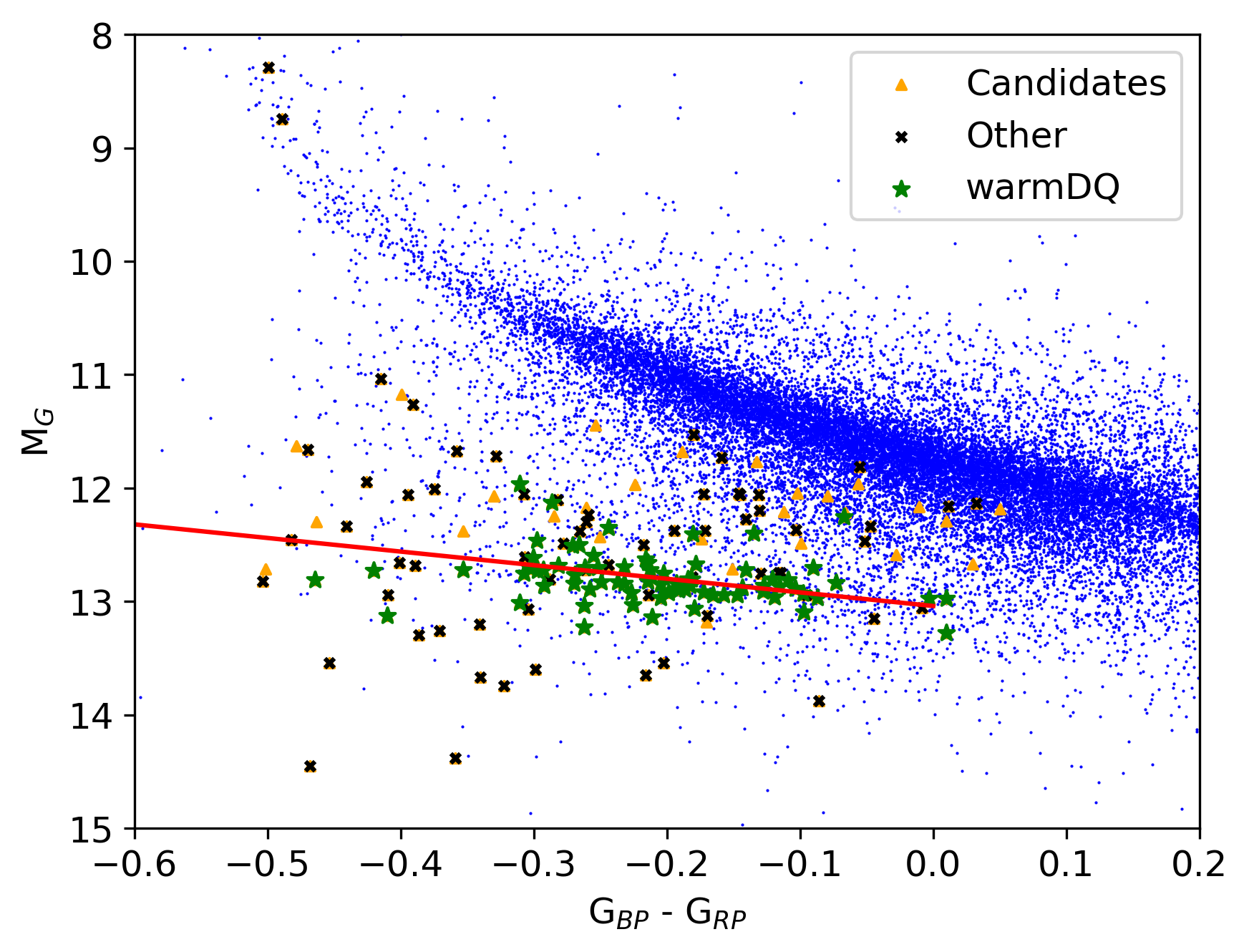}
\includegraphics[width=3.4in]{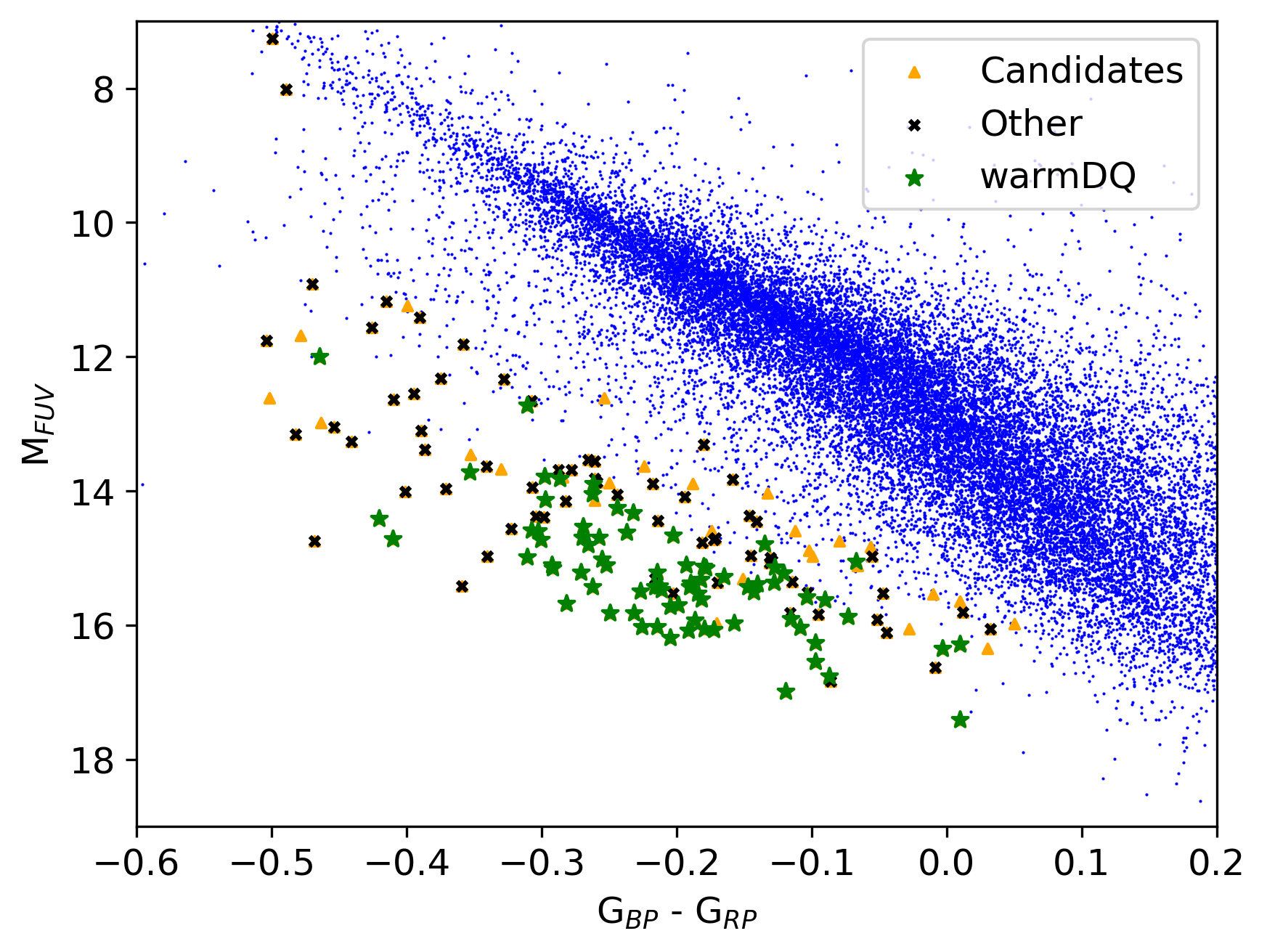}
\caption{Optical (top) and UV (bottom) color-magnitude diagram of our sample of spectroscopically confirmed warm DQs (green stars) compared to the
rest of the Gaia EDR3 white dwarfs \citep{gentile21} with GALEX data. The red line in the top panel marks  $\zeta=M_G - 1.2\ (G_{\rm BP} - G_{\rm RP}) = 13.04$.}
\label{fighr} 
\end{figure}

\begin{figure*}
\centering
\includegraphics[width=6in]{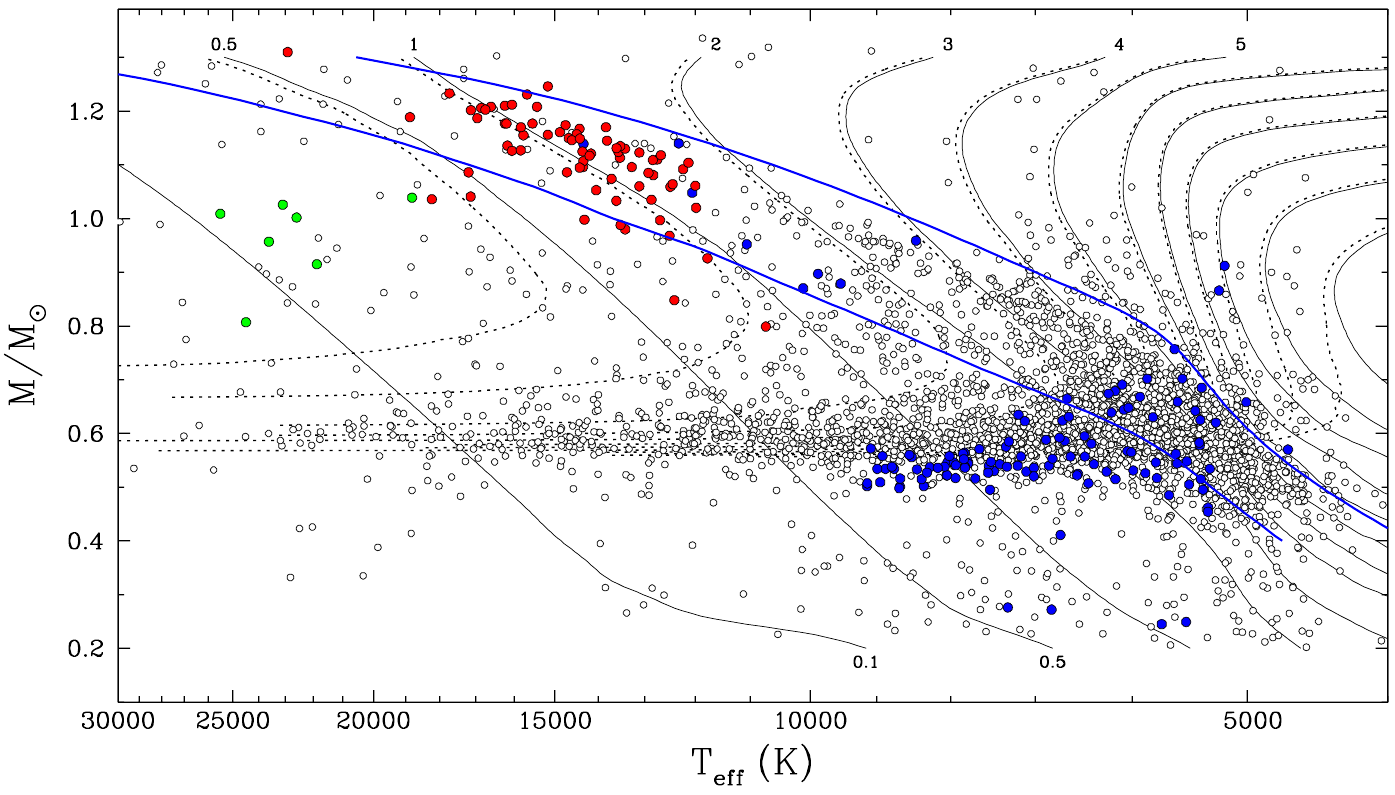} 
\caption{Stellar masses as a function of effective temperature for the spectroscopically confirmed white dwarfs in the 100 pc sample and the SDSS footprint \citep[white dots,][]{kilic20,kilic25}, including DQ white dwarfs (blue dots). Red dots mark our warm DQ sample, and the green dots show the hot DQs \citep{coutu19,koester19}. Solid curves are theoretical isochrones, labeled in units of Gyr, obtained from standard cooling sequences with CO-core compositions. Also shown are the same isochrones where the main-sequence progenitor lifetimes are taken into account (dotted lines).
The lower blue solid curve indicates the onset of crystallization at the
center of evolving models, while the upper one indicates the locations where 80\% of the total mass has solidified.}
\label{figtm}
\end{figure*}

Gaia Data Release 2 revealed an over-density of white dwarfs below the main white dwarf sequence, labeled the Q-branch \citep{gaia18}. 
\citet{tremblay19} showed that this over-density is
due to core crystallization and associated cooling delays \citep{blouin21,bedard24}. \citet{cheng19} further demonstrated that $\sim$6\% of high mass white dwarfs suffer from an 8 Gyr extra cooling delay on the Q-branch. 

Figure \ref{fighr} top panel shows an optical color-magnitude diagram of our sample of spectroscopically confirmed warm DQs, along with the rest of the
Gaia white dwarfs with GALEX FUV data. \citet{camisassa21} defined the parameter $\zeta = M_G - 1.2 (G_{\rm BP} - G_{\rm RP})$ to describe the location
of the Q-branch, which is delimited by $\zeta=13.0$ to 13.4. Interestingly, warm DQs cluster on the Q-branch with $\zeta$ ranging from 12.3 to 13.6.
The sample mean (marked by the red line) and standard deviation are $\zeta = 13.04 \pm 0.22$. Hence, the location of these warm
DQs on the Gaia color-magnitude diagram can also be used to refine target selection for future surveys where parallax measurements are available.
For example, limiting our spectroscopic sample to the $\pm3\sigma$ ranges of the $G_{\rm BP} - G_{\rm RP}$, ${\rm FUV-NUV}$, and $\zeta$ parameters would increase our success
rate from 54\% to 71\%. Kinematic selection can help further with the survey efficiency (see \S \ref{kin}.). 

The bottom panel in Figure \ref{fighr}  shows a UV-optical color-magnitude diagram taking advantage of FUV photometry. Since warm DQs suffer from
additional carbon opacity in the FUV, they are the faintest targets in their absolute FUV magnitudes at a given temperature (or optical color). They occupy
a distinct parameter space at the bottom left of the white dwarf sequence in this diagram. 

\subsection{Warm DQ Temperature and Mass Distributions}

Analyzing the warm DQ population in the SDSS spectroscopy sample with Gaia parallaxes, \citet{koester19} found 26 warm DQs with temperatures
ranging from 9350 to 16,740 K and masses from 0.86 to $1.19~M_{\odot}$. They demonstrated that DQs hotter than $T_{\rm eff} \approx 10,000$ K
are  on average $0.4~M_{\odot}$ more massive than the classical (cool) DQs. Similarly, \citet{coutu19} studied the mass
distribution of DQ white dwarfs with Gaia parallaxes, and identified  31 warm DQs in the same temperature range and with masses ranging from 0.81 to $1.15~M_{\odot}$, where 19 objects are common between the two samples.  In addition, \citet{kawka23} studied the mass distribution of C-enriched DQs
in a broader temperature range, and also found that they are more massive than the classical DQs. 

We revisit the mass and temperature distribution of warm DQs based on the 75 objects identified in our survey. Figure \ref{figtm} shows the stellar
masses as a function of temperature for our warm DQ sample (red dots) along with the the spectroscopically confirmed white dwarfs in the 100 pc
sample and the SDSS footprint \citep[white dots,][]{kilic20,kilic25}. Blue dots mark the DQ white dwarfs in the 100 pc sample, and green dots mark the
hot DQs \citep{coutu19,koester19}. We include the theoretical
isochrones for C/O-core white dwarfs with thick envelopes, $q({\rm He}) = 10^{-2}$ and $q$(H) = $10^{-4}$. Also shown are the same isochrones
where the progenitor lifetimes are taken into account (dotted lines). The blue curves mark the onset of crystallization in the core (lower curve) and
where 80\% of the star has solidified (upper curve).

Figure \ref{figtm} highlights the differences between warm DQs that are massive and the classical DQs that are found below 10,000 K and at
much lower masses. The sample mean and standard deviation for our warm DQ sample are $T_{\rm eff} = 14,560 \pm 1970$ K and
$M=1.11 \pm 0.09~M_{\odot}$. Warm DQs are roughly twice as massive as the classical DQs. 
In addition, the warm DQ population is found either on or near the crystallization sequence. 

The exact location of the onset of crystallization curve depends on the core composition and the thickness of the surface helium layer. Here, we show the onset of crystallization for a core made up of 50\% carbon and 50\% oxygen. A more oxygen-rich core would
crystallize earlier \citep[e.g.,][]{blatman24}, at hotter temperatures, shifting the crystallization curve to the left.
On the other hand, \citet{bedard24} found that the location of the Q branch requires models with thin helium layers, which makes sense
for merger products and is needed to explain the DAQ phenomenon \citep[e.g.,][]{hollands20}. However, thinner layers lead to later crystallization, which
would move the crystallization curves to the right. This is a stronger effect compared to the shift due to the core-composition. 
Hence, the exact location of the onset of crystallization shown in this figure is somewhat uncertain. 
Nevertheless, it is safe to conclude that we only see warm DQs either before or during core-crystallization. 

Even though the estimated cooling ages of warm DQs are
on the order of 1 Gyr, their kinematics indicate an origin in the thick disk or halo (see below). Hence, the overdensity of warm DQs on the
crystallization sequence is likely because they are stuck there for up to $\sim$10 Gyr due to distillation of neutron-rich impurities during
crystallization \citep{blouin21,bedard24}, which requires a C/O core. 
If warm DQs had ONe cores, they would not suffer from long cooling delays, and we would expect to see a significantly larger population of cool massive
DQs in the 100 pc sample. Hence, the number density of warm DQs as a function of temperature provides indirect evidence that they must
have C/O cores. 

\begin{deluxetable*}{llcccc}
\tabletypesize{\tiny}
\tablecolumns{6} \tablewidth{0pt}
\tablecaption{Physical Parameters for Warm DQ white dwarfs.\label{tabpardq}}
\tablehead{\colhead{Name} & \colhead{Comp} & \colhead{$T_{\rm eff}$ (K)} & \colhead{Mass} ($M_{\odot}$) & \colhead{$\log{g}$} & \colhead{Cooling Age (Gyr)}}
\startdata
WDJ000638.90+081247.04   & [C/H]=2.12 & $11983 \pm 100$ & 1.020$_{-0.016}^{+0.015}$ & 8.683$_{-0.021}^{+0.019}$ & 1.31 \\
WDJ001423.65+302214.88   & [C/H]=0.60 & $14418 \pm 854$ & 1.167$_{-0.063}^{+0.053}$ & 8.953$_{-0.104}^{+0.106}$ & 1.21 \\
WDJ004430.27$-$451348.26 & [C/H]=0.61 & $13115 \pm 694$ & 1.060$_{-0.042}^{+0.040}$ & 8.748$_{-0.056}^{+0.060}$ & 1.15 \\
WDJ011550.57+502545.98   & [C/H]=1.91 & $17727 \pm 349$ & 1.233$_{-0.011}^{+0.011}$ & 9.108$_{-0.025}^{+0.027}$ & 0.89 \\
WDJ011629.20$-$150147.25 & [C/H]=1.06 & $13114 \pm 319$ & 1.123$_{-0.058}^{+0.042}$ & 8.865$_{-0.087}^{+0.071}$ & 1.36 \\
WDJ014656.90$-$042613.50 & [C/H]=1.26 & $12000 \pm$  \nodata  & 1.061$_{-0.010}^{+0.012}$ & 8.752$_{-0.014}^{+0.016}$ & 1.46 \\
WDJ015756.65$-$051541.73 & [C/H]=1.28 & $13803 \pm 577$ & 1.145$_{-0.081}^{+0.055}$ & 8.909$_{-0.126}^{+0.100}$ & 1.27 \\
WDJ020549.70+205707.96   & [C/H]=0.97 & $17133 \pm 271$ & 1.202$_{-0.007}^{+0.008}$ & 9.032$_{-0.015}^{+0.015}$ & 0.89 \\
WDJ023633.74+250348.86   & [C/H]=2.00 & $14044 \pm 341$ & 1.053$_{-0.041}^{+0.037}$ & 8.736$_{-0.055}^{+0.052}$ & 0.90 \\
WDJ024459.25+423011.43   & [C/H]=1.75 & $14744 \pm 233$ & 1.174$_{-0.011}^{+0.010}$ & 8.967$_{-0.019}^{+0.020}$ & 1.17 \\
WDJ024524.49$-$251209.70 & [C/H]=2.76 & $13524 \pm 468$ & 1.113$_{-0.051}^{+0.033}$ & 8.844$_{-0.075}^{+0.054}$ & 1.23 \\
WDJ025719.30$-$614033.08 & [C/H]=1.06 & $12745 \pm 360$ & 1.110$_{-0.019}^{+0.018}$ & 8.839$_{-0.028}^{+0.030}$ & 1.41 \\
WDJ030446.58$-$642628.33 & [C/H]=0.80 & $13412 \pm 465$ & 0.980$_{-0.029}^{+0.031}$ & 8.615$_{-0.035}^{+0.039}$ & 0.77 \\
WDJ031054.74+220137.77   & [C/H]=1.54 & $16595 \pm 246$ & 1.208$_{-0.009}^{+0.010}$ & 9.045$_{-0.018}^{+0.021}$ & 0.97 \\
WDJ032521.11+254038.62   & [C/H]=0.96 & $16857 \pm 481$ & 1.206$_{-0.015}^{+0.016}$ & 9.041$_{-0.030}^{+0.033}$ & 0.94 \\
WDJ033845.40$-$534428.60 & [C/H]=1.90 & $13511 \pm 848$ & 0.988$_{-0.057}^{+0.062}$ & 8.628$_{-0.069}^{+0.081}$ & 0.77 \\
WDJ034140.15$-$252505.02 & [C/H]=1.09 & $14670 \pm 1201$ & 1.150$_{-0.122}^{+0.082}$ & 8.917$_{-0.182}^{+0.162}$ & 1.12 \\
WDJ042728.55$-$320603.04 & [C/H]=1.28 & $12828 \pm 359$ & 1.081$_{-0.031}^{+0.025}$ & 8.786$_{-0.044}^{+0.038}$ & 1.30 \\
WDJ044413.03$-$431509.09 & [C/H]=0.72 & $13705 \pm 874$ & 1.074$_{-0.039}^{+0.038}$ & 8.773$_{-0.054}^{+0.058}$ & 1.06 \\
WDJ053407.26$-$412843.94 & [C/H]=1.38 & $15164 \pm 3860$ & 1.246$_{-0.093}^{+0.114}$ & 9.146$_{-0.192}^{+0.326}$ & 1.23 \\
WDJ082329.59$-$085629.47 & [C/H]=0.73 & $15538 \pm 474$ & 1.177$_{-0.018}^{+0.019}$ & 8.974$_{-0.032}^{+0.037}$ & 1.05 \\
WDJ083135.57$-$223133.63 & [C/H]=$-$0.03 & $13557 \pm 157$ & 1.124$_{-0.007}^{+0.006}$ & 8.866$_{-0.011}^{+0.010}$ & 1.26 \\
WDJ090157.91+575135.91   & [C/H]=2.35 [O/C]=$-0.3$ & $14193 \pm 271$ & 1.117$_{-0.017}^{+0.018}$ & 8.851$_{-0.025}^{+0.030}$ & 1.10 \\
WDJ090229.67+303958.30   & [C/H]=1.36 & $14340 \pm 399$ & 1.096$_{-0.044}^{+0.028}$ & 8.812$_{-0.062}^{+0.044}$ & 1.00 \\
WDJ091723.70+654107.17   & [C/H]=0.65 & $14160 \pm 558$ & 1.121$_{-0.042}^{+0.037}$ & 8.860$_{-0.064}^{+0.062}$ & 1.13 \\
WDJ091851.74$-$052536.93 & [C/H]=1.55 & $12927 \pm 296$ & 1.085$_{-0.043}^{+0.032}$ & 8.794$_{-0.061}^{+0.048}$ & 1.29 \\
WDJ093026.42+641436.95   & [C/H]=1.87 & $15427 \pm 904$ & 1.208$_{-0.051}^{+0.042}$ & 9.046$_{-0.098}^{+0.097}$ & 1.14 \\
WDJ095837.16+585303.16   & [C/H]=0.89 & $16961 \pm 553$ & 1.187$_{-0.019}^{+0.022}$ & 8.997$_{-0.036}^{+0.043}$ & 0.87 \\
WDJ100911.54$-$215821.87 & [C/H]=0.62 & $16168 \pm 722$ & 1.136$_{-0.041}^{+0.041}$ & 8.887$_{-0.064}^{+0.073}$ & 0.82 \\
WDJ103655.39+652252.01   & [C/H]=2.22 & $16052 \pm 418$ & 1.126$_{-0.020}^{+0.021}$ & 8.868$_{-0.031}^{+0.035}$ & 0.80 \\
WDJ104906.61+165923.72   & [C/H]=1.47 & $13418 \pm 419$ & 1.130$_{-0.042}^{+0.039}$ & 8.879$_{-0.065}^{+0.066}$ & 1.31 \\
WDJ110058.04+175807.15   & [C/H]=1.86 & $13273 \pm 260$ & 1.096$_{-0.025}^{+0.021}$ & 8.814$_{-0.037}^{+0.032}$ & 1.24 \\
WDJ111811.89$-$392705.93 & [C/H]=1.79 & $11770 \pm 109$ & 0.926$_{-0.011}^{+0.012}$ & 8.533$_{-0.013}^{+0.013}$ & 0.96 \\
WDJ112538.72$-$311247.73 & [C/H]=1.27 & $12864 \pm 472$ & 1.035$_{-0.030}^{+0.034}$ & 8.706$_{-0.039}^{+0.047}$ & 1.11 \\
WDJ113918.84$-$391857.31 & [C/H]=0.97 & $15764 \pm 739$ & 1.155$_{-0.027}^{+0.027}$ & 8.927$_{-0.045}^{+0.050}$ & 0.95 \\
WDJ120331.90+645101.41   & [C/H]=2.30 & $12837 \pm 73$ & 1.109$_{-0.006}^{+0.006}$ & 8.838$_{-0.010}^{+0.009}$ & 1.39 \\
WDJ123549.68$-$380203.98 & [C/H]=2.46 [O/C]=0.2 & $12439 \pm 305$ & 1.064$_{-0.023}^{+0.025}$ & 8.757$_{-0.032}^{+0.035}$ & 1.35 \\
WDJ125938.32$-$260158.23 & [C/H]=1.24 & $14486 \pm 361$ & 1.157$_{-0.021}^{+0.021}$ & 8.932$_{-0.036}^{+0.038}$ & 1.17 \\
WDJ130717.12$-$375839.08 & [C/H]=1.04 & $14334 \pm 1304$ & 1.107$_{-0.062}^{+0.061}$ & 8.833$_{-0.089}^{+0.102}$ & 1.04 \\
WDJ133151.40+372755.15   & [C/H]=1.48 & $16744 \pm 382$ & 1.203$_{-0.012}^{+0.013}$ & 9.034$_{-0.025}^{+0.026}$ & 0.94 \\
WDJ133941.33+015749.19   & [C/H]=1.21 & $14309 \pm 433$ & 0.998$_{-0.097}^{+0.060}$ & 8.643$_{-0.116}^{+0.080}$ & 0.67 \\
WDJ143437.84+225859.68   & [C/H]=1.69 & $14416 \pm 336$ & 1.095$_{-0.034}^{+0.032}$ & 8.810$_{-0.048}^{+0.051}$ & 0.98 \\
WDJ143534.18+043425.70   & [C/H]=1.97 & $17199 \pm 630$ & 1.086$_{-0.074}^{+0.055}$ & 8.792$_{-0.103}^{+0.087}$ & 0.53 \\
WDJ144854.90+051903.80   & [C/H]=1.16 & $15827 \pm 318$ & 1.170$_{-0.013}^{+0.012}$ & 8.960$_{-0.024}^{+0.021}$ & 0.98 \\
WDJ145524.89+420910.81   & [C/H]=2.17 & $13602 \pm 487$ & 1.033$_{-0.056}^{+0.054}$ & 8.702$_{-0.071}^{+0.076}$ & 0.91 \\
WDJ153450.85+354034.09   & [C/H]=1.52 & $17140 \pm 507$ & 1.041$_{-0.035}^{+0.035}$ & 8.713$_{-0.046}^{+0.049}$ & 0.47 \\
WDJ160514.47$-$075528.74 & [C/H]=1.28 & $15831 \pm 327$ & 1.127$_{-0.021}^{+0.020}$ & 8.871$_{-0.034}^{+0.033}$ & 0.84 \\
WDJ162236.25+300455.29   & [C/H]=1.43 & $16235 \pm 242$ & 1.177$_{-0.007}^{+0.007}$ & 8.974$_{-0.013}^{+0.013}$ & 0.94 \\
WDJ171034.72$-$200541.95 & [C/H]=1.47 & $14597 \pm 206$ & 1.146$_{-0.010}^{+0.010}$ & 8.910$_{-0.017}^{+0.016}$ & 1.12 \\
WDJ175253.90$-$663452.57 & [C/H]=3.00 & $12675 \pm 136$ & 1.118$_{-0.007}^{+0.007}$ & 8.856$_{-0.011}^{+0.010}$ & 1.46 \\
WDJ175631.55+372827.41   & [C/He]=$-$3.00 & $10730 \pm 260$ & 0.799$_{-0.047}^{+0.041}$ & 8.341$_{-0.052}^{+0.045}$ & 0.90 \\
WDJ175821.12+590644.92   & [C/H]=1.82 & $18873 \pm 453$ & 1.189$_{-0.009}^{+0.009}$ & 9.000$_{-0.017}^{+0.017}$ & 0.65 \\
WDJ182531.11$-$514834.11 & [C/H]=1.24 & $15672 \pm 2337$ & 1.231$_{-0.066}^{+0.062}$ & 9.103$_{-0.133}^{+0.173}$ & 1.14 \\ 
WDJ190619.92+202137.67   & [C/H]=1.61 & $16235 \pm 513$ & 1.210$_{-0.015}^{+0.013}$ & 9.050$_{-0.029}^{+0.029}$ & 1.03 \\
WDJ190956.00$-$533734.45 & [C/H]=1.72 & $13827 \pm 3136$ & 1.170$_{-0.165}^{+0.140}$ & 8.960$_{-0.248}^{+0.347}$ & 1.33 \\
WDJ192555.20$-$034626.55 & [C/H]=1.82 & $14876 \pm 138$ & 1.161$_{-0.006}^{+0.005}$ & 8.940$_{-0.010}^{+0.009}$ & 1.11 \\
WDJ193431.84$-$653025.80 & [C/H]=1.32 & $13517 \pm 468$ & 1.135$_{-0.019}^{+0.020}$ & 8.888$_{-0.031}^{+0.034}$ & 1.30 \\
WDJ194919.09$-$420025.11 & [C/H]=1.76 & $14712 \pm 837$ & 1.086$_{-0.042}^{+0.045}$ & 8.794$_{-0.060}^{+0.069}$ & 0.89 \\
WDJ203508.73$-$574026.84 & [C/H]=1.13 & $12689 \pm 995$ & 0.997$_{-0.074}^{+0.078}$ & 8.643$_{-0.090}^{+0.105}$ & 0.99 \\
WDJ205700.44$-$342556.37 & [C/H]=1.04 & $22916 \pm 1117$ & 1.310$_{-0.015}^{+0.017}$ & 9.352$_{-0.048}^{+0.051}$ & 0.62 \\
WDJ210901.84+255821.12   & [C/H]=1.39 & $16052 \pm 391$ & 1.212$_{-0.013}^{+0.013}$ & 9.056$_{-0.028}^{+0.027}$ & 1.06 \\
WDJ214023.96$-$363757.44 & [C/H]=2.04 & $12233 \pm 78$ & 1.092$_{-0.004}^{+0.005}$ & 8.807$_{-0.007}^{+0.007}$ & 1.50 \\
WDJ214713.66+111005.68   & [C/H]=1.94 & $15166 \pm 447$ & 1.156$_{-0.048}^{+0.043}$ & 8.929$_{-0.078}^{+0.081}$ & 1.05 \\
WDJ215755.84$-$114844.06 & [C/H]=3.00 & $18230 \pm 547$ & 1.036$_{-0.049}^{+0.044}$ & 8.702$_{-0.063}^{+0.062}$ & 0.39 \\
WDJ222658.76$-$375629.39 & [C/H]=1.45 & $12132 \pm 376$ & 1.104$_{-0.031}^{+0.031}$ & 8.829$_{-0.046}^{+0.049}$ & 1.56 \\
WDJ223421.25+144521.41   & [C/H]=1.86 & $12500 \pm 198$ & 0.968$_{-0.115}^{+0.107}$ & 8.597$_{-0.134}^{+0.141}$ & 0.91 \\
WDJ224234.11$-$480404.19 & [C/H]=1.70 & $12486 \pm 500$ & 1.059$_{-0.048}^{+0.051}$ & 8.749$_{-0.065}^{+0.074}$ & 1.31 \\
WDJ225635.07$-$213015.24 & [C/H]=1.67 & $16192 \pm 651$ & 1.177$_{-0.042}^{+0.030}$ & 8.974$_{-0.074}^{+0.060}$ & 0.95 \\
WDJ230930.47+282540.41   & [C/H]=1.28 & $14354 \pm 512$ & 1.125$_{-0.038}^{+0.035}$ & 8.867$_{-0.058}^{+0.059}$ & 1.10 \\
WDJ233410.23$-$262257.02 & [C/H]=2.05 & $12407 \pm 153$ & 0.848$_{-0.023}^{+0.021}$ & 8.412$_{-0.025}^{+0.023}$ & 0.69 \\
WDJ234501.17$-$062543.22 & [C/H]=1.42 & $14414 \pm 271$ & 1.149$_{-0.012}^{+0.012}$ & 8.915$_{-0.020}^{+0.021}$ & 1.16 \\
WDJ234933.31$-$232036.52 & [C/H]=1.00 & $13598 \pm 329$ & 1.131$_{-0.018}^{+0.017}$ & 8.880$_{-0.029}^{+0.028}$ & 1.27 
\enddata
\end{deluxetable*}

\subsection{Kinematics}
\label{kin}

\begin{figure}
\includegraphics[width=3.4in]{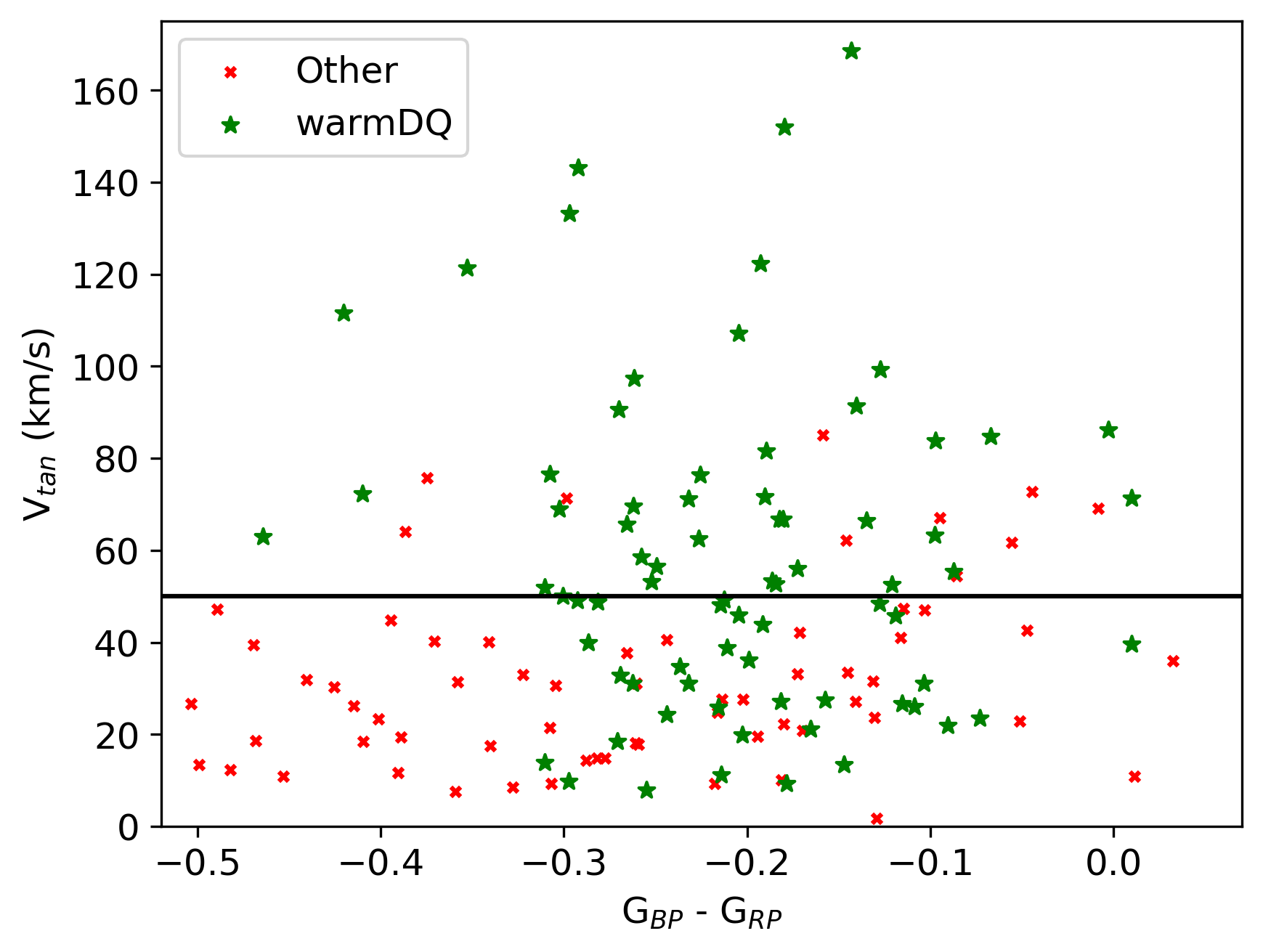}
\caption{Tangential velocity versus $G_{\rm BP} - G_{\rm RP}$ color for our sample of DQ (green stars) and
non-DQ white dwarfs (red crosses). The black line shows the 50 km s$^{-1}$ limit.}
\label{figvtan}
\end{figure}

Figure \ref{figvtan} shows the tangential velocity
distribution of the DQ and non-DQ white dwarfs in our sample, along with the 50 km s$^{-1}$ limit, which \citet{wegg12} designated as
the smoking gun signature of merger remnants.  Out of the 140 objects in our sample with spectral classifications, the 13 fastest objects are warm DQs. In addition, 41 of the 75 DQs are above the $v_{\rm tan}\geq50$ km s$^{-1}$ limit. This is in contrast
to 10 non-DQs above the same velocity limit. Hence, warm DQs make up 80\% of the stars with V$_{\rm tan}\geq50$ km s$^{-1}$ in our FUV-selected
sample. 

\citet{coutu19} found that 10 out of 22, or 45\%, of their warm DQ population has tangential velocities higher than 50 km s$^{-1}$,
compared to a fraction of only 5\% for the DA white dwarfs with similar atmospheric parameters. Hence, warm DQs are clearly unusual in their kinematics.
Their kinematics, high masses, and unusual atmospheric compositions indicate a merger origin \citep{dunlap15,coutu19,kawka23}.

Out of the 167 FUV-selected targets, we were not able to observe 27 candidates, 11 of which are more massive than $0.9~M_{\odot}$ based
on pure H atmosphere model fits. Among these massive candidates, five have tangential velocities higher than 50 km s$^{-1}$. Since
our survey has an 80\% success rate above this velocity limit, those five candidates, WDJ002404.89$-$102917.89, WDJ003511.67$-$134440.78, WDJ010802.37$-$245227.57, WDJ013624.41+445017.10, and WDJ233429.08$-$321618.14, are the best targets for follow-up spectroscopy.

\subsection{Future Prospects: UVEX and CASTOR}

The Ultraviolet Explorer (UVEX) is a medium-class explorer mission with a target launch date of 2030. UVEX will have FUV and NUV imaging
bands similar to GALEX. It will perform an all-sky survey that is 50/100 times deeper than GALEX in the NUV/FUV \citep{kulkarni21}. 
In a single visit, UVEX will approach a 5$\sigma$ limiting magnitude of $\sim$24.7 in the FUV for a typical field outside of the Galactic plane.
After 10 repeated visits, the limiting magnitude will approach 26. Hence, a combination of UVEX FUV photometry and optical photometry
from the Legacy Survey of Space and Time (LSST) at the Rubin Observatory would provide an unprecedented opportunity to identify warm DQs.

The Cosmological Advanced Survey Telescope for Optical and ultraviolet Research \citep[CASTOR,][]{cheng24} is another proposed mission that
will provide wide-field optical and UV photometry with a target launch date in late 2020s. CASTOR will have a broadband UV filter covering the
1500-3000 \AA\ range, which includes the numerous C absorption lines in the FUV. With a target limiting magnitude of $\sim27$ in $\sim600$ s,
CASTOR will image $\sim$5\% of the sky within the first 5 years. Hence, UV photometry from CASTOR can also be used to identify warm DQs in
its wide-field survey footprint. 

The absolute FUV magnitudes of the warm DQs in our sample range from 12.0 to 17.4 (see Figure \ref{fighr}). UVEX's and CASTOR's limiting
magnitudes of $\geq26$ will expand the discovery horizon for the faintest warm DQs to beyond 500 pc. Even though these faint white dwarfs will not have
precise parallax measurements available, they can still be identified through spectroscopic follow-up of the outliers in FUV-optical color-color
diagrams from a combination of UVEX and wide-field optical surveys like the LSST.

\begin{acknowledgements}

This work is supported in part by the NSF under grant  AST-2205736, the NASA under grants 80NSSC22K0479, 80NSSC24K0380, and 80NSSC24K0436, the NSERC Canada, the Fund FRQ-NT (Qu\'ebec), and the Smithsonian Institution.

Based on observations obtained at the MMT Observatory, a joint facility of the Smithsonian  Institution and the University of Arizona.

This paper includes data gathered with the 6.5 meter Magellan Telescopes located at Las Campanas Observatory, Chile.

Based on observations obtained at the international Gemini Observatory, a program of NSF's NOIRLab, which is managed by the Association of Universities for Research in Astronomy (AURA) under a cooperative agreement with the National Science Foundation on behalf of the Gemini Observatory partnership: the National Science Foundation (United States), National Research Council (Canada), Agencia Nacional de Investigaci\'{o}n y Desarrollo (Chile), Ministerio de Ciencia, Tecnolog\'{i}a e Innovaci\'{o}n (Argentina), Minist\'{e}rio da Ci\^{e}ncia, Tecnologia, Inova\c{c}\~{o}es e Comunica\c{c}\~{o}es (Brazil), and Korea Astronomy and Space Science Institute (Republic of Korea).

\end{acknowledgements}

\facilities{Gemini:Gillett (GMOS spectrograph), Gemini:South (GMOS-S spectrograph), Magellan: Baade (MagE spectrograph), MMT (Blue Channel spectrograph)}

%\bibliographystyle{aasjournal}
%\bibliography{/Users/nova/AAStex/ref}

\appendix

\section{DA White Dwarfs}
\label{app}

\begin{figure*}[h]
\centering
\includegraphics[width=3.0in]{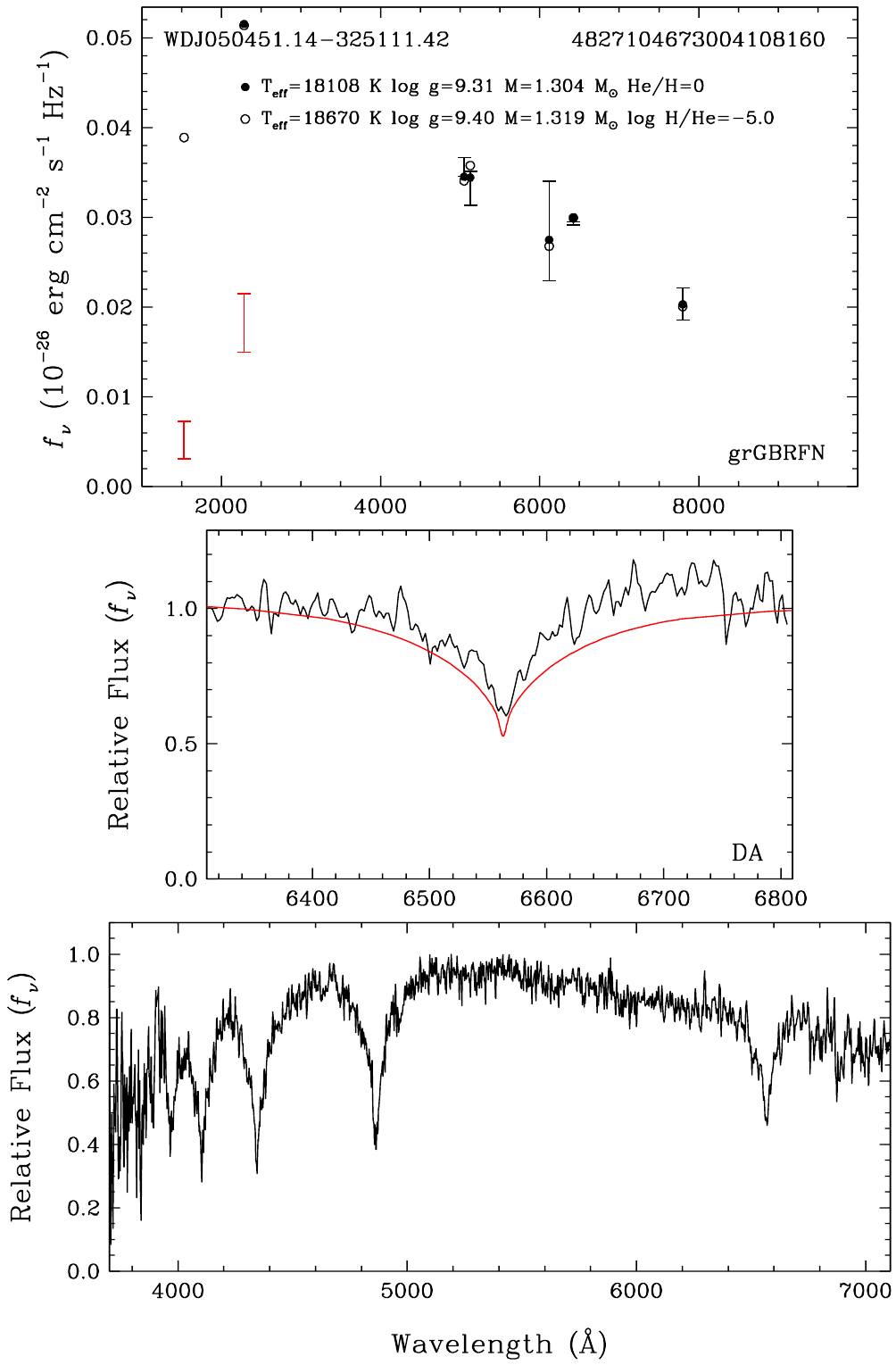}
\includegraphics[width=3.0in]{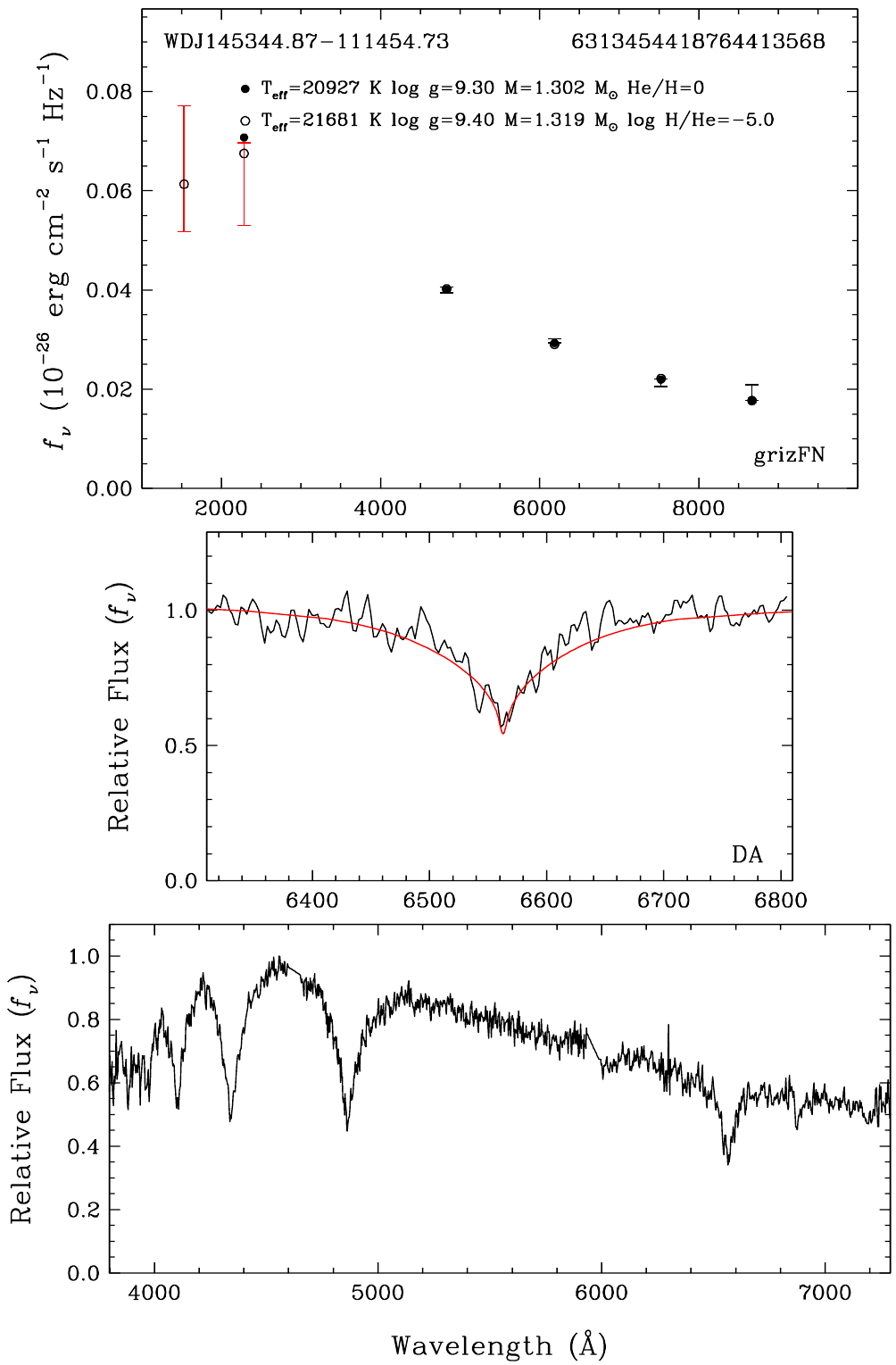}
\caption{Model atmosphere fits to the three most massive DA white dwarfs in our sample. For each star, the top panel shows the best-fitting pure H (filled dots)
and pure He (open circles) atmosphere white dwarf models to the photometry (error bars). This panel also includes the object name,
Gaia Source ID,  and the photometry used in the fitting. The middle panel shows the predicted spectrum (red line) based on the pure H solution.
The bottom panel shows a broader wavelength range.}
\label{figda} 
\end{figure*}

There are 21 normal DA white dwarfs in our sample, including 14 new DAs presented here for the first time. Figure \ref{figda} shows our model fits for
two of the most massive DAs in our sample. 
For each star, the top panel shows the available photometry (error bars) along with the predicted fluxes
from the best-fitting pure H (filled dots) and He-dominated (open circles) atmosphere models with $\log$ H/He = $-5$. The labels in the same panel
give the name, Gaia Source ID, and the photometry used in the fitting. Any excluded photometric bandpasses are shown in red. The middle panel shows
the predicted spectrum based on the pure H solution, along with the observed H$\alpha$ line. This is not a fit to the line profile: we over-plot the predicted
H line (red line) from the photometric fit. The bottom panel shows a broader spectral range. The photometric fits provide an excellent match to the observed
H$\alpha$ line profiles for these stars, confirming a pure H composition and masses $\sim$$1.3~M_{\odot}$. 
We provide the model fits for all targets in our sample (except for two DAs and three DQs that do not have their spectra available in the MWDD) on Zenodo, which can be accessed via the DOI \href{https://doi.org/10.5281/zenodo.15733428}{10.5281/zenodo.15733429}.
The model grids used for each spectral type are described in \citet{kilic25}. Table \ref{tabparot} presents the physical parameters for all non-DQ white dwarfs in our sample. 

\begin{figure}
\centering
\includegraphics[width=3.4in]{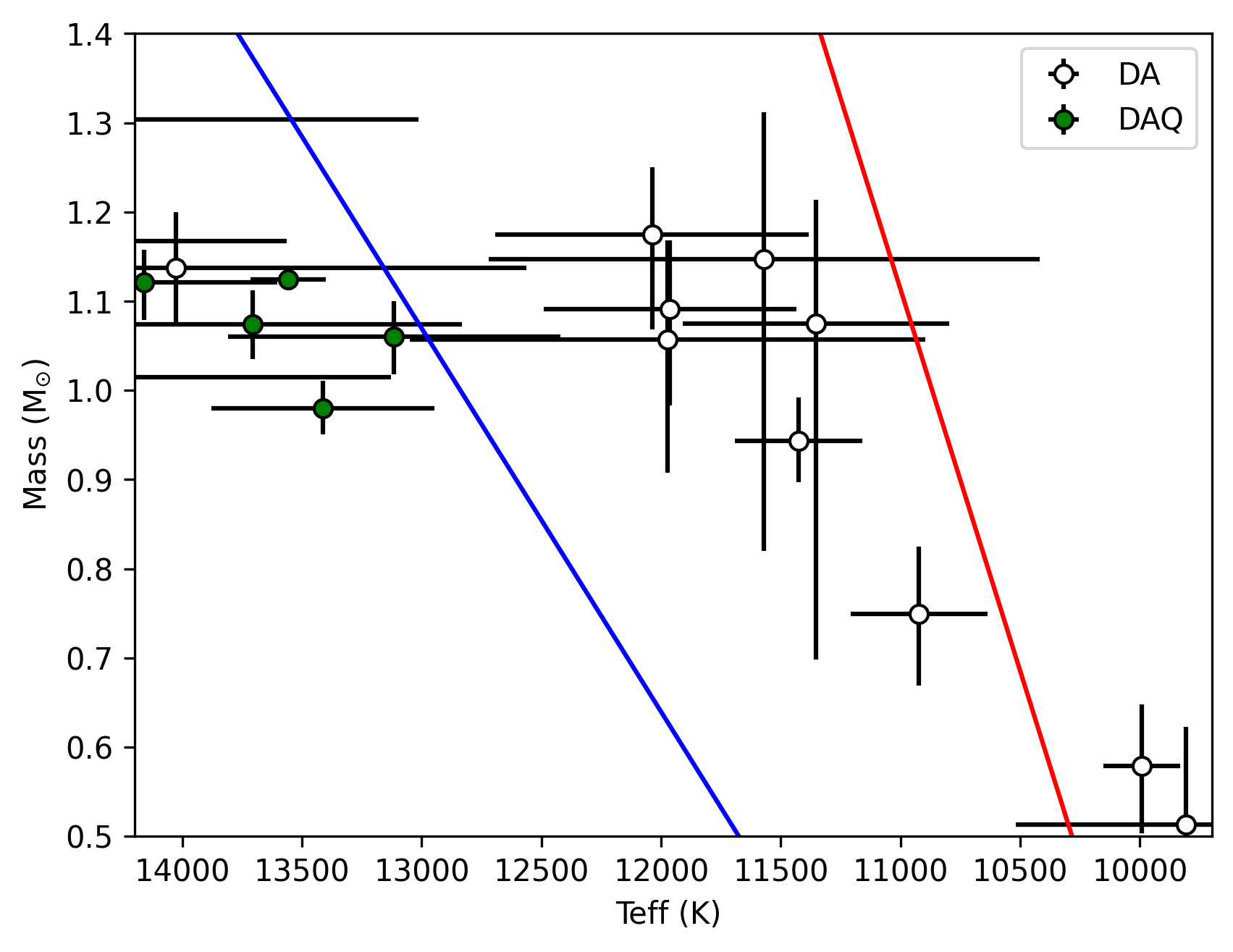}
\caption{Masses and effective temperatures for DA and DAQ white dwarfs in our sample along with the empirical boundaries of the ZZ Ceti instability strip
from \citet{vincent20}.}
\label{figzzceti} 
\end{figure}

Due to our relatively blue color-selection of $G_{\rm BP} - G_{\rm RP}=-0.5$ to +0.05, all of the DAs in our sample are relatively hot with $T_{\rm eff}\gtrsim10,000$ K.
Interestingly, 11 of them are more massive than $1~M_{\odot}$. In addition, seven of these DAs are within the ZZ Ceti instability strip, hence
may show pulsations. Figure \ref{figzzceti} shows the boundaries of the ZZ Ceti instability strip along with the DA and DAQ white dwarfs in our sample.
The seven ZZ Ceti candidates are: J0247+3327, J0947+3724, J1058$-$0005, J1539+4953, J1846+5138, J2015+0338, J2323+1221.
Follow-up high-cadence photometry of these seven targets would be essential for confirming their nature. J1058$-$0005 is the most massive
ZZ Ceti candidate in our sample with $M=1.175^{+0.075}_{-0.107}~M_{\odot}$. Among the DAQ white dwarfs, J0044$-$4513 is closest to the
blue edge of the ZZ Ceti instability strip. Even though it is slightly less massive than J0551+4135, the only pulsating DAQ currently known
\citep{hollands20,vincent20,kilic24}, follow-up time series photometry would be helpful to search for pulsations in this system.

\begin{figure*}
\includegraphics[width=2.3in]{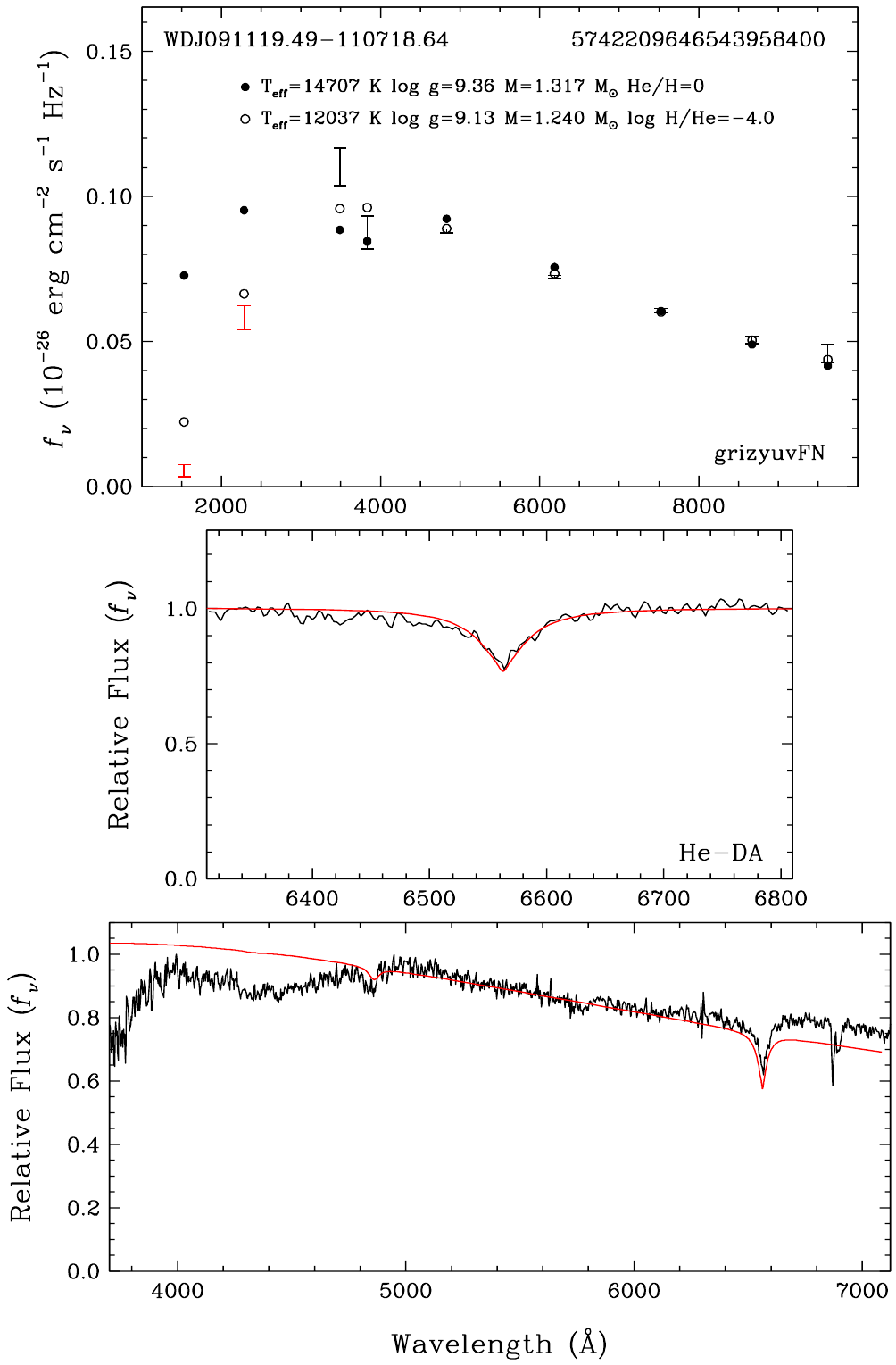}
\includegraphics[width=2.3in]{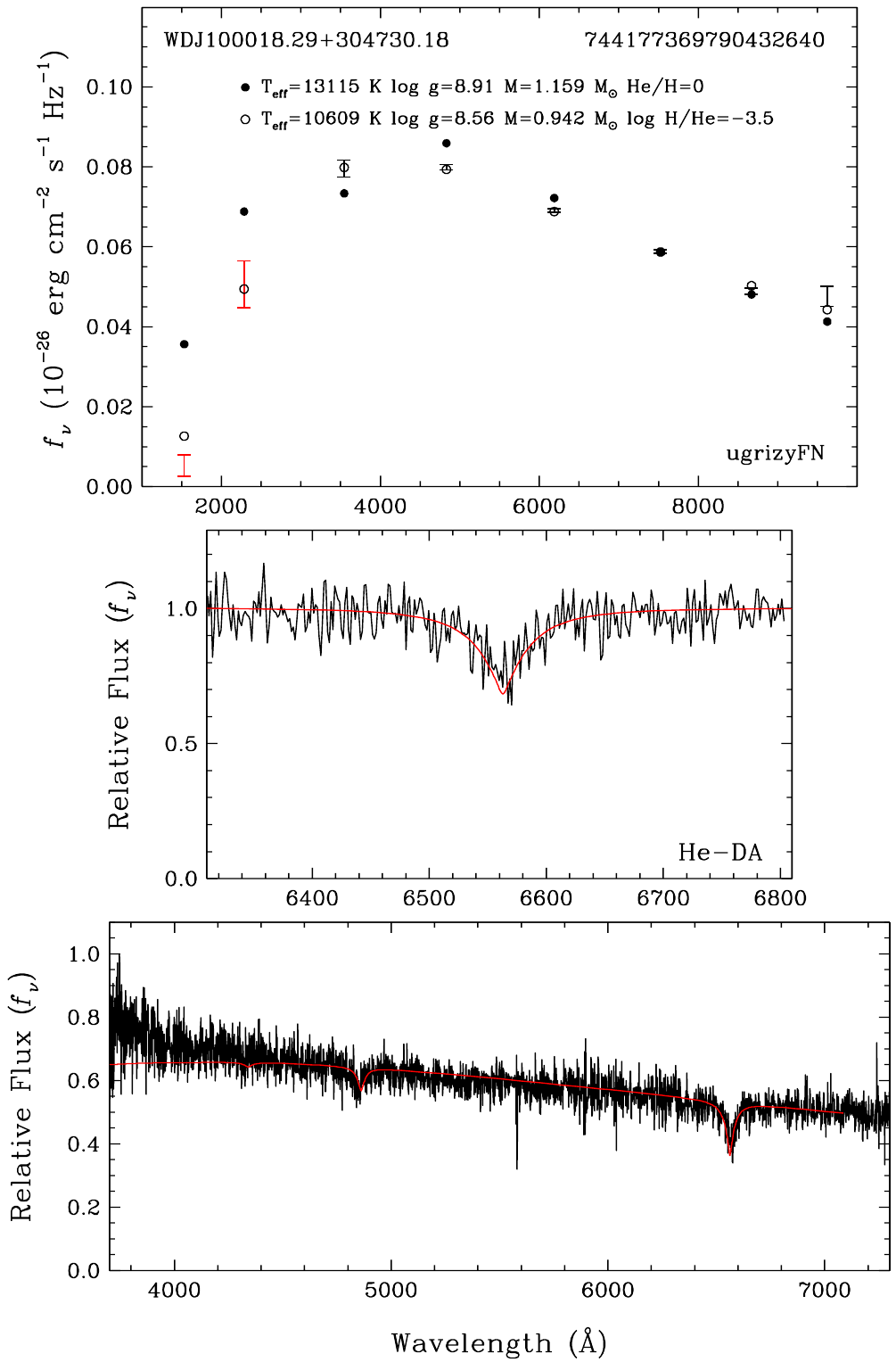}
\includegraphics[width=2.3in]{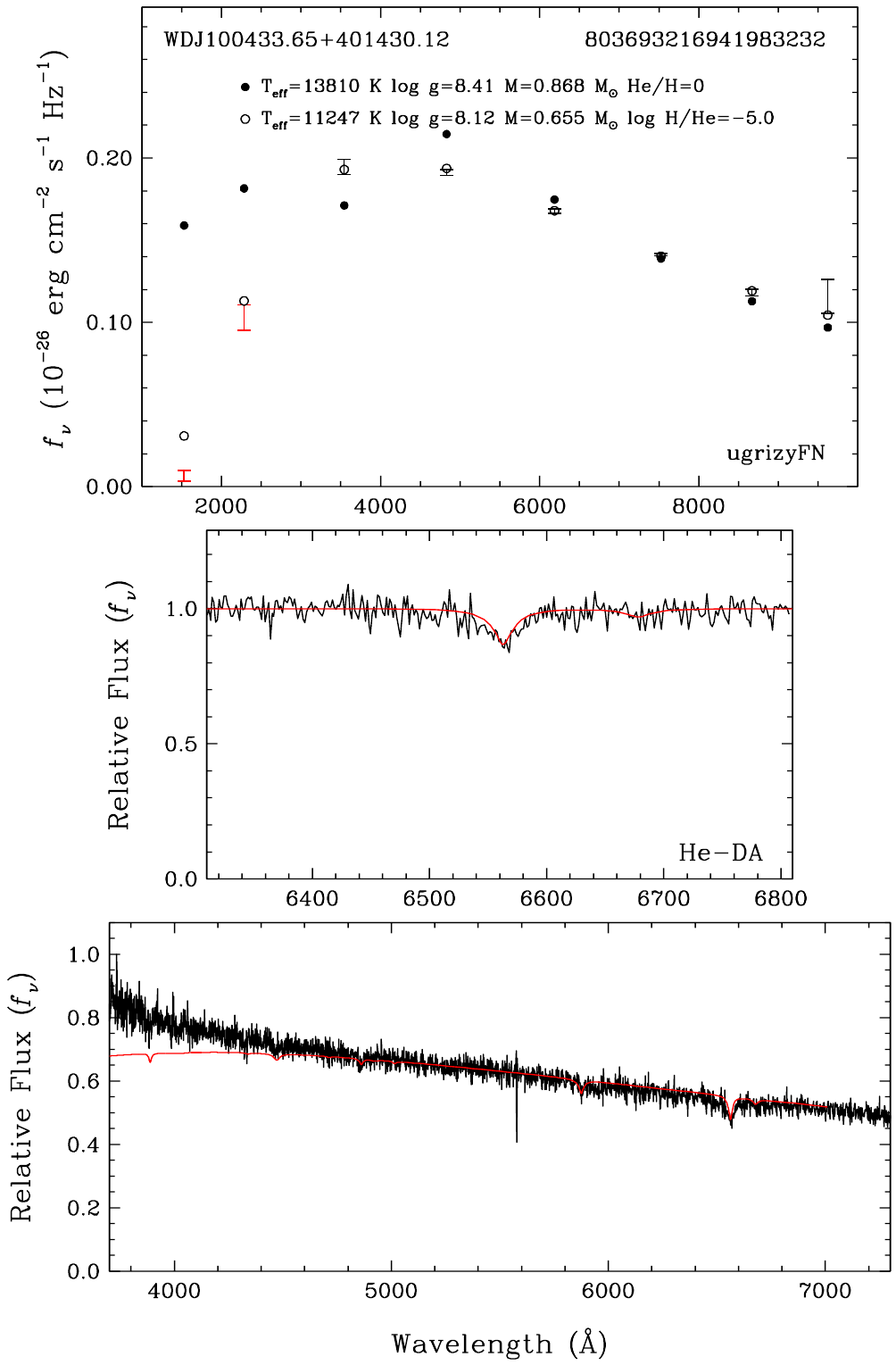}
\caption{Model atmosphere fits to He-DAs. The middle and bottom panels show the spectroscopic fits using mixed H/He atmospheres.}
\label{figheda} 
\end{figure*}

We identify three additional DA white dwarfs where the spectral energy distributions are better reproduced by He-dominated atmospheres. Figure
\ref{figheda} shows our model fits to these He-DA white dwarfs using mixed H/He atmospheres (middle and bottom panels). Balmer lines in these objects are relatively weak since they are heavily broadened
through van der Waals interactions. Both the UV-optical photometry and the observed Balmer line profiles clearly require He-dominated atmospheres
in these stars. Our model fits with $T_{\rm eff}\sim11,000$-12,000 K and trace amounts of hydrogen ($\log$ H/He $=-5$ to $-3.5$) provide excellent fits to the data. 

These three He-DAs are relatively massive with $M=0.87, 1.16$, and $1.29~M_{\odot}$, respectively. \citet{kilic25} found 11 He-DAs in the 100 pc sample within
the SDSS footprint, including J0050+3138 with $T_{\rm eff} = 12,518 \pm 207$ K and $M=1.215 \pm0.008~M_{\odot}$. The rest of their He-DA sample
has stars with $T_{\rm eff}$ in the range 7200-10,200 K and $M=0.59-0.82~M_{\odot}$. Hence, the hottest He-DAs appear to be
relatively massive, though this statement is based on only four stars. As the surface convection zone deepens in DA white dwarfs below about 15,000 K, we
expect the ones with thin H layers to mix first and turn into He-rich atmosphere white dwarfs. With decreasing temperature and a deepening convection
zone, DAs with thicker layers mix later \citep[see][and references therein]{bedard24b}. This would indicate that the warm He-DAs in our sample likely
had thin surface H layers. Two of the most massive He-DAs in our sample, J0911$-$1107 and J1000+3047, have $v_{\rm tan}>50$ km s$^{-1}$, which
favors a kinematically old population and a likely merger remnant, that could explain a relatively thin H envelope.

\section{Magnetic DAs}

There are 24 magnetic DAs in our sample. We first calculate a model atmosphere using the parameters obtained from photometry under the assumption of
a pure H composition.  We construct a grid of magnetic spectra with varying  viewing angle $i$, dipole field strength $B_{d}$, and dipole offset
$a_{z}$, given in units of the stellar radius. We calculate the total line opacity as the sum of the individual Stark-broadened Zeeman components. The specific intensities at the surface, $I(\nu,\mu,\tau_{\nu}=0)$, are obtained by solving the radiative transfer equation for various field strengths and values
of $\mu$ ($\mu= \cos \theta$), where $\theta$ is the angle between the propagation of light and the normal to the surface of the star. Further details of these magnetic models are provided in \citet{moss24}. For $B\sim1$ MG fields with relatively small Zeeman split features, we fit the spectrum in the H$\alpha$
region, which provides the best constraints on the field strength. For larger fields, we use the entire spectrum. We use the Pikaia genetic algorithm
\citep{carbonneau95} to find the best fit within a specified range of magnetic fields (usually from 0 to 800 MG), viewing angles (0 to 90 degrees),
and dipole offsets ($a_z$ from $-0.8$ to +0.8).

\begin{figure*}[h]
\includegraphics[width=2.3in]{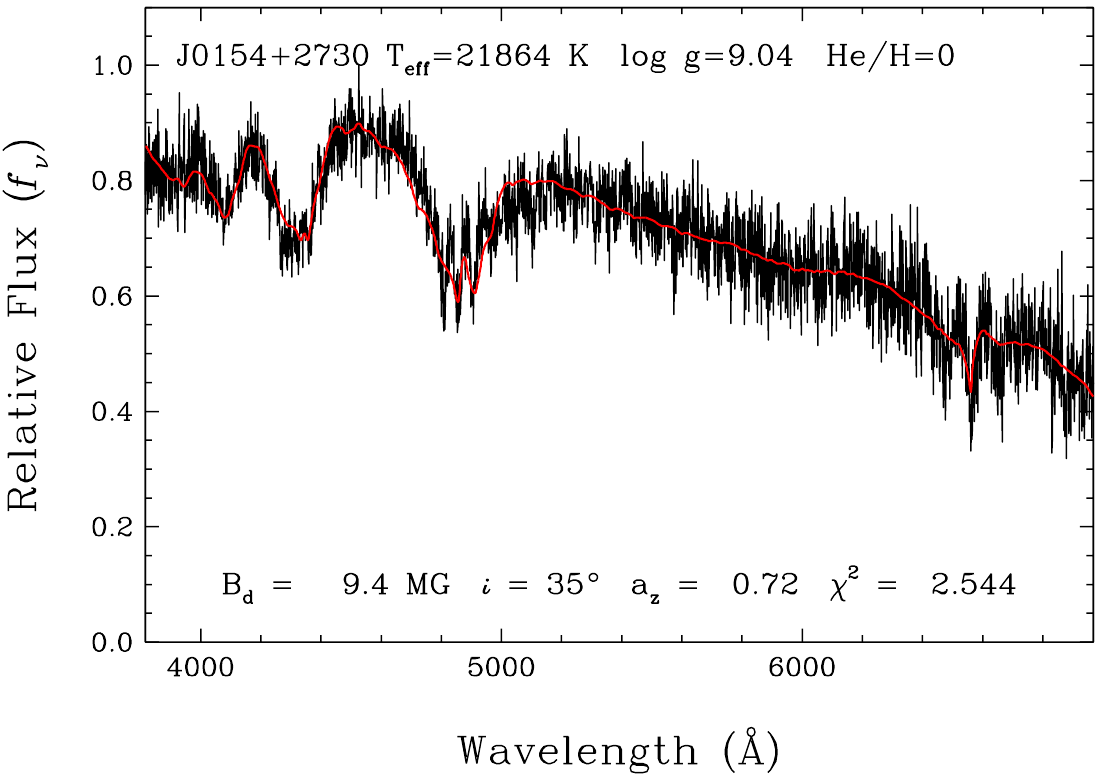}
\includegraphics[width=2.3in]{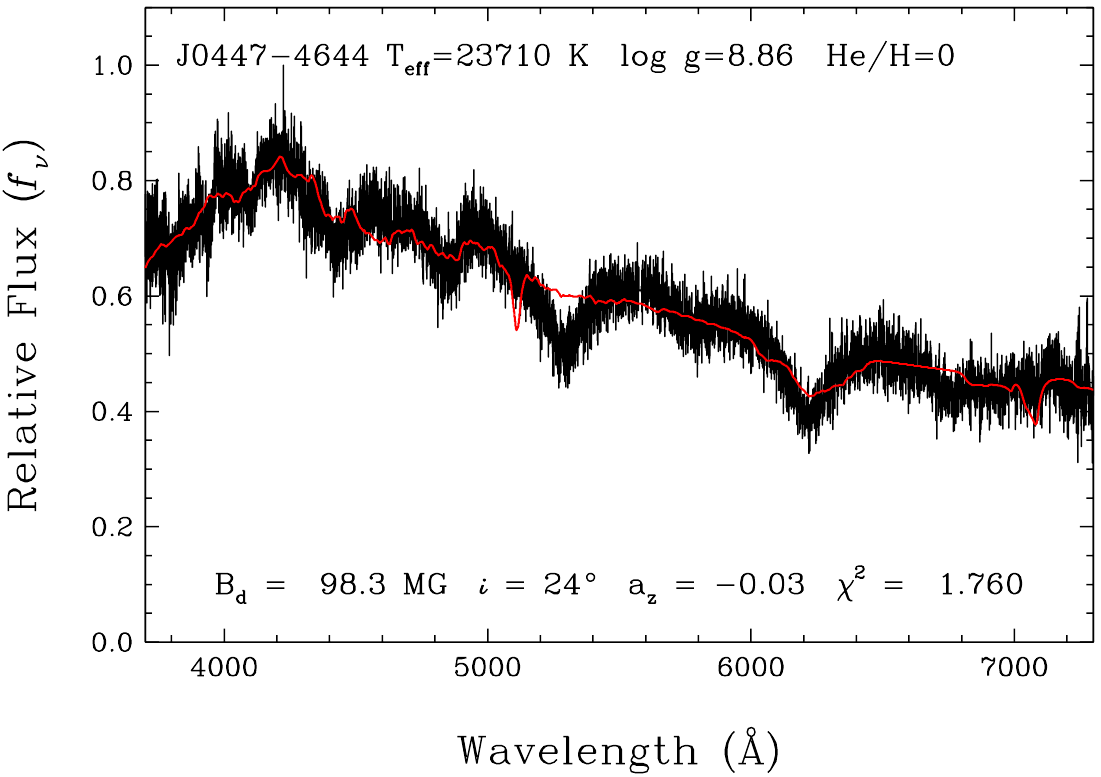}
\includegraphics[width=2.3in]{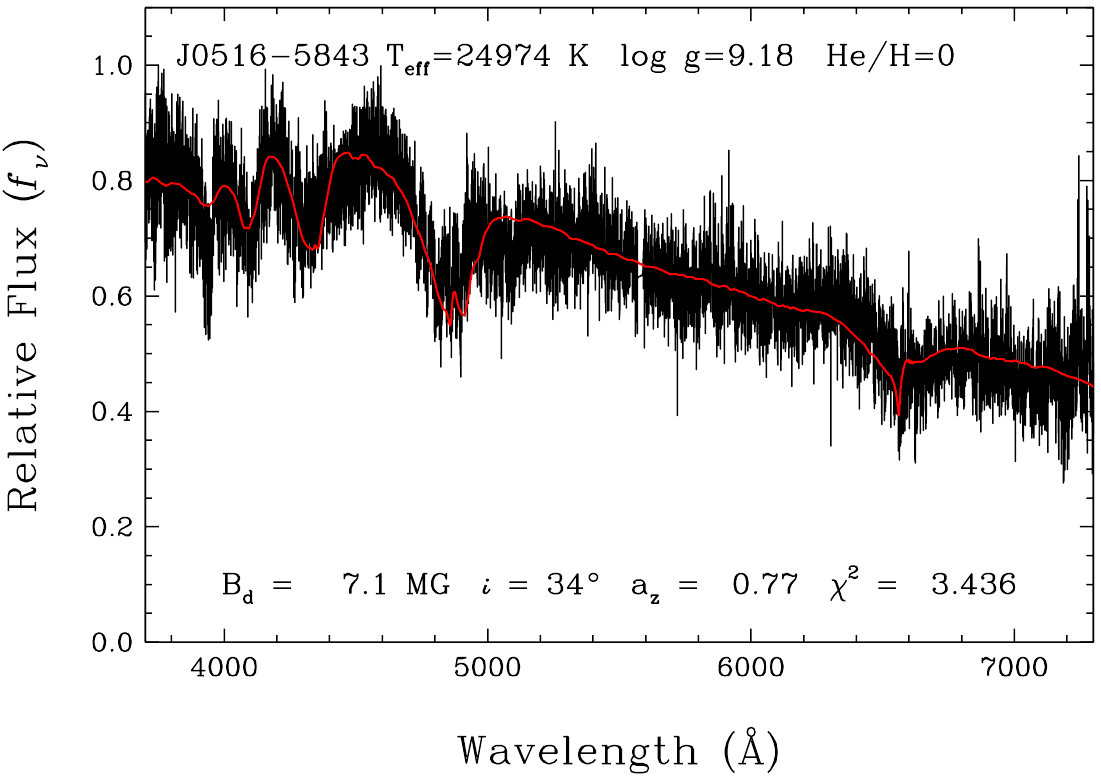}
\includegraphics[width=2.3in]{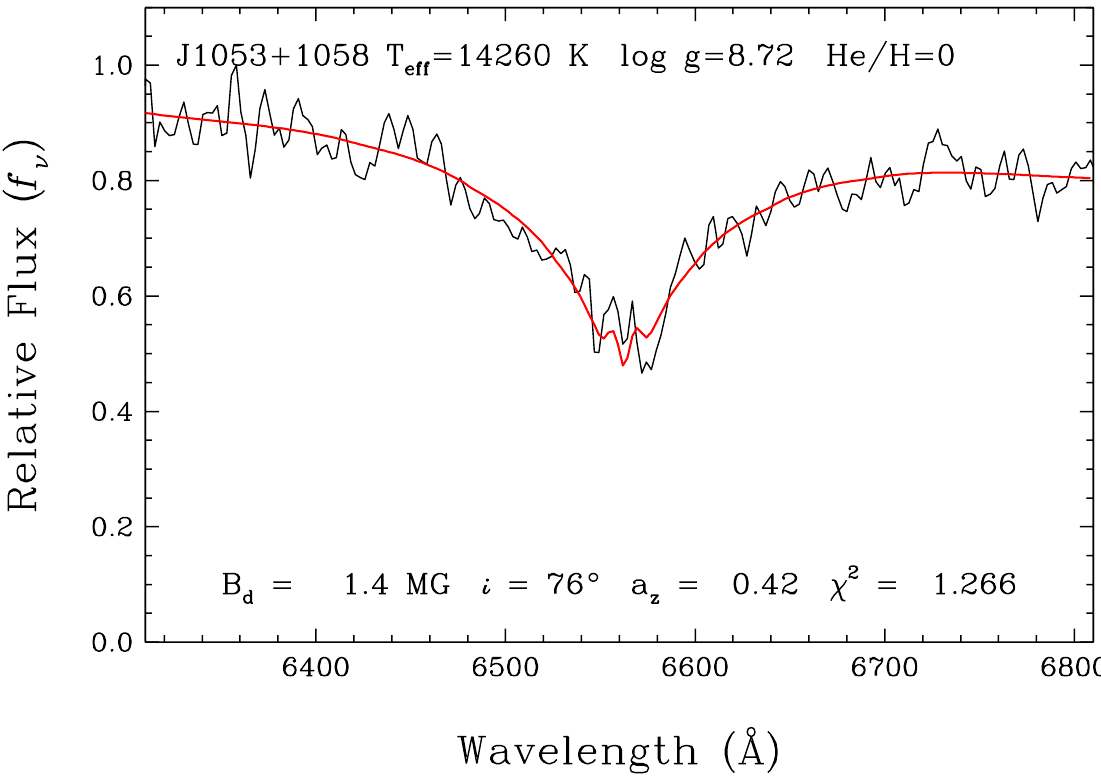}
\includegraphics[width=2.3in]{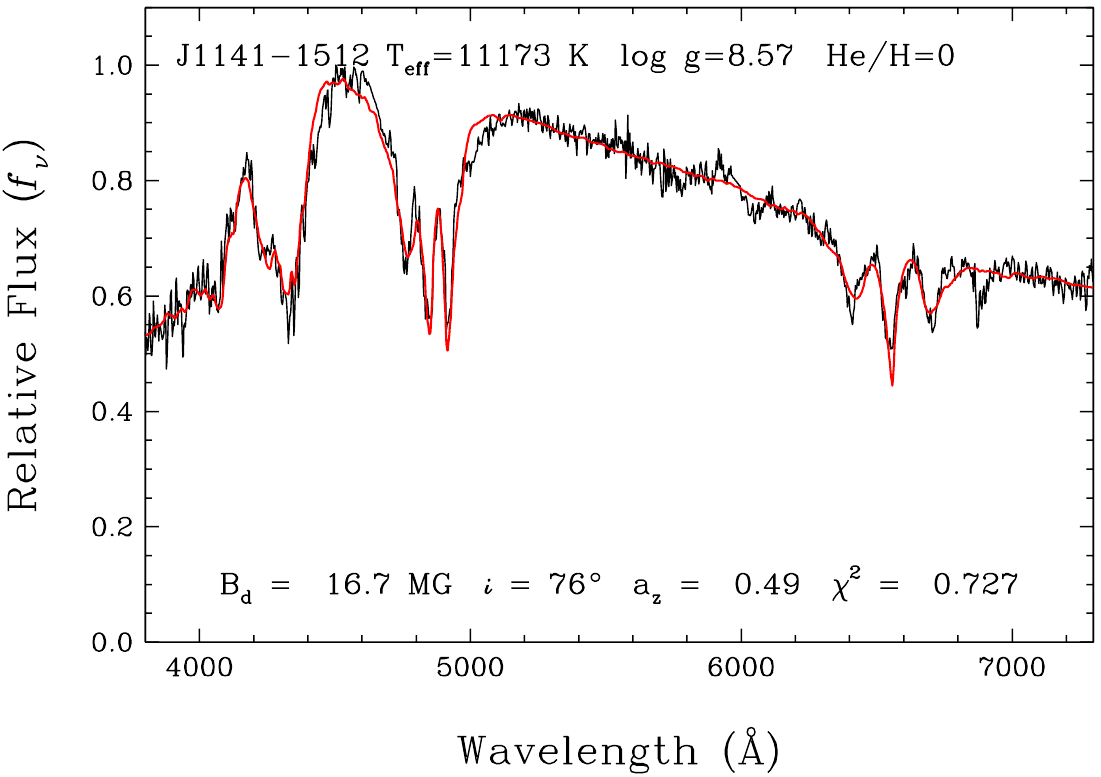}
\includegraphics[width=2.3in]{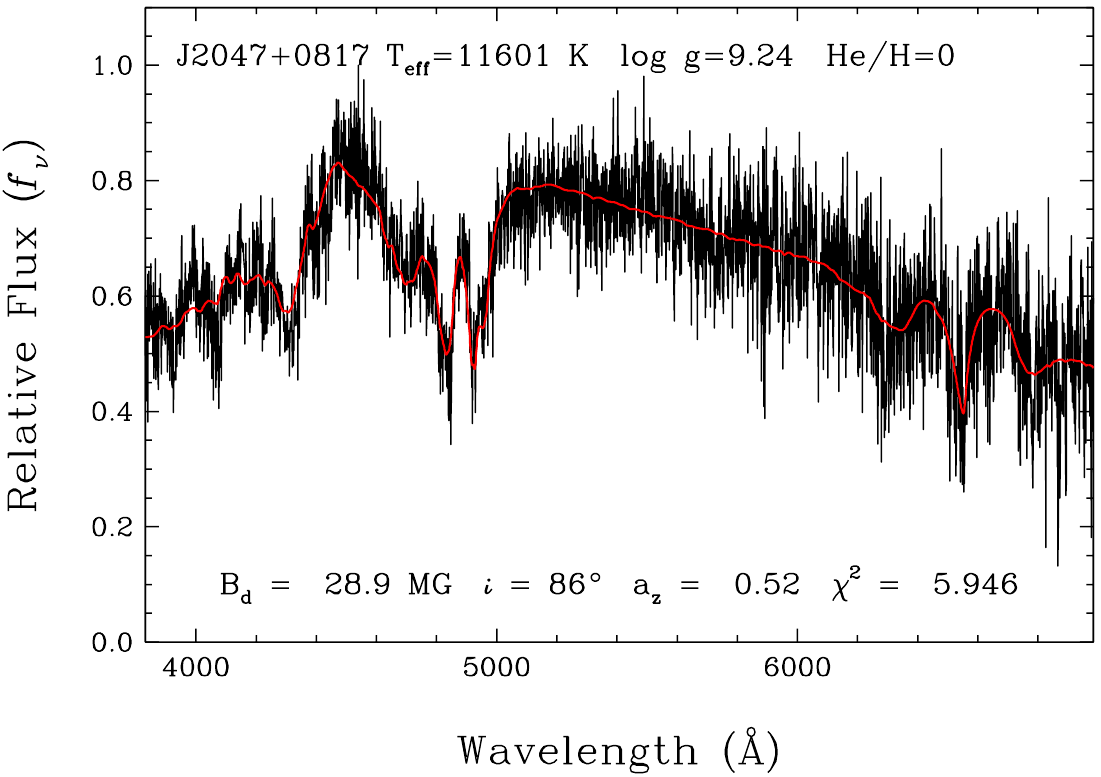}
\caption{Model atmosphere fits to six magnetic DAs.}
\label{figmagda} 
\end{figure*}

Figure \ref{figmagda} displays our model fits to six of the magnetic DAs with effective temperatures ranging from about 11,000 K to 25,000 K. 
Our magnetic models provide an excellent fit to five of these stars with $B_d\leq30$ MG. The fit to J0447$-$4644 is also good, indicating
a field strength $B_d\sim98$ MG. However, the model predicts additional features that are not observed (e.g., around 5100 \AA) and misses
a broad feature at 5300 \AA. Fitting the spectra of highly magnetic white dwarfs is challenging, as the lines are usually shifted and broadened beyond
recognition. Hence, the fit for J0447$-$4644 should be treated as tentative. 

We are not able to find reliable magnetic fits for the rest of the magnetic DAs in our sample, mostly because the spectral features are weak, noisy, or our adopted geometry is insufficient. Hence, we only show non-magnetic fits for those targets in the model fits provided on Zenodo. Note that there are also
two additional magnetic DXH white dwarfs in our sample, where the atmospheric composition is uncertain. It is possible that some of the magnetic ``DAH'' in
our sample may also have unusual compositions. Since we are not able to find good solutions for the majority of the magnetic white dwarfs in our sample,
those spectral classifications should be considered tentative.

\section{DB and DC White Dwarfs}

There are two DB white dwarfs in our sample, including the metal-rich DBZA J0738+1835 \citep{dufour12}. A detailed model atmosphere analysis of
J0738+1835 is beyond the scope of this paper. We assume a composition of $\log$ H/He = $-6$ \citep[adopted from][]{dufour12} in our fits for this star. 
The other DB, J2252+1308, does not show any evidence of hydrogen in its spectrum, and we assume a pure helium atmosphere composition.

There are seven DC white dwarfs in our sample with featureless spectra, one (J0158+0522) with $T_{\rm eff} = 13,072 \pm 372$ K and $M=1.249_{-0.027}^{+0.023}~M_{\odot}$,
and six with $M<1.0~M_{\odot}$ and effective temperatures between 9600 and 11,300 K. We use He-dominated atmospheres with trace amounts of
H ($\log$ H/He $=-5$) to fit the spectral energy distributions of these stars. A comparison between model fits using pure H and mixed H/He atmospheres
shows that the optical and UV photometry of all seven DCs are best explained by He-dominated atmospheres. Even though strongly magnetic DAs
can also appear as a DC above $T_{\rm eff}=12,000$ K, the UV photometry of J0158+0522 clearly favors the He-dominated solution. J0158+0522's
optical spectrum is relatively noisy. Hence, a higher quality spectrum would be helpful to confirm its nature.

\section{DZ White Dwarfs}

\begin{figure*}[h]
\includegraphics[width=2.3in]{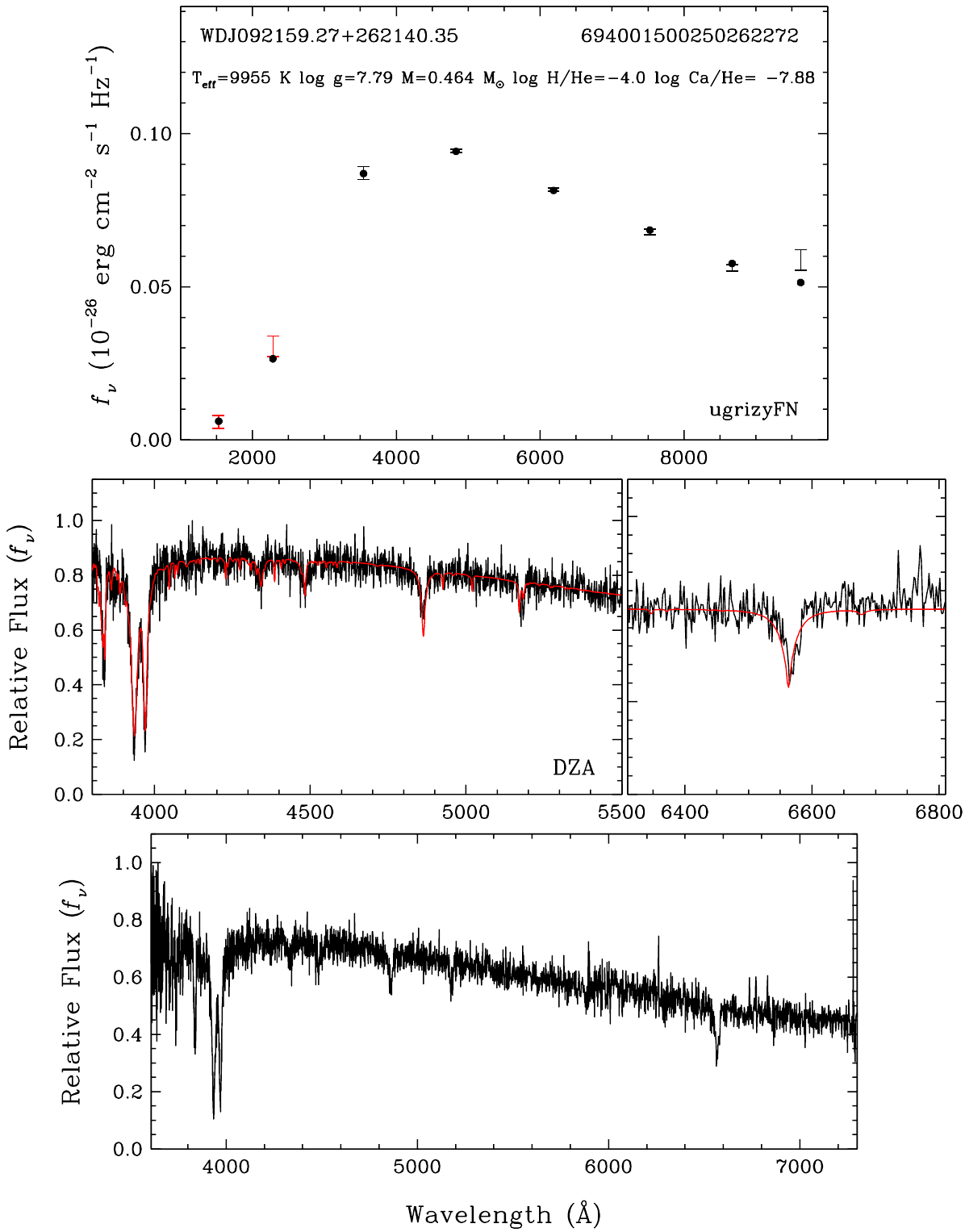}
\includegraphics[width=2.3in]{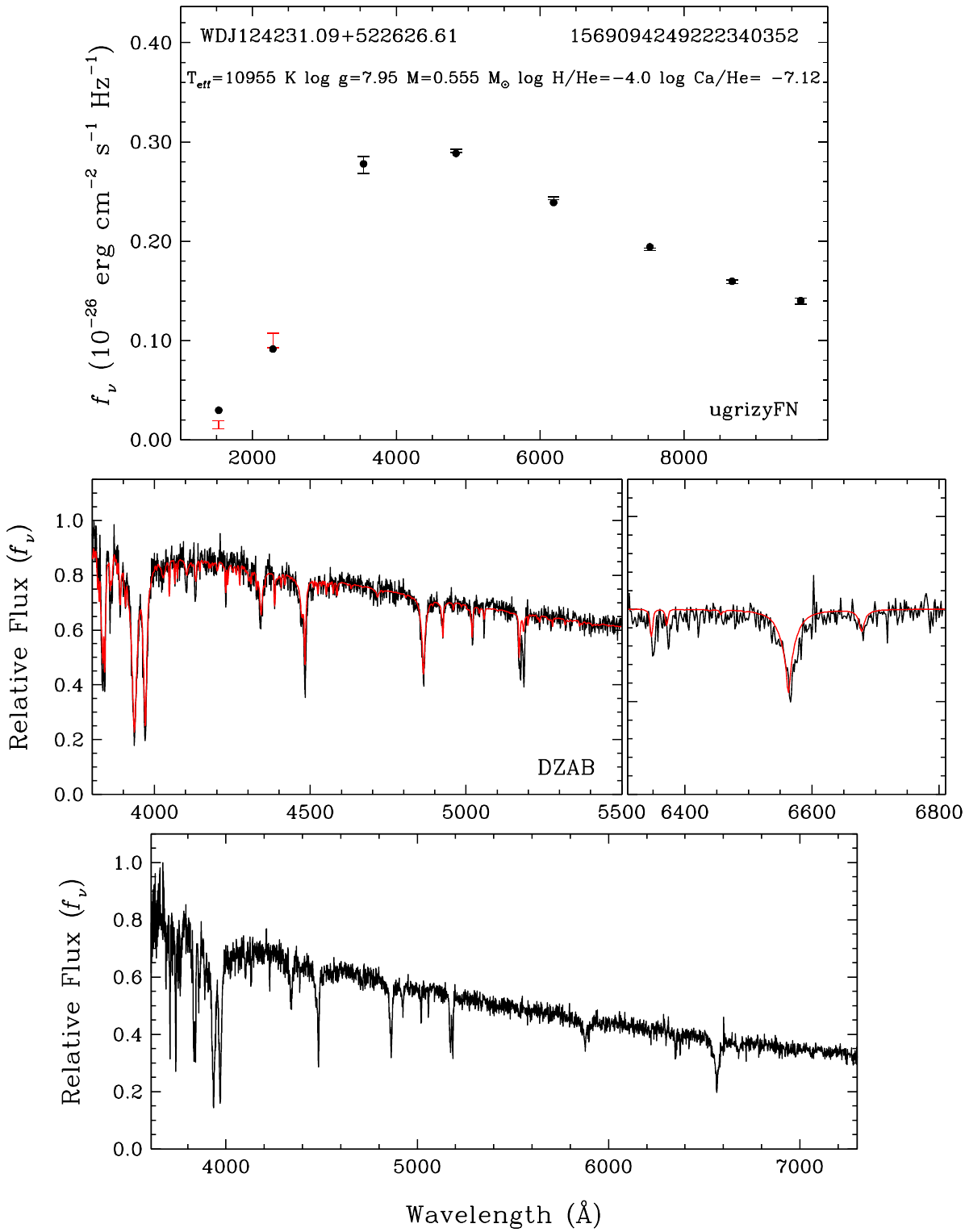}
\includegraphics[width=2.3in]{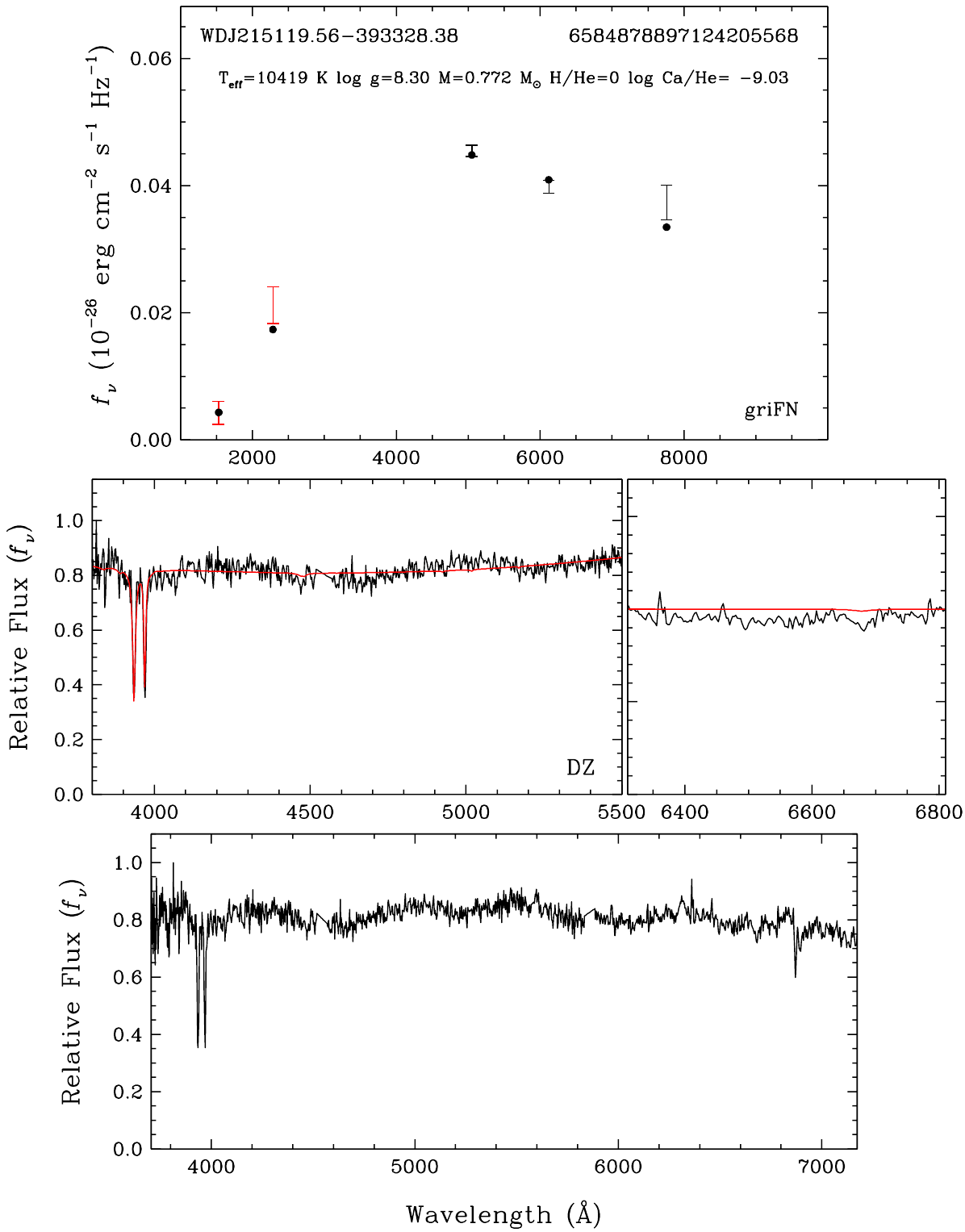}
\caption{Model atmosphere fits to three DZ white dwarfs. The top and middle panels show the photometric and spectroscopic model fits, respectively. The bottom panel shows a broader wavelength range for each star.}
\label{figdz} 
\end{figure*}

There are 6 DZ white dwarfs in our sample that display metal absorption lines, mainly \ion{Ca}{2} H and K. We rely on the
photometric technique to determine the temperature and surface gravity, and fit the blue portion of the spectrum
to constrain Ca/He. We scale the abundances of the other heavy elements to match the abundance ratios seen in CI chondrites. 
Three of the objects do not show an H$\alpha$ line; we use H-free atmosphere models for those stars. For the remaining three
stars, we use the H$\alpha$ region to constrain the H/He ratio. Figure \ref{figdz} shows the model fits to three of these stars with spectral types
DZA, DZAB, and DZ, respectively. Models
with $T_{\rm eff}=$ 9880-11,650 K and $\log$ Ca/He = $-9.0$ to $-7.1$ provide an excellent fit to the observed spectral energy
distributions of all six DZs. Remarkably, the models also provide an excellent match to the GALEX FUV and NUV
photometry, which is not used in the fits. This suggests that the accreted material in these stars likely has abundances similar to
CI chondrites.

\begin{deluxetable*}{lccccc}
\tabletypesize{\tiny}
\tablecolumns{6} \tablewidth{0pt}
\tablecaption{Physical Parameters for Other White Dwarfs.\label{tabparot}}
\tablehead{\colhead{Name} & \colhead{Comp} & \colhead{Spectral Type} & \colhead{$T_{\rm eff}$} & \colhead{Mass} & \colhead{$\log{g}$}\\
 &  & & (K) & ($M_{\odot}$) & (cm s$^{-2}$) }
\startdata
WDJ000732.26+331727.61   & H & DA & $56887 \pm 2302$ & 0.614$_{-0.012}^{+0.011}$ & 7.809$_{-0.017}^{+0.017}$ \\
WDJ012143.90$-$670335.46 & H & DA & $22972 \pm 4740$ & 1.279$_{-0.059}^{+0.054}$ & 9.214$_{-0.152}^{+0.199}$ \\
WDJ024730.78+332711.28   & H & DA & $11973 \pm 1077$ & 1.057$_{-0.149}^{+0.111}$ & 8.720$_{-0.188}^{+0.175}$ \\
WDJ030843.02+032811.98   & H & DA & $16026 \pm 788$ & 1.200$_{-0.060}^{+0.045}$ & 9.001$_{-0.111}^{+0.100}$ \\
WDJ050451.14$-$325111.42 & H & DA & $18108 \pm 5095$ & 1.304$_{-0.058}^{+0.069}$ & 9.309$_{-0.174}^{+0.273}$ \\
WDJ075455.50+164217.03   & H & DA & $15308 \pm 327$ & 0.482$_{-0.037}^{+0.039}$ & 7.750$_{-0.057}^{+0.057}$ \\
WDJ094711.69+372424.57   & H & DA & $11963 \pm 528$ & 1.091$_{-0.108}^{+0.077}$ & 8.781$_{-0.146}^{+0.127}$ \\
WDJ101823.73+072713.01   & H & DA & $9992 \pm 161$ & 0.579$_{-0.076}^{+0.069}$ & 7.964$_{-0.099}^{+0.082}$ \\
WDJ105840.08$-$000503.01 & H & DA & $12037 \pm 655$ & 1.175$_{-0.107}^{+0.075}$ & 8.949$_{-0.173}^{+0.161}$ \\
WDJ105930.78+673404.52   & H & DA & $14028 \pm 1465$ & 1.137$_{-0.065}^{+0.063}$ & 8.867$_{-0.101}^{+0.115}$ \\
WDJ140443.25$-$164706.48 & H & DA & $18815 \pm 360$ & 0.830$_{-0.091}^{+0.06}$ & 8.341$_{-0.106}^{+0.071}$ \\
WDJ145344.87$-$111454.73 & H & DA & $20927 \pm 2290$ & 1.302$_{-0.083}^{+0.042}$ & 9.295$_{-0.227}^{+0.173}$ \\
WDJ153907.93+495346.18   & H & DA & $10923 \pm 287$ & 0.749$_{-0.080}^{+0.076}$ & 8.235$_{-0.09}^{+0.086}$ \\
WDJ184606.54+513815.50   & H & DA & $11426 \pm 265$ & 0.943$_{-0.046}^{+0.049}$ & 8.534$_{-0.055}^{+0.061}$ \\
WDJ194419.87$-$522422.08 & H & DA & $9807 \pm 712$ & 0.513$_{-0.083}^{+0.110}$ & 7.845$_{-0.119}^{+0.138}$ \\ 
WDJ194936.96$-$385101.59 & H & DA & $16441 \pm 3313$ & 1.015$_{-0.132}^{+0.132}$ & 8.641$_{-0.163}^{+0.197}$ \\
WDJ201534.41+033848.20   & H & DA & $11570 \pm 1151$ & 1.147$_{-0.327}^{+0.165}$ & 8.889$_{-0.425}^{+0.401}$ \\
WDJ232324.81+122126.13   & H & DA & $11352 \pm 557$ & 1.075$_{-0.377}^{+0.139}$ & 8.754$_{-0.452}^{+0.239}$ \\
WDJ233402.19$-$471426.65 & H & DA & $37388 \pm 1430$ & 0.358$_{-0.013}^{+0.013}$ & 7.204$_{-0.027}^{+0.029}$ \\
\hline
WDJ091119.49$-$110718.64 & [H/He]=$-$4.0 & He-DA & $12037 \pm 364$ & 1.240$_{-0.017}^{+0.016}$ & 9.130$_{-0.040}^{+0.040}$ \\
WDJ100018.29+304730.18 & [H/He]=$-$3.5 & He-DA & $10609 \pm 173$ & 0.942$_{-0.042}^{+0.043}$ & 8.558$_{-0.048}^{+0.053}$ \\
WDJ100433.65+401430.12 & [H/He]=$-$5.0 & He-DA & $11247 \pm 212$ & 0.655$_{-0.027}^{+0.029}$ & 8.118$_{-0.031}^{+0.033}$ \\
\hline
WDJ073842.57+183509.71   & [H/He]=$-$6.00 & DBZA & $14182 \pm 247$ & 0.625$_{-0.027}^{+0.029}$ & 8.062$_{-0.032}^{+0.034}$ \\
WDJ225206.13+130831.19   & He & DB & $11974 \pm 261$ & 0.504$_{-0.052}^{+0.057}$ & 7.855$_{-0.073}^{+0.074}$ \\
\hline
WDJ015810.54+052201.64   & [H/He]=$-$5.0 & DC & $13072 \pm 372$ & 1.249$_{-0.027}^{+0.023}$ & 9.153$_{-0.062}^{+0.064}$ \\
WDJ021553.69$-$155542.84 & [H/He]=$-$5.0 & DC & $9664 \pm 247$ & 0.575$_{-0.055}^{+0.055}$ & 7.992$_{-0.069}^{+0.064}$ \\
WDJ130606.76+592610.28   & [H/He]=$-$5.0 & DC & $10677 \pm 159$ & 0.572$_{-0.02}^{+0.021}$ & 7.983$_{-0.025}^{+0.026}$ \\
WDJ135959.08$-$083843.21 & [H/He]=$-$5.0 & DC & $11292 \pm 242$ & 0.549$_{-0.079}^{+0.068}$ & 7.940$_{-0.104}^{+0.083}$ \\
WDJ185033.69+645255.61   & [H/He]=$-$5.0 & DC & $10244 \pm 122$ & 0.479$_{-0.023}^{+0.026}$ & 7.816$_{-0.033}^{+0.034}$ \\
WDJ211401.28$-$004907.31 & [H/He]=$-$5.0 & DC & $10876 \pm 354$ & 0.776$_{-0.248}^{+0.162}$ & 8.306$_{-0.288}^{+0.183}$ \\
WDJ215413.55+070026.11   & [H/He]=$-$5.0 & DC & $9831 \pm 393$ & 0.978$_{-0.224}^{+0.122}$ & 8.616$_{-0.255}^{+0.165}$ \\
\hline
WDJ080137.97+100759.97   & [Ca/He]=$-$7.76 [H/He]=$-$5.0 & DZ & $11646 \pm 264$ & 0.753$_{-0.063}^{+0.066}$ & 8.269$_{-0.07}^{+0.074}$ \\
WDJ092159.27+262140.35   & [Ca/He]=$-$7.88 [H/He]=$-$4.0 & DZA & $9955 \pm 171$ & 0.464$_{-0.077}^{+0.081}$ & 7.789$_{-0.117}^{+0.106}$ \\
WDJ124231.09+522626.61   & [Ca/He]=$-$7.12 [H/He]=$-$4.0 & DZAB & $10955 \pm 205$ & 0.555$_{-0.021}^{+0.02}$ & 7.952$_{-0.026}^{+0.026}$ \\
WDJ140820.06+153507.11   & [Ca/He]=$-$8.17  & DZ & $10545 \pm 177$ & 0.706$_{-0.041}^{+0.044}$ & 8.201$_{-0.047}^{+0.048}$ \\
WDJ151245.42+364739.48   & [Ca/He]=$-$8.90  & DZ & $9876 \pm 234$ & 0.564$_{-0.109}^{+0.101}$ & 7.973$_{-0.144}^{+0.118}$ \\
WDJ215119.56$-$393328.38 & [Ca/He]=$-$9.03  & DZ & $10419 \pm 1256$ & 0.772$_{-0.16}^{+0.185}$ & 8.302$_{-0.181}^{+0.209}$ \\
\hline
WDJ011528.38+434858.75   & H & DAH & $21813 \pm 1036$ & 1.147$_{-0.064}^{+0.054}$ & 8.876$_{-0.102}^{+0.103}$ \\
WDJ015400.97+273017.89   & H & DAH & $21864 \pm 928$ & 1.218$_{-0.049}^{+0.039}$ & 9.038$_{-0.098}^{+0.094}$ \\
WDJ021409.34$-$121202.14 & H & DAH & $31160 \pm 1076$ & 1.241$_{-0.014}^{+0.013}$ & 9.089$_{-0.032}^{+0.035}$ \\ 
WDJ042043.24+645011.13   & H & DAH & $26009 \pm 615$ & 0.964$_{-0.021}^{+0.022}$ & 8.540$_{-0.026}^{+0.029}$ \\
WDJ044733.02$-$464402.77 & H & DAH & $23710 \pm 2038$ & 1.141$_{-0.049}^{+0.048}$ & 8.862$_{-0.079}^{+0.089}$ \\
WDJ051610.94$-$584338.11 & H & DAH & $24974 \pm 2372$ & 1.269$_{-0.041}^{+0.036}$ & 9.179$_{-0.104}^{+0.116}$ \\
WDJ063134.77$-$310921.58 & H & DAH & $28245 \pm 911$ & 1.223$_{-0.015}^{+0.015}$ & 9.044$_{-0.033}^{+0.036}$ \\
WDJ074531.24+062150.67   & H & DAH & $21959 \pm 965$ & 1.013$_{-0.027}^{+0.028}$ & 8.629$_{-0.036}^{+0.038}$ \\
WDJ075450.39+372757.75   & H & DAH & $17895 \pm 710$ & 1.267$_{-0.059}^{+0.027}$ & 9.178$_{-0.142}^{+0.082}$ \\
WDJ082247.61+120146.85   & H & DXH & $19768 \pm 627$ & 1.035$_{-0.063}^{+0.053}$ & 8.669$_{-0.082}^{+0.077}$ \\
WDJ090201.07+511111.72   & H & DAH & $25693 \pm 1385$ & 1.269$_{-0.036}^{+0.032}$ & 9.177$_{-0.090}^{+0.101}$ \\
WDJ090632.65+080716.16   & H & DAH & $16977 \pm 442$ & 1.201$_{-0.018}^{+0.014}$ & 9.001$_{-0.034}^{+0.03}$ \\
WDJ105328.55+105801.70   & H & DAH & $14260 \pm 525$ & 1.059$_{-0.132}^{+0.092}$ & 8.721$_{-0.170}^{+0.142}$ \\
WDJ113515.63$-$155350.59 & H & DAH & $16491 \pm 779$ & 1.124$_{-0.048}^{+0.041}$ & 8.838$_{-0.074}^{+0.071}$ \\
WDJ114154.92$-$151231.64 & H & DAH & $11173 \pm 594$ & 0.966$_{-0.13}^{+0.125}$ & 8.571$_{-0.153}^{+0.167}$ \\
WDJ130604.67$-$231126.08 & H & DAH & $24840 \pm 685$ & 1.181$_{-0.027}^{+0.024}$ & 8.948$_{-0.050}^{+0.048}$ \\
WDJ145558.39+181252.36   & H & DXH & $25179 \pm 772$ & 1.236$_{-0.038}^{+0.025}$ & 9.082$_{-0.084}^{+0.063}$ \\
WDJ145725.14$-$232854.86 & H & DAH & $26224 \pm 893$ & 1.145$_{-0.028}^{+0.025}$ & 8.868$_{-0.047}^{+0.045}$ \\
WDJ170802.58+542017.65   & H & DAH & $20323 \pm 1759$ & 0.949$_{-0.053}^{+0.053}$ & 8.526$_{-0.064}^{+0.067}$ \\
WDJ171920.03$-$144639.09 & H & DAH? & $16898 \pm 155$ & 1.164$_{-0.004}^{+0.005}$ & 8.919$_{-0.008}^{+0.009}$ \\
WDJ172721.29+235142.75   & H & DAH & $26471 \pm 1112$ & 1.229$_{-0.042}^{+0.032}$ & 9.064$_{-0.092}^{+0.079}$ \\
WDJ180042.39+472725.42   & H & DAH & $17016 \pm 746$ & 1.034$_{-0.067}^{+0.048}$ & 8.673$_{-0.088}^{+0.067}$ \\
WDJ190501.71$-$394757.34 & H & DAH & $21021 \pm 742$ & 1.126$_{-0.019}^{+0.019}$ & 8.836$_{-0.032}^{+0.031}$ \\
WDJ204728.93+081722.66   & H & DAH & $11601 \pm 728$ & 1.285$_{-0.143}^{+0.054}$ & 9.243$_{-0.316}^{+0.203}$ \\
WDJ211952.10$-$623758.85 & H & DAH & $16283 \pm 1999$ & 1.087$_{-0.072}^{+0.073}$ & 8.768$_{-0.101}^{+0.118}$ \\
WDJ224459.60+331017.37   & H & DAH & $20938 \pm 885$ & 1.201$_{-0.035}^{+0.028}$ & 8.998$_{-0.068}^{+0.060}$ \\
\enddata
\end{deluxetable*}

\end{document}